%% file: Main.tex
\RequirePackage[l2tabu,orthodox]{nag}		% for deprecated commands
\documentclass[reqno]{amsart}
\usepackage[margin=1.5in,bottom=1.25in]{geometry}		% for tighter margins (80 CPL)

\usepackage{amsmath}		% for AMS macros
\usepackage{amssymb}		% for AMS symbols
\usepackage{amsfonts}		% for AMS fonts
\usepackage{amsthm}		% for theorems
\usepackage[foot]{amsaddr}		% for author footnotes

\usepackage{mathtools}		% for advanced math
\mathtoolsset{%
}

\usepackage[utf8]{inputenc}		% for source encoding
\usepackage[T1]{fontenc}		% for font encoding

\usepackage[proportional,tabular,lining,sf,mono=false]{libertine}
\usepackage{dsfont}		% for blackboard bold font
\usepackage[%		% for math font selection
cal=cm,
]
{mathalfa}

\usepackage[labelfont={bf,small},labelsep=colon,font=small]{caption}	% for caption control
\usepackage[dvipsnames,svgnames]{xcolor}		% for color
\colorlet{MyBlue}{MediumBlue!80!Black}
\colorlet{MyGreen}{DarkGreen!85!Black}
\colorlet{MyRed}{Crimson!80!Black}

\newcommand{\afterhead}{.}		% for changing headings
\newcommand{\para}[1]{\medskip\paragraph{\textbf{#1\afterhead}}}

\usepackage{wasysym}		% for extra symbols

\usepackage{subcaption}		% for subfigures
\usepackage{tikz}		% for figures
\usetikzlibrary{calc,patterns}

\usepackage{array}		% for flexible tables
\usepackage{booktabs}		% for better tables
\usepackage[inline,shortlabels]{enumitem}		% for list control
\setenumerate{itemsep=\smallskipamount,topsep=\medskipamount}

\usepackage{microtype}		% for microtypography

\usepackage{acronym}		% for acronyms
\usepackage{latexsym}		% for symbols
\usepackage{xspace}		% for flexible spaces

\usepackage[numbers,sort&compress]{natbib}		% for citations

\bibpunct[, ]{[}{]}{,}{n}{,}{,}

\usepackage{hyperref}
\hypersetup{
colorlinks=true,
linktocpage=true,
pdfstartview=FitH,
breaklinks=true,
pdfpagemode=UseNone,
pageanchor=true,
pdfpagemode=UseOutlines,
plainpages=false,
bookmarksnumbered,
bookmarksopen=false,
bookmarksopenlevel=1,
hypertexnames=true,
pdfhighlight=/O,
urlcolor=MyBlue,linkcolor=MyBlue,citecolor=MyBlue,	% for on-screen
pdftitle={},
pdfauthor={},
pdfsubject={},
pdfkeywords={},
pdfcreator={pdfLaTeX},
pdfproducer={LaTeX with hyperref}
}

\newcommand{\EMAIL}[1]{\email{\href{mailto:#1}{#1}}}

\usepackage[sort&compress,capitalize,nameinlink]{cleveref}		% for cleveref formatting
\crefname{assumption}{Assumption}{Assumptions}
\crefformat{footnote}{#2\footnotemark[#1]#3}		% for multiple footnotes

\usepackage[textwidth=30mm]{todonotes}		% for todo's
\newcommand{\debug}[1]{#1}		% for removing macro coloring
\newcommand{\revise}[1]{#1}		% for revision markup

\def\beginrev{}		% for revision markup
\def\endrev{}		% for revision markup

\usepackage{algorithm}		% for algorithm environments
\usepackage{algpseudocode}		% for algorithm macros
\theoremstyle{plain}
\newtheorem{theorem}{Theorem}		% for theorems
\newtheorem{corollary}{Corollary}		% for corollaries
\newtheorem{lemma}{Lemma}		% for lemmas
\newtheorem{proposition}{Proposition}		% for propositions

\newtheorem*{corollary*}{Corollary}		% for corollaries (unnumbered)

\theoremstyle{definition}
\newtheorem{definition}{Definition}		% for definitions
\newtheorem{assumption}{Assumption}		% for assumptions
\newtheorem{example}{Example}		% for examples

\newtheorem*{definition*}{Definition}		% for definitions (unnumbered)
\newtheorem*{assumption*}{Assumptions}		% for assumptions (unnumbered)
\newtheorem*{example*}{Example}		% for examples (unnumbered)

\theoremstyle{remark}
\newtheorem*{remark*}{Remark}		% for remarks (unnumbered)

\newcommand{\envend}{\hfill{\small\S}}

\newenvironment{Proof}[1][Proof]{\begin{proof}[#1]}{\end{proof}}
\newcounter{proofpart}

\numberwithin{example}{section}		% for example numbering

\newcommand{\newmacro}[2]{\newcommand{#1}{\debug{#2}}}		% for shorthand definitions
\newcommand{\newop}[2]{\DeclareMathOperator{#1}{\debug{#2}}}		% for shorthand definitions

\DeclarePairedDelimiter{\braces}{\{}{\}}		% for braces
\DeclarePairedDelimiter{\bracks}{[}{]}		% for brackets
\DeclarePairedDelimiter{\parens}{(}{)}		% for parentheses

\DeclarePairedDelimiter{\abs}{\lvert}{\rvert}		% for absolute value
\DeclarePairedDelimiter{\ceil}{\lceil}{\rceil}		% for ceiling
\DeclarePairedDelimiterX{\setdef}[2]{\{}{\}}{#1:#2}		% for set builder notation
\DeclarePairedDelimiterX{\window}[2]{[}{]}{#1\,.\,.\,#2}		% for set builder notation
\DeclarePairedDelimiterXPP{\exclude}[1]{\mathopen{}\setminus}{\{}{\}}{}{#1}

\newcommand{\cf}{cf.\xspace}		% for consistency
\newcommand{\eg}{e.g.,\xspace}		% for consistency
\newcommand{\ie}{i.e.,\xspace}		% for consistency

\newcommand{\textpar}[1]{\textup(#1\textup)}		% for upshape parentheses

\newcommand{\txs}{\textstyle}		% for forcing inline style

\newcommand{\alt}[1]{#1'}		% for alternates

\newcommand{\N}{\mathbb{N}}		% for naturals
\newcommand{\R}{\mathbb{R}}		% for reals
\DeclareMathOperator{\bigoh}{\mathcal O}		% for Landau O
\newcommand{\dd}{\:d}		% for integrators
\newcommand{\eps}{\varepsilon}		% for better epsilon
\DeclareMathOperator*{\intersect}{\bigcap}		% for intersections
\DeclareMathOperator*{\union}{\bigcup}		% for unions

\DeclareMathOperator{\dom}{dom}		% for domain
\newcommand{\defeq}{\coloneqq}		% for direct definition
\newcommand{\from}{\colon}		% for function definition
\newmacro{\set}{\mathcal{S}}		% for generic set

\newmacro{\points}{\mathcal{K}}		% for point set
\newmacro{\intpoints}{\points^{\circ}}		%for point set interior
\newmacro{\point}{x}		% for generic point
\newmacro{\pointalt}{\alt\point}		% for alternate point

\newmacro{\dpoints}{\mathcal{Y}}		% for second point set (duals, etc.)
\newmacro{\dpoint}{y}		% for second generic point
\newmacro{\dpointalt}{\alt\dpoint}		% for second alternate variable

\newmacro{\base}{p}		% for base point
\newmacro{\basealt}{q}		% for alternate base point

\newcommand{\test}[1][\point]{\hat{#1}}		% for test point (\point by default)
\DeclareMathOperator{\bd}{bd}		% for boundary
\DeclareMathOperator{\diam}{diam}		% for diameter
\DeclareMathOperator{\dist}{dist}		% for distance
\DeclareMathOperator{\intr}{int}		% for interior

\newmacro{\region}{\mathcal{D}}		% for open sets
\newmacro{\open}{\mathcal{U}}		% for open sets
\newmacro{\cpt}{\mathcal{C}}		% for compacts
\newmacro{\nhd}{\mathcal{U}}		% for neighborhoods

\newmacro{\start}{1}		% for start index
\newmacro{\runstart}{\tau_{\mathrm{start}}}		% for running start
\newmacro{\runend}{\tau_{\mathrm{end}}}		% for running start
\newmacro{\running}{1,2,\dotsc}		% for running index
\newmacro{\run}{t}		% for main sequence index
\newmacro{\runalt}{s}		% for alternate sequence index
\newmacro{\nRuns}{T}		% for total number of runs
\newmacro{\runtime}{\Gamma}		% for runtime

\newmacro{\runs}{\mathcal{\nRuns}}		% for set of indices

\newcommand{\new}[1]{#1^{+}}		% for new iterate

\newmacro{\state}{X}		% for main state
\newmacro{\dstate}{Y}		% for dual state
\newmacro{\aux}{\tilde\state}		% for auxiliary state

\newmacro{\step}{\gamma}		% for step-size
\newmacro{\stepgap}{\chi}		% for step-size
\newmacro{\coef}{\lambda}		% for coefficient

\newmacro{\vecspace}{\mathcal{X}}		% for generic vector space
\newmacro{\vdim}{d}		% for dimension
\newmacro{\coord}{k}		% for coordinate index
\newmacro{\bvec}{e}		% for basis vectors
\newmacro{\bvecs}{\mathcal{E}}		% for basis vectors

\newmacro{\subspace}{\mathcal{W}}		% for subspace
\newmacro{\tanspace}{\mathcal{Z}}		% for tangent space
\newmacro{\tanvec}{z}		% for tangent vectors

\DeclarePairedDelimiterX{\braket}[2]{\langle}{\rangle}{#1,#2}		% for duality pairing
\newcommand{\dual}[1]{#1^{\ast}}		% for dual variables

\newmacro{\dspace}{\mathcal{Y}}		% for dual space
\newmacro{\dvec}{v}		% for dual basis vectors
\newmacro{\dbvec}{\eps}		% for dual basis vectors

\newmacro{\ones}{\mathbf{1}}		% for vector of ones
\newmacro{\mat}{M}		% for generic matrix

\DeclareMathOperator{\relint}{ri}		% for relative interior

\newmacro{\cvx}{\mathcal{C}}		% for generic convex set
\newmacro{\subd}{\partial}		% for subdifferential

\newmacro{\strong}{\mu}		% for convexity modulus
\DeclareMathOperator{\Eucl}{\Pi}		% for Euclidean projection
\newmacro{\hreg}{h}		% for regularizer
\newmacro{\breg}{D}		% for Bregman divergence
\newmacro{\proxmap}{\mathcal{P}}		% for mirror map
\newmacro{\mirror}{Q}		% for mirror map
\newmacro{\fench}{F}		% for Fenchel coupling
\newmacro{\hstr}{K}		% for strong convexity constant
\newmacro{\depth}{H}		% for regularizer depth
\newmacro{\zone}{\mathbb{D}}		% for Bregman zone

\newmacro{\subpoints}{\points_{\hreg}}		% for prox-domain
\newmacro{\subcvx}{\cvx_{\hreg}}		% for prox-domain

\DeclarePairedDelimiterXPP{\prox}[2]{\proxmap}{(}{)}{}{#1;#2}		% for prox-mapping
\DeclarePairedDelimiterXPP{\proxplay}[2]{\proxmap_{\play}}{(}{)}{}{#1;#2}		% for prox-mapping

\DeclareMathOperator*{\argmax}{arg\,max}		% for argmax
\DeclareMathOperator*{\argmin}{arg\,min}		% for argmin

\newop{\Opt}{Opt}		% for value of problem
\newop{\Sol}{Sol}		% for solution of problem
\newop{\gap}{Gap}		% for gap function
\newop{\err}{err} 	% for error

\newmacro{\obj}{f}		% for objective function
\newmacro{\sobj}{F}		% for stochastic objective
\newmacro{\gvec}{g}		% for gradient vector
\newmacro{\gbound}{G}		% for gradient bound

\newmacro{\oper}{A}		% for operator
\newmacro{\vecfield}{v}		% for vector field
\newmacro{\vbound}{G}		% for field bound
\newmacro{\lips}{L}		% for Lipschitz constant

\newcommand{\sol}[1][\point]{#1^{\ast}}		% for solutions (x by default)
\newcommand{\sols}{\sol[\points]}		% for set of solutions

\newmacro{\minmax}{\Phi}		% for minmax objective

\newmacro{\minvar}{\point_{1}}		% for minimization variable
\newmacro{\minvaralt}{\alt\minvar}		% for alternate minvar
\newmacro{\minvars}{\points_{1}}		% for minvar space

\newmacro{\maxvar}{\point_{2}}		% for maximization variable
\newmacro{\maxvars}{\points_{2}}		% for maxvar space
\newmacro{\maxvaralt}{\alt\maxvar}		% for alternate maxvar

\newop{\Nash}{NE}		% for Nash equilibrium
\newop{\BR}{BR}		% for best responses
\newop{\reg}{Reg}		% for regret
\newop{\val}{val}		% for value of game

\newcommand{\eq}{\sol}		% for Nash equilibria
\newcommand{\eqs}{\sols}		% for Nash equilibria

\newmacro{\play}{i}		% for main player index
\newmacro{\playalt}{j}		% for alternate player index
\newmacro{\playaltalt}{k}		% for alternate player index
\newmacro{\nPlayers}{N}		% for number of players
\newmacro{\players}{\mathcal{\nPlayers}}		% for set of players

\newmacro{\pure}{a}		% for main strategy index
\newmacro{\purealt}{\beta}		% for alternate strategy index
\newmacro{\nPures}{A}		% for number of strategies
\newmacro{\pures}{\mathcal{\nPures}}		% for set of strategies

\newmacro{\cost}{c}		% for cost function
\newmacro{\loss}{\ell}		% for loss function
\newmacro{\pay}{u}		% for payoff function
\newmacro{\payv}{v}		% for payoff vector
\newmacro{\pot}{F}		% for potential function

\newmacro{\game}{\mathcal{G}}		% for game
\newmacro{\fingame}{\Gamma}		% for finite game

\DeclareMathOperator{\ex}{\mathbb{E}}		% for expectations
\DeclareMathOperator{\prob}{\mathbb{P}}		% for probability
\DeclareMathOperator{\simplex}{\Delta}		% for simplices

\newmacro{\sample}{\omega}		% for samples
\newmacro{\samples}{\Omega}		% for sample space
\newmacro{\filter}{\mathcal{F}}		% for filtrations
\newmacro{\probspace}{(\samples,\filter,\prob)}		% for probability space

\newmacro{\mean}{\mu}		% for mean of distribution
\newmacro{\sdev}{\sigma}		% for mean of distribution
\newmacro{\variance}{\sdev^{2}}		% for mean of distribution

\newmacro{\dkl}{D_{\mathrm{KL}}}		% for Kullback Leibler
\newcommand{\as}{\textpar{a.s.}\xspace}		% for almost surely

\providecommand\given{}		% empty command for conditionals

\DeclarePairedDelimiterXPP{\exof}[1]{\ex}{[}{]}{}{%		% for conditional expectations
\renewcommand\given{\nonscript\:\delimsize\vert\nonscript\:\mathopen{}} #1}

\DeclarePairedDelimiterXPP{\exofsample}[1]{\ex_{\samples}}{[}{]}{}{%		% for conditional expectations
\renewcommand\given{\nonscript\:\delimsize\vert\nonscript\:\mathopen{}} #1}

\DeclarePairedDelimiterXPP{\probof}[1]{\prob}{(}{)}{}{%		% for conditional probabilities
\renewcommand\given{\nonscript\:\delimsize\vert\nonscript\:\mathopen{}} #1}

\newcommand{\est}[1]{\hat #1}		% for estimates

\newmacro{\signal}{V}		% for input signal
\newmacro{\error}{Z}		% for error variable
\newmacro{\noise}{U}		% for noise
\newmacro{\bias}{b}		% for bias
\newmacro{\diff}{r}		% for diff to limit

\newmacro{\sbound}{M}		% for signal bound
\newmacro{\bbound}{B}		% for bias bound
\newmacro{\totbound}{S}		% for bias bound
\newmacro{\dbound}{\bar\breg}		% for Bregman bound
\newmacro{\difbound}{R}		% for Bregman bound

\newmacro{\snoise}{\psi}		% for scalar noise
\newmacro{\sbias}{\beta}		% for scalar bias
\newmacro{\sdiff}{\rho}		% for scalar diff

\newmacro{\noisedev}{\sigma}		% for noise stdev
\newmacro{\noisevar}{\noisedev^{2}}		% for noise variance

\newmacro{\mix}{\delta}		% for query radius

\newmacro{\unitvec}{w}		% for unit vectors
\newmacro{\unitvar}{W}		% for query direction
\newmacro{\perturb}{\xi}		% for perturbation

\newcommand{\query}{\est\point}		% for query point
\newcommand{\pivot}{\point}		% for pivot point
\newmacro{\radius}{r}		% for Bregman radius

\newmacro{\resource}{s}
\newmacro{\nResources}{S}
\newmacro{\resources}{\mathcal{\nResources}}

\newmacro{\graph}{\mathcal{G}}
\newmacro{\vertices}{\mathcal{V}}
\newmacro{\edges}{\mathcal{E}}

\DeclarePairedDelimiterX{\product}[2]{\langle}{\rangle}{#1,#2}		% for scalar product
\DeclarePairedDelimiter{\norm}{\lVert}{\rVert}		% for norm
\DeclarePairedDelimiterXPP{\dnorm}[1]{}{\lVert}{\rVert}{_{\ast}}{#1}		% for dual norm
\newmacro{\gmat}{g}		% for metric tensor
\newmacro{\dfun}{\dist_{\gmat}}
\newmacro{\ball}{\mathcal{B}}		% for balls
\newmacro{\sphere}{\mathbb{S}}		% for spheres

\newop{\vbudget}{VB}
\newop{\tvar}{V}
\newop{\bvar}{B}
\newop{\svar}{S}
\newop{\varreg}{VReg}
\newop{\dynreg}{DynReg}
\newop{\preg}{\overline{Reg}}		% for pseudo-regret

\newmacro{\budget}{V}
\newmacro{\batch}{\Delta}
\newmacro{\iBatch}{k}
\newmacro{\nBatches}{m}

\newmacro{\bexp}{b}		% for bias exponent
\newmacro{\pexp}{p}		% for step exponent
\newmacro{\vexp}{r}		% for tracking exponent
\newmacro{\qexp}{q}		% for batch exponent
\newmacro{\sexp}{s}		% for variance exponent

\newmacro{\const}{c}
\newmacro{\energy}{E}
\newmacro{\event}{\mathcal{N}}
\newmacro{\gamevar}{\Sigma^{2}}
\newmacro{\temp}{\alpha}		% for learning rate
\newmacro{\weight}{\lambda}

\newmacro{\bregbound}{H}
\newmacro{\bdiam}{\mathcal{D}}

\newmacro{\termOne}{\mathrm{I}}
\newmacro{\termTwo}{\mathrm{I\hspace{-.2ex}I}}
\newmacro{\termThree}{\mathrm{I\hspace{-.2ex}I\hspace{-.2ex}I}}

\begin{document}

%*************************************************************
%*****    FRONT MATTER AND METADATA
%*************************************************************

%----------------------------------------------------------------------
%%% TITLE & AUTHORS
%----------------------------------------------------------------------
\title{\beginrev Multi-agent online learning in time-varying games}

\author
[B.~Duvocelle]
{Benoit Duvocelle$^{\sharp}$}
\address{$^{\sharp}$\,%
Maastricht University, Department of Quantitative Economics, P.O. Box 616, NL\textendash 6200 MD Maastricht, The Netherlands.}
\EMAIL{b.duvocelle@maastrichtuniversity.nl}

%-------------------------------------------------------------------
\author
[P.~Mertikopoulos]
{Panayotis Mertikopoulos$^{\star,\ddag}$}
\address{$^{\star}$\,%
Univ. Grenoble Alpes, CNRS, Inria, LIG, 38000, Grenoble, France.}
\address{$^{\ddag}$\,%
Criteo AI Lab.}
\EMAIL{panayotis.mertikopoulos@imag.fr}

%-------------------------------------------------------------------
\author
[M.~Staudigl]
{\\Mathias Staudigl$^{\diamond}$}
\address{$^{\diamond}$\,%
Maastricht University, Department of Data Science and Knowledge Engineering, P.O. Box 616, NL\textendash 6200 MD Maastricht, The Netherlands}
\EMAIL{m.staudigl@maastrichtuniversity.nl}

%-------------------------------------------------------------------
\author
[D.~Vermeulen]
{Dries Vermeulen$^{\sharp}$}
\EMAIL{d.vermeulen@maastrichtuniversity.nl}

%----------------------------------------------------------------------
%%% KEYWORDS
%----------------------------------------------------------------------
\subjclass[2010]{%
Primary 91A10, 91A26;
secondary 68Q32, 68T02.}

\keywords{%
%Dynamic regret;
Nash equilibrium;
mirror descent;
time-varying games}

%----------------------------------------------------------------------
%%% THANKS
%----------------------------------------------------------------------
%\thanks{\beginrev
%The authors are deeply grateful to the associate editor and the two anonymous referees for providing many insightful comments and remarks that greatly improved the manuscript.
%\endrev}

%----------------------------------------------------------------------
%%% ACRONYMS
%----------------------------------------------------------------------
\newcommand{\acli}[1]{\textit{\acl{#1}}}		% for italicized acro
\newcommand{\aclip}[1]{\textit{\aclp{#1}}}		% for italicized acro (plural)
\newcommand{\acdef}[1]{\textit{\acl{#1}} \textup{(\acs{#1})}\acused{#1}}		% for acro def
\newcommand{\acdefp}[1]{\emph{\aclp{#1}} \textup(\acsp{#1}\textup)\acused{#1}}	% for acro def (plural)

\newacro{LHS}{left-hand side}
\newacro{RHS}{right-hand side}
\newacro{wp1}[w.p.$1$]{with probability $1$}

\newacro{APT}{asymptotic pseudotrajectory}
\newacro{DGA}{dampened gradient approximation}
\newacro{ODE}{ordinary differential equation}
\newacro{DI}{differential inclusion}
\newacro{TV}{tracking variation}
\newacro{MG}{mirror gradient}
\newacro{EGD}{entropic gradient descent}
\newacro{lsc}[l.s.c.]{lower semi-continuous}
\newacro{SFO}{stochastic first-order oracle}
\newacro{MDS}{martingale difference sequence}
\newacro{KL}{Kullback\textendash Leibler}
\newacro{PM}{prox-method}
\newacro{BPM}{Bregman proximal method}
\newacro{FTRL}{``follow the regularized leader''}
\newacro{DGF}{distance-generating function}
\newacro{SP}{saddle-point}
\newacro{SPSA}{single-point stochastic approximation}
\newacro{NE}{Nash equilibrium}
\newacroplural{NE}[NE]{Nash equilibria}
\newacro{DC}{diagonal concavity}
\newacro{DSC}{diagonal strict concavity}
\newacro{VI}{variational inequality}
\newacroplural{VI}{variational inequalities}
\newacro{iid}[i.i.d.]{independent and identically distributed}
\newacro{EW}{exponential weights}
\newacro{MW}{multiplicative weights}
\newacro{OGD}{online gradient descent}
\newacro{GMD}{generalized mirror descent}
\newacro{OMD}{online mirror descent}
\newacro{MD}{mirror descent}
\newacro{SINR}{signal-to-interference-plus-noise ratio}
\newacro{SMD}{stochastic mirror descent}
\newacro{DA}{dual averaging}

%----------------------------------------------------------------------
%%% ABSTRACT
%----------------------------------------------------------------------
\begin{abstract}
\input{Abstract}
\end{abstract}
\maketitle

%*************************************************************
%*****    BODY TEXT
%*************************************************************
%\allowdisplaybreaks		% for breaking long displays
\acresetall		% for resetting acros
\acused{wp1}

%----------------------------------------------------------------------
%%% INTRODUCTION
%----------------------------------------------------------------------
\section{Introduction}
\label{sec:introduction}
\input{Introduction}

%----------------------------------------------------------------------
%%% PRELIMS
%----------------------------------------------------------------------
\section{Preliminaries}
\label{sec:prelims}
\input{Prelims}

%----------------------------------------------------------------------
%%% SETUP
%----------------------------------------------------------------------
\section{The learning model}
\label{sec:setup}
\input{Setup}

%%----------------------------------------------------------------------
%%%% LEARNING
%%----------------------------------------------------------------------
%\section{No-regret learning}
%\label{sec:learning}
%\input{Learning}

%----------------------------------------------------------------------
%%% RESULTS
%----------------------------------------------------------------------
\section{Equilibrium tracking and convergence analysis}
\label{sec:results}
\input{Results}

%----------------------------------------------------------------------
%%% BANDIT
%----------------------------------------------------------------------
\section{Learning with payoff-based information}
\label{sec:bandit}
\input{Bandit}

%----------------------------------------------------------------------
%%% DISCUSSION
%----------------------------------------------------------------------
\section{Further results and discussion}
\label{sec:discussion}
\input{Discussion}

%----------------------------------------------------------------------
%%% CONCLUSIONS
%----------------------------------------------------------------------
\section{Concluding remarks}
\label{sec:conclusion}
\input{Conclusion}

%*************************************************************
%*****    APPENDICES
%*************************************************************
\numberwithin{lemma}{section}		% for numbering  in the appendix
\numberwithin{proposition}{section}		% for numbering  in the appendix
\numberwithin{equation}{section}		% for numbering in the appendix
\appendix

%----------------------------------------------------------------------
%%% APP: AUXILIARY
%----------------------------------------------------------------------
\section{Basic properties of Bregman proximal mappings}
\label{app:Bregman}
\input{App-Bregman}

%----------------------------------------------------------------------
%%% APP: REGRET
%----------------------------------------------------------------------
%\section{Regret minimization}
%\label{app:regret}
%\input{App-Regret}

%----------------------------------------------------------------------
%%% ACKNOWLEDGMENTS
%----------------------------------------------------------------------
\section*{Acknowledgments}
\begingroup
\small
\input{Thanks}
\endgroup

%*************************************************************
%*****    BIBLIOGRAPHY
%*************************************************************
\bibliographystyle{ormsv080}
\bibliography{IEEEabrv,bibtex/Bibliography-PM,bibtex/Bibliography-TVG}

\end{document}

%% file: Abstract.tex
%----------------------------------------------------------------------
%%% THANKS
%----------------------------------------------------------------------
% !TEX root = ./Main.tex
%
%
We examine the long-run behavior of multi-agent online learning in games that evolve over time.
%----------------------------------------------------------------------
\beginrev
Specifically, we focus on a wide class of policies based on \acl{MD}, and we show that the induced sequence of play
\begin{enumerate*}
[(\itshape a\upshape)]
\item
converges to \acl{NE} in time-varying games that stabilize in the long run to a strictly monotone limit;
and
\item
it stays asymptotically close to the evolving equilibrium of the sequence of stage games (assuming they are strongly monotone).
\end{enumerate*}
Our results apply to both gradient-based and payoff-based feedback \textendash\ \ie when players only get to observe the payoffs of their chosen actions.\!
\endrev
%----------------------------------------------------------------------

%% file: Introduction.tex
%----------------------------------------------------------------------
%%%  INTRODUCTION
%----------------------------------------------------------------------
% !TEX root = ./Main.tex

%----------------------------------------------------------------------
\beginrev

Consider a repeated multi-agent decision process that unfolds as follows:
\begin{enumerate}[left=0pt]
\item
At each stage $\run=\running$, every agent selects an action from some continuous set.
\item
Each agent receives a reward based on their chosen action and the actions of all other players.
These rewards are determined by a normal form game $\game_{\run}$
%possibly evolving with time and a priori unknown  to the players.
that evolves over time and is a priori unknown to the players.
%a priori unknown to the players, and evolving over time.
\item
Based on the reward that they received (and/or any other payoff-relevant information), the players update their actions and the process repeats.
\end{enumerate}
The main questions that we seek to address in this general context are the following:
First,
\emph{are there online learning policies that allow players to track a \acl{NE} over time \textpar{or to converge to one if the stage games stabilize}?}
And, if so,
\emph{what is the impact of the information available to the players and the variability of the sequence of stage games?}

%----------------------------------------------------------------------
%%%  BACKGROUND
%----------------------------------------------------------------------
\subsection*{Background\afterhead}

One of the most widely used policies for learning in games is the \acdef{MD} class of algorithms and its variants, \cf \citet{SS11,BCB12}, and references therein.
This family of first-order methods dates back to \citet{NY83},
and contains as special cases
standard (sub)gradient descent methods,
\acl{EGD} \citep{BecTeb03},
the ``Hedge'' (or exponential/multiplicative weights) algorithm in finite games \cite{Vov90,LW94,ACBFS95},
and, in games with a linear payoff structure,
the \acf{FTRL} class of policies \cite{SSS06}.
These methods have been applied to a wide range of games \textendash\ from min-max to potential games \textendash\ leading to a vast corpus of literature that is impossible to survey here;
for an appetizer, see \cite{NJLS09,JNT11,RS13-NIPS,Nes09,LC05,KDB15,BBF20} and references therein.

In the single-player case, the standard figure of merit is the minimization of the learner's \emph{regret}, \ie the cumulative payoff difference between the player's chosen policy and the ``best policy in hindsight'' (static or dynamic, depending on the precise notion of regret under consideration).
%For this reason, much of the literature on online learning revolves around no-regret algorithms that are min-max optimal, both in terms of the horizon $\nRuns$ of the process as well as the dimensionality of the player's action spaces.
In this context, when the payoff functions encountered by the learner are concave, \ac{MD} methods guarantee an $\bigoh(\sqrt{\nRuns})$ static regret bound which is well known to be order-optimal \citep{ABRT08};
moreover, if the problem has a favorable geometry (\eg when the learner's action set is a simplex or a spectrahedron), these bounds are ``almost'' dimension-free, a fact which is of crucial importance in practical applications.

In view of these appealing guarantees, one might expect this picture to carry over effortlessly to multi-agent decision problems as well.
However, game-theoretic learning can be considerably more involved because, in addition to the \emph{exogenous} variability of the stage game $\game_{\run}$ as a function of $\run$, the players' individual reward functions also vary \emph{endogenously} as a function of the actions chosen by the other players at any given time $\run$.
Moreover, the standard solution concept in game theory is that of a \acli{NE} \textendash\ not the players' regret (external, internal, or dynamic).
As a result, even though the algorithms under study are essentially the same in both single- and multi-agent environments,
the analysis and the results obtained in these two settings are often markedly different.

Our paper focuses on multi-agent problems and aims to analyze the equilibrium tracking and convergence properties of \ac{MD}-based policies in time-varying games.
In so doing, we seek to partially fill a gap in the existing literature on game-theoretic learning, which has focused almost exclusively on the case where there are no exogenous variations in the players' payoff functions \textendash\ 
\ie when the stage game $\game_{\run}$ remains \emph{fixed} for all $\run$.
To provide the necessary context, we begin by discussing below some relevant works, and we outline our main contributions right after.

%----------------------------------------------------------------------
%%%  RELATED
%----------------------------------------------------------------------
\subsection*{Related work\afterhead}

Starting with mixed-strategy learning in finite games, a ``folk'' result in the field states that the empirical frequency of no-regret play converges to the game's \emph{Hannan set} (also known as the set of coarse correlated equilibria).
However, as was shown by \citet{VZ13}, the Hannan set of a game may contain strategies that assign positive weight \emph{only} to dominated strategies, a point which is clearly incompatible with Nash play. 
More to the point, the impossibility result of \citet{HMC03} shows that there are no uncoupled dynamics leading to \acl{NE} in all games:
since no-regret dynamics are unilateral by construction \textendash\ and hence uncoupled a fortiori \textendash\ it is not possible to establish a blanket causal link between no-regret play and convergence to \acl{NE}.
%In fact, even in the relatively simple context of two-player zero-sum games,
%the analysis of \citet{HSV09} and \citet{MPP18} shows that no-regret learning may cycle indefinitely without converging, always remaining a uniform distance away from the game's \aclp{NE}.

For this reason, deriving the equilibrium convergence properties of multi-agent learning processes require a more specialized look, typically zooming in on specific classes of games.
%requires finer, more specialized scrutiny, typically focusing on special classes of games.
In the case of mixed extensions of finite games, \citet{LC03,LC05} and \citet{CGM15} showed that a variant of the \acl{EW} algorithm converges to an $\eps$-perturbed equilibrium with probability $1$ in potential and $2\times 2\times\dotsm\times2$ games.
More recently, in the case of \emph{continuous} potential games, \citet{PL14} showed that a lifted variant of \ac{MD}-based methods converges weakly to an $\eps$-neighborhood of the game's set of \aclp{NE}.
Importantly, in all these works, convergence is established by first showing that a naturally associated continuous-time dynamical system converges, and then using the so-called \ac{ODE} method of stochastic approximation \citep{Ben99,BHS05} to translate this result to discrete time.

\endrev
%----------------------------------------------------------------------

More relevant for our purposes is the recent work of \citet{MZ19} who focused on the class of \emph{monotone games}, \ie continuous games that satisfy the so-called \acdef{DSC} condition of \citet{Ros65}.
Specifically, using the same \ac{ODE} stochastic approximation tools discussed above, \cite{MZ19} showed that the sequence of play generated by a specific version of the \acl{DA} algorithm of \citet{Nes09} converges to \acl{NE} with probability $1$, even with in the presence of noise and uncertainty.
The analysis of \cite{MZ19} was subsequently extended by \citet{BLM18} to learning with payoff-based, ``bandit feedbak'' \textendash\ \ie when players observe only the payoff of the action that they played. 
%----------------------------------------------------------------------
\beginrev
At around the same time \textendash\ and always in the context of monotone games \textendash\ \citet{TK19,TK19-CDC} used a Tikhonov regularization approach to obtain a series of comparable results for ``merely monotone'' games (\ie monotone games that are not necessarily \emph{strictly} monotone).
Finally, in a very recent paper, \citet{BBF20} used stochastic approximation methodologies to prove the convergence of a payoff-based, \acdef{DGA} scheme in two other classes of one-dimensional concave games \textendash\ 
games with strategic complements, and ordinal potential games with isolated equilibria.

%----------------------------------------------------------------------
%%%  CONTRIBUTIONS
%----------------------------------------------------------------------
\subsection*{Our contributions\afterhead}
In all the works described above, the game faced by the players remains \emph{fixed} throughout the learning process, and the variation in the players' individual payoff functions is strictly \emph{endogenous} \textendash\ \ie it is only due to the other players' evolving action choice.
By contrast, our paper seeks to tackle problems where the sequence of games encountered by the players also evolves \emph{exogenously} \textendash\ \ie players encounter a \emph{time-varying game}.
\endrev
%----------------------------------------------------------------------

In this general context,
%\footnote{In the sequel, we do not use the term ``descent'' in order to avoid the clash with game-theoretic tradition which typically focuses on payoffs instead of losses (as in online learning and optimization).}
we consider two distinct regimes:
\begin{enumerate*}
[(\itshape a\upshape)]
\item
when the sequence of stage games converges to some well-defined limit (in our case, a strictly monotone game);
and
\item
when $\game_{\run}$ evolves over time without converging.
\end{enumerate*}
In terms of feedback, we consider a flexible oracle model which provides noisy payoff gradient estimates to the players based on the actions that they chose at each stage of the process.
We then show that, if the sequence of stage games stabilizes to some well-defined limit, the induced sequence of play converges to a \acl{NE} of the limit game with probability $1$, irrespective of the magnitude of the noise entering the players' gradient signals.
On the other hand, if the stage games do not stabilize, there is no equilibrium state to converge to (either static or in the mean);
in this case, we focus on the players' ability to track the equilibrium of $\game_{\run}$ as it evolves over time.
More precisely, we show that
%the player's tracking error (defined as the average distance from equilibrium at each stage)
the average distance from equilibrium vanishes over time, and we provide an explicit estimate for this ``tracking error'' in terms of the variation of the sequence of stage games (assuming they are strongly monotone).

Finally, to account for environments where gradient information is not available to the players, we also consider the case of learning with \emph{payoff-based} feedback.
By considering a one-shot gradient estimation process based on \acl{SPSA} techniques \citep{Spa97,FKM05,BLM18}, we map the problem of payoff-based learning to our generic oracle model, and we show that our convergence and equilibrium tracking results still apply in this case (though the corresponding rates are reduced as a consequence of the players' having even less information at their disposal).

%These results comprise a first step towards understanding the behavior of utility-maximizing agents in unknown, online environments where the top-down, ``rationalistic'' viewpoint of dynamic/stochastic games does not apply.
%Specifically, even though the standard rationality postulates do not hold in our setting (knowledge of the game being played, common knowledge of rationality, etc.), our results show that no-regret learning based on \acl{MD} can still lead to equilibrium in dynamic environments.
%We find this property particularly appealing, as it provides an important link between online learning and the emergence of rational behavior in strategic environments that evolve over time.

%----------------------------------------------------------------------
\beginrev

In terms of proof techniques, the exogenous dependence of $\game_{\run}$ on $\run$ means that the continuous-time limit of the players' learning process is likewise non-autonomous (\ie it also depends on $\run$).
As a result, there is no longer a well-defined ``mean field equation'' to approximate, so it is not possible to employ the ODE method of \citet{Ben99} that underlies the series of papers discussed above.
Instead, to establish convergence to an equilibrium in the ``stable limit'' regime, we work directly in discrete time and we employ a mix of submartingale limit theory and quasi-Fejér arguments.
Finally, our equilibrium tracking result relies on decomposing the horizon of play into batches of appropriately chosen length and subsequently utilizes a batch comparison technique that was introduced by \citet{BGZ15} to analyze the dynamic regret of \emph{single-agent} online learning algorithms.

\endrev
%----------------------------------------------------------------------

%% file: Prelims.tex
%----------------------------------------------------------------------
%%% PRELIMINARIES
%----------------------------------------------------------------------
% !TEX root = ./Main.tex

%----------------------------------------------------------------------
%%% Definitions
%----------------------------------------------------------------------
\subsection{Notation\afterhead}
\label{sec:prelims-notation}

Let $\vecspace$ be a $\vdim$-dimensional real space with norm $\norm{\cdot}$, and let $\cvx$ be a compact convex subset of $\vecspace$.
In what follows, we will write
$\dspace \defeq \dual\vecspace$ for the dual of $\vecspace$,
$\braket{\dpoint}{\point}$ for the duality pairing between $\dpoint\in\dspace$ and $\point\in\vecspace$,
and
$\dnorm{\dpoint} = \sup\setdef{\braket{\dpoint}{\point}}{\norm{\point}\leq 1}$ for the dual norm of $\dpoint\in\dspace$.
We will also write
$\relint(\cvx)$ for the relative interior of $\cvx$,
$\bd(\cvx)$ for its boundary,
and
$\diam(\cvx) = \sup\setdef{\norm{\pointalt - \point}}{\point,\pointalt\in\cvx}$ for its diameter.
Finally, for concision, we will write $\window{a}{b} = \{a,a+1,\dotsc,b\}$ for the set of positive integers spanned by $a,b\in\N$.

%----------------------------------------------------------------------
%%% Definitions
%----------------------------------------------------------------------
\subsection{Continuous games\afterhead}
\label{sec:prelims-setup}

Throughout our paper, we focus on games with a finite number of players and continuous action sets.
Specifically, every player $\play\in\players = \{1,\dotsc,\nPlayers\}$ is assumed to select an \emph{action} $\point_{\play}$ from a compact convex subset $\points_{\play}$ of a finite-dimensional normed space $\vecspace_{\play}$;
subsequently, every player receives a \emph{reward} based on each player's individual objective and the \emph{action profile} $\point = (\point_{\play};\point_{-\play}) \equiv (\point_{1},\dotsc,\point_{\play},\dotsc,\point_{\nPlayers})$ of all players' actions.
In more detail, writing $\points \defeq \prod_{\play\in\players} \points_{i}$ for the game's \emph{action space},
%and $\vecspace \defeq \prod_{\play\in\players}\vecspace_{\play}$ for its corresponding ambient space,%
%\footnote{Unless explicitly mentioned otherwise, we will assume that $\vecspace$ is endowed with the norm $\norm{\point}^{2} = \sum_{\play} \norm{\point_{\play}}^{2}$.
%Also, to simplify notation, we will use the same notation for the norm of each factor space $\vecspace_{\play}$ and rely on the context to resolve any ambiguities.}
we assume that each player's reward is determined by an associated \emph{payoff} (or \emph{utility}) \emph{function} $\pay_{\play}\from\points\to\R$.
The tuple $\game \equiv \game(\players,\points,\pay)$ will then be referred to as a \emph{continuous game}.
\smallskip

%----------------------------------------------------------------------
\beginrev

In terms of regularity, we will assume throughout that the players' payoff functions are continuously differentiable, and we will write $\vecfield_{\play}(\point)$ for the individual payoff gradient of the $\play$-th player,
\ie
\begin{align}
\label{eq:payv}
\vecfield_{\play}(\point)
	&= \nabla_{\point_{\play}} \pay_{\play}(\point_{\play};\point_{-\play})
\shortintertext{or, putting all players together,}
\label{eq:payv-profile}
\vecfield(\point)
	&= (\vecfield_{1}(\point),\dotsc,\vecfield_{\nPlayers}(\point)).
\end{align}
In the above, we are tacitly assuming that $\pay_{\play}$ is defined on an open neighborhood of $\points$ in the ambient space $\vecspace \defeq \prod_{\play\in\players}\vecspace_{\play}$ of the game;
none of our results depend on this device, so we do not make this assumption explicit.
We will also adopt the established convention of treating $\vecfield_{\play}(\point)$ as an element of the dual space $\dspace_{\play} \defeq \dual\vecspace_{\play}$ of $\vecspace_{\play}$.
Finally, we will assume that $\vecspace$ is endowed with the norm $\norm{\point}^{2} = \sum_{\play} \norm{\point_{\play}}^{2}$ where, for ease of notation, we write $\norm{\cdot}$ for the norm of each factor space $\vecspace_{\play}$ and rely on the context to resolve any ambiguities.

%we do so in order to emphasize the pairing between $\vecfield_{\play}(\point)$ and displacement vectors $\tanvec_{\play}\in\vecspace_{\play}$ via the evaluation mapping $\tanvec_{\play}\mapsto \braket{\vecfield_{\play}(\point)}{\tanvec_{\play}} = \pay_{\play}'(\point;\tanvec_{\play}) = \left.d/dt\right\vert_{t=0}\pay_{\play}(\point_{\play}+t\tanvec_{\play};\point_{-i})$.

\endrev
%----------------------------------------------------------------------

%----------------------------------------------------------------------
%%% Nash equilibrium
%----------------------------------------------------------------------
\subsection{\aclp{NE} and monotonicity\afterhead}
\label{sec:prelims-Nash}

The most prevalent solution concept in game theory is that of a \acdef{NE}.
This is an action profile $\eq\in\points$ that is resilient to unilateral deviations, \ie
\begin{equation}
\label{eq:NE}
\tag{NE}
\pay_{\play}(\eq_{\play};\eq_{-\play})
	\geq \pay_{\play}(\point_{\play};\eq_{-\play})
	\quad
	\text{for all $\point_{\play}\in\points_{\play}$ and all $\play\in\players$}.
\end{equation}
The set of \aclp{NE} of $\game$ will be denoted in the sequel as $\eqs \defeq \Nash(\game)$.

%----------------------------------------------------------------------
\beginrev

By virtue of their definition, it is straightforward to check that \aclp{NE} satisfy the Stampacchia \acl{VI}
%first-order optimality condition
%\begin{equation}
%\label{eq:NE-var}
%\braket{\vecfield_{\play}(\eq)}{\point_{\play} - \eq_{\play}}
%	\leq 0
%	\quad
%	\text{for all $\point_{\play}\in\points_{\play}$, $\play\in\players$},
%\end{equation}
%More concisely, this can be written as a Stampacchia \acl{VI} of the form
\begin{equation}
\label{eq:SVI}
\tag{SVI}
\braket{\vecfield(\eq)}{\point - \eq}
	\leq 0
	\quad
	\text{for all $\point\in\points$}.
\end{equation}
As a result, finding a \acl{NE} of a continuous game typically involves solving the Stampacchia problem \eqref{eq:SVI}.
This observation forms the basis of an important link between game theory and optimization,
\cf \citet{FP03}, \citet{FK07}, \citet{LRS19}, and references therein.

Now, starting with the seminal work of \citet{Ros65}, much of the literature has focused on games that satisfy the \acdef{DC} condition
\begin{equation}
\label{eq:DC}
\tag{DC}
\braket{\vecfield(\alt \point) - \vecfield(\point)}{\alt \point - \point}
	\leq 0
	\quad
	\text{for all $\point,\alt \point\in\points$}.
\end{equation}
Owing to the link between \eqref{eq:DC} and the theory of monotone operators in optimization, games that satisfy \eqref{eq:DC} are commonly referred to as \emph{monotone games}.%
\footnote{More precisely, \citet{Ros65} uses the name \acdef{DSC} for a weighted variant of \eqref{eq:DC} which holds as a strict inequality when $\pointalt\neq\point$.
\citet{HS09} use the term ``stable'' to refer to a class of population games that satisfy a condition similar to \eqref{eq:DC}, while \citet{San15} and \citet{SW16} respectively call such games ``contractive'' and ``dissipative''.
We use the term ``monotone'' throughout to underline the connection of \eqref{eq:DC} with operator theory and \aclp{VI}.}
In particular, mirroring the corresponding terminology from convex analysis, we will say that $\game$ is:
\begin{enumerate}
%[\indent\itshape a\upshape)]
\item
\emph{Strictly monotone} if \eqref{eq:DC} holds as a strict inequality when $\pointalt\neq\point$.
\item
\emph{Strongly monotone} if there exists a positive constant $\strong>0$ such that
\begin{equation}
\label{eq:DC-strong}
\braket{\vecfield(\pointalt) - \vecfield(\point)}{\pointalt - \point}
	\leq -\strong \norm{\pointalt - \point}^{2}
	\quad
	\text{for all $\point,\pointalt\in\points$}.
\end{equation}
\end{enumerate}
Obviously, we have the inclusions $\text{``strongly monotone''} \subsetneq \text{``strictly monotone''} \subsetneq \text{``monotone''}$, mirroring the corresponding chain of inclusions $\text{``strongly concave''} \subsetneq \text{``strictly concave''} \subsetneq \text{``concave''}$ for concave functions.

%If $\game$ is monotone, $\Nash(\game)$ coincides with the solution set of \eqref{eq:SVI}, which is itself convex and compact \citep{FP03};
%in particular, if the game is strictly or strongly monotone, $\Nash(\game)$ is a singleton.
%Moreover, \aclp{NE} of monotone games can also be characterized as solutions of the \emph{Minty} \acl{VI}
%\begin{equation}
%\label{eq:MVI}
%\tag{MVI}
%\braket{\vecfield(\point)}{\point - \eq}
%	\leq 0
%	\quad
%	\text{for all $\point\in\points$}.
%\end{equation}
%This property of \aclp{NE} of monotone games will play a crucial role in our analysis and we will use it freely in the sequel;
%for a detailed discussion, the reader is referred to \citet{FP03}, \citet{LRS19}, and \citet{MZ19}.

\endrev
%----------------------------------------------------------------------

Examples of monotone games include
Cournot oligopolies \citep{MS96},
Kelly auctions and Tullock competitions \citep{NOR10},
signal covariance and power control problems in wireless communications \citep{DMMP15,MM16},
atomic splittable congestion games in networks with parallel links \citep{ORShi93,SW16},
and many other problems where online decision-making is the norm.
For an extensive list of applications in different contexts, see \citet{FK07} and \citet{SFPP10}.

%% file: Setup.tex
%----------------------------------------------------------------------
%%% SETUP
%----------------------------------------------------------------------
% !TEX root = ./Main.tex

%----------------------------------------------------------------------
\beginrev

To account for the possibility of exogenous variations in the game-theoretic setup of the previous section, we will assume that the players face a different stage game $\game_{\run}$ at each decision opportunity.
More explicitly, the envisioned sequence of play unfolds as follows:
\begin{enumerate}
\item
At each stage $\run=\running$, every agent $\play\in\players$ selects an action $\state_{\play,\run} \in \points_{\play}$.
\item
%Based on the $\run$-th stage game $\game_{\run}$,
Each player receives their associated reward based on $\game_{\run}$, and they observe \textendash\ or otherwise construct \textendash\ an estimate $\signal_{\play,\run} \in \dspace_{\play}$ of their individual payoff gradients.
\item
Subsequently, players update their actions and the process repeats.
\end{enumerate}

%\MS{I do not see why players need to learn the rewards}
%\PM{They don't, it's just part of the process (it's the sequence of events, not an algorithm).}
%%----------------------------------------------------------------------
%%% TVG description begins here
%
%\begin{algorithm}[h]
%\small
%\ttfamily
%\caption*{\textbf{Sequence of events in a time-varying game}}
%\input{Algorithms/TVG}
%\end{algorithm}
%
%%% TVG description ends here
%%----------------------------------------------------------------------
%In the above, $\state_{\play,\run} \in \points_{\play}$ denotes the action chosen by the $\play$-th player at stage $\run$, while $\signal_{\play,\run} \in \dspace_{\play}$ is a ``gradient signal'' that is observed (or otherwise inferred) by the players after collecting their rewards.
%In the rest of this section, we discuss in detail our blanket assumptions for the sequence of stage games $\game_{\run}$ and the gradient signals $\signal_{\run}$.

\noindent
The core ingredients of the above framework are
\begin{enumerate*}
[(\itshape a\upshape)]
\item
the sequence of stage games $\game_{\run}$ encountered by the players;
\item
the sequence of gradient signals $\signal_{\play,\run} \in \dspace_{\play}$ observed (or inferred) at each stage;
and
\item
the way that players update their actions as a function of the observed information.
\end{enumerate*}
We discuss each of these elements in detail below.

\endrev
%----------------------------------------------------------------------

%----------------------------------------------------------------------
%%% GAME SEQUENCE
%----------------------------------------------------------------------
\subsection{The stage game sequence\afterhead}
\label{sec:stage}

The only blanket assumption that we will make for the sequence of stage games $\game_{\run}$ is that the players' payoff functions are Lipschitz continuous and smooth.
More precisely, we will posit the following requirement for the players' $\run$-th stage payoff field $\vecfield_{\run}(\point) = (\vecfield_{\play,\run}(\point))_{\play\in\players}$:

\begin{assumption}
\label{asm:payv}
%Let $\vecfield_{\run}(\point) = (\vecfield_{\play,\run}(\point))_{\play\in\players}$ denote the players' individual gradient field in the game $\game_{\run}$.
%There exist constants $\vbound,\lips > 0$ such that
The game's payoff functions are $C^{2}$-smooth;
in particular, there exist constants $\vbound_{\play},\lips_{\play} > 0$ such that
\begin{subequations}
\begin{align}
\label{eq:vbound}
\norm{\vecfield_{\play,\run}(\point)}_{\ast}
	&\leq \vbound_{\play}
	\\
\label{eq:lips}
\norm{\vecfield_{\play,\run}(\pointalt) - \vecfield_{\play,\run}(\point)}_{\ast}
	&\leq \lips_{\play} \norm{\pointalt - \point}
\end{align}
\end{subequations}
for all $\run=\running$, and all $\play\in\players$, $\point,\pointalt\in\points$.
\end{assumption}

For posterity, we will also write $\vbound \defeq \max_{\play}\vbound_{\play}$ and $\lips_{\play} \defeq \max_{\play} \lips_{\play}$.
Beyond this mild regularity assumption, the sequence of stage games is assumed arbitrary.
For instance,
the evolution of $\game_{\run}$ could be random (\ie $\game_{\run}$ could be determined by some randomly drawn parameter $\theta_{\run}$ at each stage),
%it could depend on the players' actions (\eg as in the literature on dynamic/repeated games),
it could be governed by an underlying (hidden) Markov chain model,
etc.
In particular, we do not assume that the stage game $\game_{\run}$ is revealed to the players before choosing an action:
from their individual viewpoint, the players are involved in a repeated decision process where the choice of an action returns a reward, but they have no knowledge of the game generating this reward.
This ``agnostic'' approach is motivated by the fact that the standard rationality postulates of game theory (full rationality, common knowledge of rationality, etc.) are not satisfied in many cases of practical interest.
We briefly discuss two concrete examples of this framework below:

%----------------------------------------------------------------------
\beginrev

\begin{example}[Repeated Kelly auctions]
\label{ex:auctions}
Consider a Kelly auction where a splittable resource (advertising time on a website, a catch of fish in a fish market, etc.) is auctioned off, day after day, to a set of $\nPlayers$ buyers \citep{KMT98,Tul80}.
In more detail, each player can place a monetary bid $\point_{i} \in [0,b_{\play}]$ to acquire a unit of said resource, up to the player's total budget $b_{\play}$.
Then, once all bids are in, the resource is allocated proportionally to each player's bid, \ie the $\play$-th player gets a fraction $\rho_{\play} = \point_{\play} / [c + \sum_{\playalt\in\nPlayers} \point_{\playalt}]$ of the auctioned resource (with $c>0$ denoting an ``entry barrier'' for participating in the auction).
Thus, if $g_{\play,\run}$ denotes the marginal gain that the $\play$-th player acquires per resource unit, the player's prorated utility at the $\run$-th epoch will be
\begin{equation}
\label{eq:Kelly}
\pay_{\play,\run}(\point_{\play};\point_{-\play})
	= \frac{g_{\play,\run} \point_{\play}}{c + \sum_{\playalt\in\players} \point_{\playalt}}
	- \point_{\play}.
\end{equation}
Clearly, the players' utility functions evolve as a function of the intrinsic value $g_{\play,\run}$ associated to a unit of the auctioned resource.
Since this value may be subject to arbitrary exogenous fluctuations (for instance, depending on the traffic coming to the website at any given time in the advertising example), we obtain a time-varying game as above.
\envend
\end{example}

\begin{example}[Power control]
\label{ex:power}
As another example, consider $\nPlayers$ wireless users transmitting a stream of packets to a common receiver over a shared wireless channel \citep{TV05}.
If the channel gain for the $\play$-th user at the $\run$-th frame is $g_{\play,\run}$ and the user transmits with power $p_{\play} \in [0,P_{\max}]$, the user's information transmission rate is given by the celebrated Shannon formula
\begin{equation}
\label{eq:SINR}
R_{\play,\run}(p_{\play};p_{-\play})
	= \log\parens*{ 1 + \frac{g_{\play,\run} p_{\play}}{\sigma + \sum_{\playalt\neq\play} g_{\playalt,\run} p_{\playalt}}},
\end{equation}
where $\sigma>0$ denotes the ambient noise in the channel \citep{SW49}.
Since the users' channel gains evolve over time (\eg due to fading, user mobility, or other fluctuations in the wireless medium), we obtain a time-varying game where each user seeks to maximize their individual communication rate.
\envend
\end{example}

\endrev
%----------------------------------------------------------------------

%\begin{remark*}
%We should also note here that
%%our choice of notation suggests that the set of players $\players$ and their action spaces $\points_{\play}$ remain unchanged for all $\run$.
%%However,
%the framework described above is sufficiently flexible to account for games where the set of players (and/or their action sets) \emph{also} vary with time.
%One way to incorporate such variations is as follows:
%if player $\play\in\players$ does not participate in the game at stage $\run$, then
%\begin{enumerate*}
%[\itshape a\upshape)]
%\item
%their payoff function $\pay_{\play,\run}$ is set identically equal to zero;
%and
%\item
%the payoff function $\pay_{\playalt,\run}$ of any other player $\playalt\in\players$ is taken independent of $\point_{\play}$.
%\end{enumerate*}
%In this case, the underlying player set $\players$ essentially corresponds to the \emph{entire} population of players that participate in the game even once during the horizon of play.
%This allows us to treat games with variable sets of players (or actions), a case of significant interest for applications to operations research.
%%\footnote{Variations in the game's action space can be handled similarly so we do not provide the details.}
%\envend
%\end{remark*}

%----------------------------------------------------------------------
%%% SIGNALS
%----------------------------------------------------------------------
\subsection{The feedback signal\afterhead}
\label{sec:signal}

The second basic ingredient of our model is the feedback available to the players after choosing an action.
In tune with the limited information setting outlined above,
%we do not assume that players can observe the actions of other players, their payoffs, or any other such information.
%Instead, we take a ``partial monitoring'' approach in the spirit of \citet{Rus99b}, \citet{CBLS06}, and \citet{LMS08}, and
we only posit that, at each stage $\run=\running$, every player $\play\in\players$ receives \textendash\ or otherwise constructs \textendash\ a ``gradient signal'' $\signal_{\play,\run} \in \dspace_{\play}$.
Analytically, this signal will be treated as if generated from a \acdef{SFO}, \ie an abstract mechanism that provides an estimate of each player's individual payoff gradient at the chosen action profile.
Specifically, if called at $\state_{\run} = (\state_{1,\run},\dotsc,\state_{\nPlayers,\run}) \in\points$, we assume that $\signal_{\play,\run}$ is of the form
% = (\signal_{1,\run},\dotsc,\signal_{\nPlayers,\run})$ is of the form
\begin{equation}
\label{eq:signal}
\tag{SFO}
\signal_{\play,\run}
	= \vecfield_{\play,\run}(\state_{\run})
	+ \error_{\play,\run}
%	+ \noise_{\run}
%	+ \bias_{\run}
\end{equation}
where
the ``observational error'' $\error_{\play,\run}$ captures all sources of uncertainty in the received input.

To differentiate further between ``random'' (zero-mean) and ``systematic'' (non-zero-mean) errors in $\signal_{\play,\run}$, it will be convenient to decompose the error process $\error_{\play,\run}$ as
\begin{equation}
\label{eq:error}
\error_{\play,\run}
	= \noise_{\play,\run} + \bias_{\play,\run}
\end{equation}
where $\noise_{\play,\run}$ is zero-mean and $\bias_{\play,\run}$ denotes the mean of $\error_{\play,\run}$.
Formally, writing $\filter_{\run} = \sigma(\state_{\start},\dotsc,\state_{\run})$ for the natural filtration of $\state_{\run}$, we set
\begin{equation}
\bias_{\play,\run}
	= \exof{\error_{\play,\run} \given \filter_{\run}}
	\qquad
	\text{and}
	\qquad
\noise_{\play,\run}
	= \error_{\play,\run} - \bias_{\play,\run}
\end{equation}
so, by definition, $\exof{\noise_{\play,\run} \given \filter_{\run}} = 0$.
In this way, the oracle feedback received by each player $\play\in\players$ can be classified according to the following statistics:
\smallskip
\begin{subequations}
\label{eq:SFO-stats}
\begin{enumerate}
\addtolength{\itemsep}{\medskipamount}
\item
\emph{Bias:}
\begin{equation}
\label{eq:bbound}
%\bbound_{\run}
%	= \exof{\dnorm{\bias_{\run}} \given \filter_{\run}}
\dnorm{\bias_{\play,\run}}
	\leq \bbound_{\play,\run}.
\end{equation}

\item
\emph{Variance:}
\begin{equation}
\label{eq:variance}
%\sdev_{\run}^{2}
%	= \exof{\dnorm{\noise_{\run}}^{2}}
\exof{\dnorm{\noise_{\play,\run}}^{2} \given \filter_{\run}}
	\leq \sdev_{\play,\run}^{2}.
\end{equation}

\item
\emph{Second moment:}
\begin{equation}
\label{eq:sbound}
%\sbound_{\run}^{2}
%	= \exof{\dnorm{\signal_{\run}}^{2}}
\exof{\dnorm{\signal_{\play,\run}}^{2} \given \filter_{\run}}
	\leq \sbound_{\play,\run}^{2}.
\end{equation}
\end{enumerate}
Finally, to simplify notation later on, we will also consider the ``signal plus noise'' error bound
\begin{equation}
\label{eq:totbound}
\totbound_{\play,\run}^{2}
	= \sbound_{\play,\run}^{2} + \noisedev_{\play,\run}^{2}.
\end{equation}
\end{subequations}
%By assumption, $\exof{\noise_{\run} \given \filter_{\run}} = 0$ for all $\run$.
In the above, $\bbound_{\play,\run}$, $\noisedev_{\play,\run}$ and $\sbound_{\play,\run}$ are to be construed as deterministic upper bounds on the bias, variance, and magnitude of the oracle signal $\signal_{\play,\run}$ that player $\play\in\players$ received at time $\run$.
We will also assume throughout that $\bbound_{\play,\run}$ is non-increasing while $\noisedev_{\play,\run}$ and $\sbound_{\play,\run}$ are non-decreasing.
Finally, in obvious notation, we will write $\signal_{\run}$, $\bias_{\run}$, $\noise_{\run}$ and so forth for the corresponding profiles $\signal_{\run} = (\signal_{\play,\run})_{\play\in\players}$ and the like.
%Also, in obvious terminology, when we want to focus on a single agent $\play\in\players$, we will write $\signal_{\play,\run}$, $\noise_{\play,\run}$, $\sbound_{\play,\run}$, $\sdev_{\play,\run}$, and so forth.
%In obvious terminology, an oracle with $\bbound_{\run}=0$ will be called \emph{unbiased},
%%and an oracle with $\lim_{\run\to0}\bbound_{\run} = 0$ will be called \emph{asymptotically unbiased};
%and an unbiased oracle with $\sdev_{\run} = 0$  will be called \emph{perfect}.%
%\footnote{We stress here that players are never assumed to observe the actions of other players:
%having access to a ``perfect oracle'' is very different from having access to ``perfect information''.}
%For now, we only note that the motivation for the black-box model \eqref{eq:signal} is rooted in the case where players can only observe their realized, in-game payoffs (the so-called \emph{bandit} setting).

\begin{remark*}
To streamline our presentation, we will first present our results in a model-agnostic manner, \ie without specifying the origins of the oracle model \eqref{eq:signal};
subsequently, in \cref{sec:bandit}, we provide an explicit construction of such an oracle from payoff-based observations, and we discuss in detail what this entails for our analysis and results.
\envend
\end{remark*}

%----------------------------------------------------------------------
%%% Mirror descent
%----------------------------------------------------------------------
\subsection{Learning via \acl{MD}\afterhead}
\label{sec:mirror}

%----------------------------------------------------------------------
\beginrev

The last element of the players' learning process concerns the way that players update their actions based on the  received feedback.
For concreteness, we will focus throughout on the widely used family of algorithms known as \acdef{MD}, which posits that players updates their actions by taking a ``proximal'' gradient step from their current action.%
\footnote{The terminology ``descent'' alludes to the fact that \eqref{eq:MD} was originally studied in the context of convex minimization (as opposed to reward \emph{maximization}).
We should also mention here that ``\acl{MD}'' is sometimes used synonymously with the popular \acdef{FTRL} protocol of \citet{SSS06}.
The two methods coincide in linear problems, but not otherwise;
in general, \ac{FTRL} requires access to a best-response oracle, so it is beyond the scope of this paper.}
Formally, this can be modeled via the basic recursion
\begin{equation}
\label{eq:MD}
\tag{MD}
\state_{\play,\run+1}
	= \proxplay{\state_{\play,\run}}{\step_{\play,\run} \signal_{\play,\run}}
\end{equation}
where:
\begin{enumerate}
[\indent 1.]
\addtolength{\itemsep}{.5ex}
\item
$\run=\running$ denotes the stage of the process.
\item
$\state_{\play,\run}$ denotes the action chosen by player $\play$ at stage $\run$.
%For concreteness, we also assume throughout that $\state_{\run}$ is initialized at the so-called ``prox-center'' $\point_{c} \defeq \argmin\hreg$ of $\points$.
\item
$\signal_{\play,\run}$ is the oracle signal of player $\play$ at stage $\run$.
%assumed throughout to be of the general form \eqref{eq:signal}.
\item
$\step_{\play,\run} > 0$ is a player-specific step-size sequence (assumed non-increasing).
\item
$\proxmap_{\play}$ denotes the ``prox-mapping'' of player $\play\in\nPlayers$
%which transforms the player's gradient signal $\signal_{\play,\run}$ into a new action
(see below for a detailed definition).
\end{enumerate}
For a pseudocode implementation from the viewpoint of a generic player, see \cref{alg:prox}.
\endrev
%----------------------------------------------------------------------

%----------------------------------------------------------------------
%% Prox algorithm begins here

\begin{algorithm}[t]
\small
\ttfamily
\caption{Learning via \acl{MD}
\hfill
\small
[player indices suppressed]}
\input{Algorithms/Prox}
\label{alg:prox}
\end{algorithm}

%% Prox algorithm ends here
%----------------------------------------------------------------------

Methods based on \acl{MD} have received intense scrutiny ever since the pioneering work of \citet{NY83};
for an appetizer, see \cite{BecTeb03,NJLS09,Nes09,SS11,BCB12,Teb18,MZ19} and references therein.
For intuition, the archetypal example of the method is based on the Euclidean prox-mapping
\begin{flalign}
\prox{\point}{\dpoint}
	= \Eucl_{\cvx}(\point + \dpoint)
	&= \argmin_{\pointalt\in\cvx}
		\braces*{ \norm{\point +  \dpoint - \pointalt}_{2}^{2} }
%	\notag\\
	= \argmin_{\pointalt\in\cvx}
		\braces*{ \braket{\dpoint}{\point - \pointalt} + \tfrac{1}{2} \norm{\pointalt - \point}_{2}^{2} }
\label{eq:prox-Eucl}
\end{flalign}
where $\Eucl_{\cvx}$ denotes the closest-point projection onto a given convex set $\cvx$.
%\ie the closest-point projection of $\point + \dpoint$ onto a given convex set $\cvx$.
%\footnote{In writing $\point + \dpoint$, we are blurring the lines between primal and dual vectors.
%This distinction is reinstated in the second line of \eqref{eq:prox-Eucl} where $\dpoint\in\dspace$ is paired properly to $\point-\pointalt\in\vecspace$.}
Going beyond this familiar example, the key novelty of \acl{MD} is to replace the quadratic term in \eqref{eq:prox-Eucl} by the so-called \emph{Bregman divergence}
\begin{equation}
\label{eq:Bregman}
\breg(\pointalt,\point)
	= \hreg(\pointalt) - \hreg(\point) - \braket{\nabla \hreg(\point)}{\pointalt - \point},
\end{equation}
induced by a ``\acli{DGF}'' $\hreg$ on $\cvx$.
This function plays the role of the squared Euclidean norm in \eqref{eq:Bregman-Eucl} and, following \citet{JNT11}, we define it as follows:

\begin{definition}
\label{def:Bregman}
Let $\cvx$ be a compact convex subset of $\vecspace \cong \R^{\vdim}$.
A convex function $\hreg\from\vecspace\to\R\cup\{\infty\}$ is said to be a \acdef{DGF} on $\cvx$ if
%----------------------------------------------------------------------
\beginrev

\begin{enumerate}
%\addtolength{\itemsep}{\smallskipamount}
\item
$\hreg$ is continuous and supported on $\cvx$, \ie $\dom\hreg \defeq \setdef{\point\in\vecspace}{\hreg(\point) < \infty} = \cvx$.

\item
$\hreg$ is $\hstr$-strongly convex relative to $\norm{\cdot}$ on $\cvx$, \ie
\begin{equation}
\label{eq:hreg-str}
\hreg(\coef\point + (1-\coef)\pointalt)
	\leq \coef \hreg(\point)
	+ (1-\coef) \hreg(\pointalt)
	- \tfrac{1}{2} \hstr \coef(1-\coef) \norm{\pointalt - \point}^{2}
%\hreg(\pointalt)
%	\geq \hreg(\point)
%	+ \braket{\nabla\hreg(\point)}{\pointalt - \point}
%	+ \tfrac{\hstr}{2} \norm{\pointalt - \point}^{2}
\end{equation}
for all $\point,\pointalt\in\cvx$ and all $\coef\in[0,1]$.

\item
The subdifferential $\subd\hreg$ of $\hreg$ admits a \emph{continuous selection}, \ie there exists a continuous mapping $\nabla\hreg\from\dom\subd\hreg \to \dspace$ such that $\nabla\hreg(\point) \in \subd\hreg(\point)$ for all $\point\in\dom\subd\hreg$.%
\footnote{We recall here that the subdifferential $\subd\hreg$ of $\hreg$ at $\point$ is defined as $\subd\hreg(\point) = \setdef{\dpoint\in\dspace}{\hreg(\pointalt) \geq \hreg(\point) + \braket{\dpoint}{\pointalt - \point} \textrm{ for all } \pointalt\in\vecspace}$.
The notation $\dom\subd\hreg \defeq \setdef{\point\in\dom\hreg}{\subd\hreg(\point)\neq\varnothing}$ stands for the domain of subdifferentiability of $\hreg$ and, by standard results in convex analysis, we have $\relint\dom\hreg \subseteq \dom\subd\hreg \subseteq \dom\hreg$.}
\end{enumerate}
For concision, given a \ac{DGF} $\hreg$ on $\cvx$, we will refer to $\subcvx \defeq \dom\subd\hreg$ as the \emph{prox-domain} of $\hreg$.
\endrev
%----------------------------------------------------------------------
The \emph{Bregman divergence} $\breg\from\subcvx\times\cvx\to\R$ induced by $\hreg$ is then given by \eqref{eq:Bregman},
and
the associated \emph{prox-mapping} $\proxmap\from\subcvx\times\dspace\to\cvx$ is defined as
\begin{equation}
\label{eq:proxmap}
\prox{\point}{\dpoint}
	= \argmin_{\pointalt\in\cvx} \ \braces*{\braket{\dpoint}{\point - \pointalt} + \breg(\pointalt,\point)}
	\quad
	\text{for all $\point\in\subcvx$, $\dpoint\in\dspace$}.
\end{equation}
Finally, we say that $\hreg$ is \emph{Lipschitz} if $\sup_{\point\in\subcvx} \dnorm{\nabla\hreg(\point)} < \infty$.
\envend
\end{definition}

Throughout the sequel, we will assume that each player $\play\in\players$ is endowed with their individual \acl{DGF} $\hreg_{\play}\from\points_{\play}\to\R$.
In obvious notation, we will also write
$\hstr_{\play}$ for the strong convexity modulus of $\hreg_{\play}$,
$\points_{\hreg_{\play}}$ for its prox-domain,
$\breg_{\play} \from \points_{\play} \times \points_{\hreg_{\play}} \to \R$ for the associated Bregman divergence,
and
$\proxmap_{\play} \from \points_{\hreg_{\play}} \times \dspace_{\play} \to \points_{\play}$ for the induced prox-mapping.
For concreteness, we provide two standard examples below:
\smallskip

\begin{example}
[Euclidean projections]
\label{ex:prox-Eucl}
We begin by revisiting Euclidean projections on a compact convex subset $\cvx$ of $\R^{\vdim}$.
The corresponding \ac{DGF} is $\hreg(\point) = \frac{1}{2} \norm{\point}^{2}$ for $\point\in\vecspace$, so $\subcvx=\cvx$ and $\nabla\hreg(\point) = \point$ for all $\point\in\cvx$.
Hence, the associated Bregman divergence is
\begin{equation}
\label{eq:Bregman-Eucl}
\breg(\pointalt,\point)
	= \tfrac{1}{2} \norm{\pointalt}_{2}^{2}
	- \tfrac{1}{2} \norm{\point}_{2}^{2}
	- \braket{\point}{\pointalt - \point}
	= \tfrac{1}{2} \norm{\pointalt - \point}_{2}^{2}
\end{equation}
and the resulting recursion $\new\point = \Eucl(\point + \step\dvec)$ is just a standard projected forward step.
\envend
\end{example}

\begin{example}
[Entropic regularization]
\label{ex:prox-entropic}
Let $\cvx = \simplex_{\vdim} \defeq \setdef{\point\in\R_{+}^{\vdim}}{\sum_{j=1}^{\vdim} \point_{j} = 1}$ denote the unit simplex of $\vecspace=\R^{\vdim}$.
A very widely used \acl{DGF} for this geometry is the (negative) \emph{Gibbs-Shannon entropy} $\hreg(\point) = \sum_{j=1}^{\vdim} \point_{j}\log\point_{j}$ (with the standard notational convention $0\cdot\log0 = 0$).
By inspection, the prox-domain of $\hreg$ is $\subcvx \defeq \relint\cvx$, and the resulting Bregman divergence is just the \acdef{KL} divergence
\begin{equation}
\label{eq:Bregman-KL}
\breg(\pointalt,\point)
	= \dkl(\pointalt, \point)
	\defeq \sum_{j=1}^{\vdim} \pointalt_{j}\log\left(\frac{\pointalt_{j}}{\point_{j}}\right)
	\quad
	\text{for all $\point\in\subcvx$, $\pointalt\in\cvx$}.
\end{equation}
In turn, a standard calculation leads to the prox-mapping
\begin{equation}
\label{eq:prox-MW}
\prox{\point}{\dpoint}
	= \frac
		{(\point_{1}e^{\dpoint_{1}},\dotsc,\point_{n}e^{\dpoint_{n}})}
		{\point_{1}e^{\dpoint_{1}} + \dotsm + \point_{n}e^{\dpoint_{n}}}
\end{equation}
for all $\point\in\subcvx$, $\dpoint\in\dspace$.
The corresponding update rule $\new\point = \prox{\point}{\step\dvec}$ is widely known in optimization as \acli{EGD} \citep{BecTeb03,KW97}, and as
\emph{``Hedge''} (or exponential/multiplicative weights update)
%\acdef{MW} algorithm (or \emph{``Hedge''} or \emph{``exponential weights''})
in game theory and online learning \citep{Vov90,LW94,AHK12,FS99,ACBFS95}.
\envend
\end{example}

%% file: Algorithms/Prox.tex
%----------------------------------------------------------------------
%%% PROX ALGORITHM
%----------------------------------------------------------------------
% !TEX root = ../Main.tex

\begin{algorithmic}[1]
\Require
%	sequence of stage games $\game_{\run}$,
	prox-mapping $\proxmap$,
	step-size $\step_{\run} > 0$
	
%\hspace{2.5ex}%
%	sequence of stage games $\game_{\run}\equiv\game_{\run}(\players,\points,\pay_{\run})$
\State
	initialize $\state_{\start} \leftarrow \argmin\hreg$
	\Comment{initialization}%
\For{$\run=\running$}
%	\State
%		set $\game\leftarrow\game_{\run}(\players,\points,\pay_{\run})$
%		\Comment{stage game definition}
	\State
		play $\state_{\run}\in\points$
		\Comment{play action}
	\State
		get gradient signal $\signal_{\run}$
		\Comment{get feedback}
	\State
		set $\state_{\run+1} \leftarrow \prox{\state_{\run}}{\step_{\run}\signal_{\run}}$
		\Comment{update action}
%	\State
%		$\run\leftarrow\run+1$
%		\Comment{next stage}
\EndFor
\end{algorithmic}

%% file: Results.tex
%----------------------------------------------------------------------
%%% RESULTS
%----------------------------------------------------------------------
% !TEX root = ./Main.tex

We are now in a position to state our main results for the equilibrium tracking and convergence properties of \eqref{eq:MD} in time-varying games.
For concreteness, we will focus below on two distinct \textendash\ and, to a large extent, complementary \textendash\ regimes:
\begin{enumerate*}
[\itshape a\upshape)]
\item
when the sequence of stage games $\game_{\run}$ converges to some limit game $\game \equiv \game_{\infty}$;
and
\item
when $\game_{\run}$ evolves over time without converging.
\end{enumerate*}
In both cases, we will treat the process defining the time-varying game as a ``black box'' and we will not scruitinize its origins in detail;
we do so in order to focus on the interplay between the variability of the sequence $\game_{\run}$ and the induced sequence of play.

%----------------------------------------------------------------------
%%% Convergence
%----------------------------------------------------------------------
\subsection{Stabilization and convergence to equilibrium}
\label{sec:convergence}

We begin with the case where the sequence of stage games stabilizes to some monotone limit game $\game \equiv \game(\players,\points,\pay)$.
%\footnote{To be clear, we are not assuming that each stage game $\game_{\run}$ is a priori monotone.}
Formally, it will be convenient to characterize this stabilization in terms of the quantity
\begin{equation}
\label{eq:difbound}
\difbound_{\play,\run}
	= \max_{\point\in\points} \dnorm{\vecfield_{\play,\run}(\point) - \vecfield_{\play}(\point)},
\end{equation}
%\ie via the maximum difference in the (unilateral) gradient field of the stage game $\game_{\run}$ and the limit game $\game$.
and we will say that the sequence of games $\game_{\run}$, $\run=\running$, converges to $\game$ if
\begin{equation}
\lim_{\run\to\infty} \difbound_{\play,\run}
	= 0
	\quad
	\text{for all $\play\in\players$}.
\end{equation}

To state our equilibrium convergence result, we will require two further assumptions.
The first is a technical ``reciprocity condition'' for the players' \ac{DGF}, namely
\begin{equation}
\label{eq:RC}
\tag{RC}
\breg(\base,\point_{\run})
	\to 0
	\quad
	\text{whenever}
	\quad
\point_{\run}
	\to \base
\end{equation}
for every sequence of actions $\point_{\run}\in\subpoints$.
This requirement is fairly standard in the trajectory analysis of \acl{MD} algorithms \citep{BecTeb03,CT93} and, taken together with the strong convexity of $\hreg$, it implies that $\point_{\run}\to\base$ if and only if $\breg(\base,\point_{\run})\to 0$ (hence the name).%
\footnote{Indeed, $\breg(\base,\point_{\run}) = \hreg(\base) - \hreg(\point_{\run}) - \braket{\nabla\hreg(\point_{\run})}{\base - \point_{\run}} \geq (\hstr/2) \norm{\point_{\run} - \base}^{2}$, so $\point_{\run}\to\base$ whenever $\breg(\base,\point_{\run}) \to 0$.}
In particular, if $\hreg$ is Lipschitz, we have
\begin{equation}
\breg(\base,\point_{\run})
	\leq \hreg(\base) - \hreg(\point_{\run}) + \dnorm{\nabla\hreg(\point_{\run})} \norm{\point_{\run} - \base}
	= \bigoh(\norm{\point_{\run} - \base})
\end{equation}
so \eqref{eq:RC} always holds in that case.
A further easy check shows that \cref{ex:prox-entropic} also satisfies this condition, so \eqref{eq:RC} is not restrictive in this regard.

%----------------------------------------------------------------------
\beginrev

The second set of conditions concerns the players' step-size sequence.
First, we will assume throughout that
\begin{align}
\label{eq:step1}
\tag{S1}
\sum_{\run=\start}^{\infty} \step_{\play,\run}
	&= \infty
	\quad
	\text{for all $\play\in\players$},
\shortintertext{%
\ie each player's learning process cannot stop prematurely.
Second, we will assume that the step-size policies of any two players $\play,\playalt\in\players$ are \emph{mutually compatible} in the sense that
} % end intertext
\label{eq:step2}
\tag{S2}
\sum_{\run=\start}^{\infty}
	\abs{ \step_{\play,\run} - \weight_{\play\playalt} \step_{\playalt,\run} }
	&< \infty
	\quad
	\text{for some $\weight_{\play\playalt} > 0$}.
%	\text{for all $\play,\playalt\in\players$}.
\end{align}
Informally, the compatibility assumption \eqref{eq:step2} means that the players' step-size policies exhibit a comparable asymptotic behavior as $\run\to\infty$, \ie $\step_{\play,\run}/\step_{\playalt,\run} = \Theta(1)$ for all $\play,\playalt\in\players$.
%step-size ratios $\step_{\play,\run} / \step_{\playalt,\run}$ stabilize to $\weight_{\play\playalt}$;
%for example, this holds whenever all players are using the same step-size policy up to a constant factor.
The rationale for this is fairly straightforward:
if a player employs a step-size policy that vanishes much faster than that of all other players, this player would effectively become a ``constant externality'' in the time-scale of the other players.
On that account, it would make more sense to consider convergence in a ``reduced'' game where this player has been effectively removed from the game \textendash\ and so on, until only the ``slower'' time-scale players remain.
%This reasoning can be made rigorous by using a multiple time-scales analysis along the lines of \citet{LC03}.
%In this way, the player would be effectively removed from the game, so it wouldn't make sense to study the algorithm's convergence to a \acl{NE} of $\game$.
Assumption \eqref{eq:step2} rules out such cases and ensures that all players remain active throughout the horizon of play.
\smallskip

With all this in hand, we have the following equilibrium convergence result:

\begin{theorem}
\label{thm:conv}
Let $\game_{\run}$ be a time-varying game converging to a strictly monotone game $\game$.
Suppose further that each player $\play\in\players$ runs \cref{alg:prox} with
a \ac{DGF} satisfying \eqref{eq:RC} and a step-size policy satisfying \eqref{eq:step1}, \eqref{eq:step2}, and
\begin{equation}
\label{eq:step3}
\tag{S3}
\sum_{\run=\start}^{\infty} \step_{\play,\run} (\difbound_{\play,\run} + \bbound_{\play,\run})
	< \infty
	\quad
	\text{and}
	\quad
\sum_{\run=\start}^{\infty} \step_{\play,\run}^{2} \totbound_{\play,\run}^{2}
	< \infty.
%	\quad
%	\text{for all $\play\in\players$}.
\end{equation}
Then, \acl{wp1}, the sequence of realized actions $\state_{\run}$ converges to the \textpar{necessarily unique} \acl{NE} $\eq$ of $\game$.
\end{theorem}

In particular, if the feedback and stabilization metrics $\bbound_{\play,\run}$, $\totbound_{\play,\run}$ and $\difbound_{\play,\run}$ behave asymptotically as $\bbound_{\play,\run} = \bigoh(1/\run^{\bexp_{\play}})$, $\totbound_{\play,\run} = \bigoh(\run^{\sexp_{\play}})$ and $\difbound_{\play,\run} = \bigoh(1/\run^{\vexp_{\play}})$ for some $\bexp_{\play}, \sexp_{\play}, \vexp_{\play} \geq 0$, we have the following immediate corollaries:

\begin{corollary}
\label{cor:convergence}
With assumptions as above, if each player follows \cref{alg:prox} with $\step_{\play,\run} \propto 1/\run^{\pexp}$ for some $\pexp > \max\{1-\vexp_{\play},1-\bexp_{\play},1/2+\sexp_{\play}\}$, $\pexp\leq1$, the induced sequence of play $\state_{\run}$ converges to \acl{NE} \acl{wp1}.
\end{corollary}

\begin{corollary}
If \cref{alg:prox} is run with perfect oracle feedback and assumptions as above, taking $\pexp > \max_{\play} \vexp_{\play}$ guarantees that $\state_{\run}$ converges to \acl{NE} \acl{wp1}.
\end{corollary}

To streamline our discussion, we postpone the proof of \cref{thm:conv} until later in this section and we proceed below with some remarks.

%----------------------------------------------------------------------
\para{Learning in static games and stochastic approximation}

The special case $\game_{\run} \equiv \game$ for all $\run=\running$ can be seen as learning in a repeated, \emph{static} game.
As we discussed in the introduction, this case has been extensively studied in the literature, usually via the so-called \ac{ODE} method of stochastic approximation \cite{Ben99,BHS05,BMP90}.
In this literature, convergence of a learning process is typically established by showing that an underlying ``mean field'' dynamical system converges, and then using a series of \ac{APT} approximation results to infer that the same applies to the discrete-time algorithm under study as well.

In this direction, the closest result to our own is the recent paper of \citet{MZ19} where the authors showed that a specific, multi-agent version of \citeauthor{Nes09}'s \cite{Nes09} \acl{DA} algorithm converges to \acl{NE} in static, strictly monotone games.
%even with imperfect gradient observations.
However, there is a number of key obstacles that arise when trying to adapt the proof techniques of \citet{MZ19} to our setting.
First and foremost, the prox-mappings $\proxmap_{\play}$ are, in general, \emph{discontinuous} across different faces of $\points_{\play}$, so \eqref{eq:MD} cannot be seen as the discretization of an \ac{ODE} (consider for example the Euclidean case where $\proxmap_{\play}$ is the closest-point projection to $\points_{\play}$).
An approach based on the theory of \acp{DI} \citep{BHS05} could help overcome this obstacle but, even then, the exogenous dependence of $\game_{\run}$ on $\run$ means that the \ac{DI} approximation of the players' learning process would be likewise non-autonomous.
Thus, given that there is no longer a well-defined continuous-time system to approximate, it is not possible to employ a dynamical systems approach as in \cite{MZ19}.

Finally, we should also note that the use of player-specific step-size sequences complicates the discretization landscape even further.
In the stochastic approximation literature, coordinate-specific step-sizes are usually treated within a multiple time-scales framework, \eg as in \cite{Bor97,LC03,LC05,PL14}.
However, in this case, the underlying \ac{ODE} must also separate the faster from the slower time-scales, which means that the players with the smaller step-sizes end up being effectively removed from the game. 
This is an important part of the reason that the literature on learning in static games has traditionally focused on learning algorithms with the same step-size across players \textendash\ and also an important reason that the stochastic approximation approach of \cite{MZ19} does not apply in our setting.

%----------------------------------------------------------------------
\para{Step-size requirements and tuning}

In the literature on learning in games, a common choice for the step-size of iterative methods is the policy $\step_{\play,\run} \propto 1/\run$, \cf \citet{ER98,Beg05,HS09,CMS10,CGM15,BBF20}, and references therein.
In view of \cref{cor:convergence}, if the players' oracle feedback is unbiased and bounded in mean square (\ie $\bexp_{\play}=\infty$, $\sexp_{\play}=0$ for all $\play\in\players$), this step-size policy guarantees convergence to a \acl{NE} as long as the game stabilizes at a power law rate \textendash\ \ie provided that $\difbound_{\run} \defeq \max_{\play} \difbound_{\play,\run} = \bigoh(1/\run^{\vexp})$ for some $\vexp>0$.%
%\footnote{In fact, if the algorithm is run with the slightly more conservative step-size policy $\step_{\run} = 1/(\run\log\run)$, convergence is guaranteed even if the game stabilizes at a much slower, sub-logarithmic rate $\difbound_{\run} = \bigoh(1\/(\log\run)^{\eps})$ \lookout{Typo?} for some $\eps>0$.}
\footnote{More generally, the policy $\step_{\play,\run} \propto 1/\run$ guarantees convergence as long as the bias decays as $\bbound_{\play,\run} = \bigoh(1/\run^{\bexp_{\play}})$ for some $\bexp_{\play}>0$ and the variance grows at most sublinearly ($\sdev_{\play,\run}^{2} = \bigoh(\run^{2\sexp_{\play}})$ for some $\sexp_{\play}<1/2$).}
In fact, if \eqref{eq:MD} is run with $\step_{\play,\run} \propto 1/(\run\log\run)$, convergence is guaranteed even if the game stabilizes at a slower, sub-logarithmic rate $\difbound_{\run} = \bigoh(1/(\log\run)^{\eps})$ for some $\eps>0$.

The policies $\step_{\play,\run} \propto 1/\run$ and $\step_{\play,\run} \propto 1/(\run\log\run)$ should be seen as conservative ``fail-safes'':
it stands to reason that, if more information about the asymptotic behavior of $\difbound_{\play,\run}$ is available, a more aggressive step-size policy (as per \cref{cor:convergence}) might be more efficient.
Specifically, if we focus as above on the case
where the players' oracle feedback is unbiased and bounded in mean square ($\bexp=\infty$, $\sexp=0$),
the second-moment term $\sum_{\run} \step_{\play,\run}^{2} \totbound_{\play,\run}^{2}$ will be subleading in \eqref{eq:step3} relative to the stabilization error term $\sum_{\run} \step_{\play,\run} \difbound_{\play,\run}$ whenever $\pexp \geq \vexp_{\play}$ \emph{for some} $\play\in\players$.
Since the summability condition \eqref{eq:step3} further requires $\pexp > 1-\vexp_{\play}$ \emph{for all} $\play\in\players$, this would suggest taking $\pexp = \min_{\play} \vexp_{\play}$ if $\min_{\play}\vexp_{\play} > 1/2$, and $\pexp$ larger than $1/2$ by an arbitrarily small amount otherwise.

By contrast, if no prior information on $\difbound_{\play,\run}$ is available, it is not clear how to choose the exponent $\pexp$ in an optimal manner relative to the variability of $\game_{\run}$.
In particular, since
%any notion of convergence invariably relies on the \emph{tail} of the sequences involved,
$\vexp_{\play}$ depends on the entire (infinite) tail of $\difbound_{\play,\run}$, adaptive policies that rely on the (finite) history of play up to time $\run$ \textendash\ \eg in the spirit of \citet{RS13-NIPS} and \citet{SALS15} \textendash\ do not seem well-suited for this purpose.
We are not aware of any way to circumvent this difficulty in terms of almost sure convergence of the sequence of play.

%This observation highlights an important difference between regret minimization and convergence to equilibrium.
%On the one hand, a rapidly-decaying step-size policy is more robust in terms of convergence, as it guarantees convergence under the slowest possible stabilization rate of $\game_{\run}$.
%On the other hand, a rapidly vanishing step-size may be suboptimal from the point of view of regret minimization, because it may incur higher regret.
%%exceeds the range of allowable step-sizes \eqref{eq:step-allowed} for all but the smallest values of $\vexp$.
%This disparity is due to the fact that a sequence of games that converges fast to a limit game is very different relative to a sequence of games that oscillates without converging at the same time-scale;
%this can be seen more clearly in \cref{thm:track} below.
%%We explore this trade-off in more detail in the sequel.

\endrev
%----------------------------------------------------------------------

%----------------------------------------------------------------------
%%% Tracking
%----------------------------------------------------------------------
\subsection{Tracking \aclp{NE}}
\label{sec:tracking}

We now turn to the study of time-varying games that evolve \emph{without} converging.
In this case, any notion of convergence for $\state_{\run}$ is meaningless because there is no equilibrium state to converge to, either static or in the mean.
As a result, we will focus instead on whether $\state_{\run}$ is capable of ``tracking'' the game's set of \aclp{NE} over a given horizon of play.

To that end, let $\game_{\run}$ be a sequence of strongly monotone games, and consider the \emph{equilibrium tracking error}
\begin{equation}
\label{eq:err}
\err(\runs)
%	= \sum_{\play\in\players} \err_{\play}(\runs)
	\defeq \sum_{\run\in\runs} \norm{\state_{\run} - \eq_{\run}}^{2}
	= \sum_{\run\in\runs} \sum_{\play\in\players} \norm{\state_{\play,\run} - \eq_{\play,\run}}^{2}
\end{equation}
where $\eq_{\run}$ is the (unique) \acl{NE} of $\game_{\run}$ and $\runs = \window{\runstart}{\runend}$, $\runstart,\runend\in\N$, denotes the playing window of interest.%
\footnote{In games with multiple equilibria, the norm should be replaced by the Hausdorff distance of the corresponding equilibrium sets;
we focus on strongly monotone games to avoid such complications.}
By construction, if $\err(\runs)$ is small relative to $\abs{\runs} = \runend - \runstart$, the sequence of chosen action $\state_{\run}$ will be close to equilibrium for most of the window of interest.
However, if the variability of $\game_{\run}$ (and, in particular, of the equilibrium $\eq_{\run}$) is too high, it is not reasonable to expect a tracking error that grows sublinearly in $\abs{\runs}$, even in the single-player case.

To quantify this, we will also consider the game's \emph{equilibrium variation} (or \emph{drift}) as
\begin{equation}
\label{eq:eqvar}
\tvar(\runs)
	\defeq \sum_{\run\in\runs} \norm{\eq_{\run+1} - \eq_{\run}},
\end{equation}
where
%$\eq_{\run} \in \eqs_{\run} \defeq \Nash(\game_{\run})$ denotes a \acl{NE} of $\game_{\run}$ and,
as before, $\runs = \window{\runstart}{\runend}$ denotes the window of interest.
%----------------------------------------------------------------------
\beginrev
For concision, if $\runs$ is of the form $\runs = \window{\start}{\nRuns}$, we will simply write $\err(\nRuns)$ and $\tvar(\nRuns)$ instead of $\err(\runs)$ and $\tvar(\runs)$ respectively.
In this case, we will say that the equilibrium variation of $\game_{\run}$ is \emph{tame} if
\begin{equation}
\label{eq:NE-tame}
\tvar(\nRuns)
%	\defeq \sum_{\run=1}^{\nRuns} \norm{\eq_{\run+1} - \eq_{\run}}
	= o(\nRuns)
	\quad
	\text{as $\nRuns\to\infty$}
\end{equation}
and we will seek to establish conditions under which \cref{alg:prox} guarantees $\err(\nRuns) = o(\nRuns)$ when \eqref{eq:NE-tame} holds.
Our main result in this direction is as follows:

\begin{theorem}
\label{thm:track}
Let $\game_{\run}$ be a sequence of strongly monotone games satisfying \cref{asm:payv}.
Suppose further that each player $\play\in\players$ runs \cref{alg:prox} with
step-size $\step_{\play,\run} \propto \run^{-\pexp_{\play}}$, $\pexp_{\play}\in(0,1)$,
a Lipschitz \acl{DGF},
and
feedback of the form \eqref{eq:signal} with $\bbound_{\play,\run} = \bigoh(1/\run^{\bexp_{\play}})$ and $\totbound_{\play,\run}^{2} = \bigoh(\run^{2\sexp_{\play}})$ for some $\bexp_{\play},\sexp_{\play} \geq 0$, $\play\in\players$.
Then the players' tracking error is bounded as
\begin{equation}
\label{eq:err-bound}
\txs
\exof{\err(\nRuns)}
%	= \bigoh(\nRuns^{1-\pexp} + \tvar(\nRuns)\nRuns^{2\pexp})
%	+ 2\diam(\points) \sum_{\run=\start}^{\nRuns} \bbound_{\run}.
	= \bigoh\parens*{
		\nRuns^{1-\min_{\play}(\pexp_{\play}-2\sexp_{\play})}
		+ \nRuns^{1-\min_{\play}\bexp_{\play}}
		+ \nRuns^{\max_{\play}\pexp_{\play} + \min_{\play}(\pexp_{\play} - 2\sexp_{\play})}\tvar(\nRuns)}.
\end{equation}
%where $\pexp_{\min} = \min_{\play} \pexp_{\play}$ and $\pexp_{\max} = \max_{\play} \pexp_{\play}$. 
\end{theorem}

\begin{corollary}
\label{cor:track}
Suppose that the players' oracle feedback is unbiased and bounded in mean square \textpar{$\bexp_{\play} = \infty$, $\sexp_{\play} = 0$ for all $\play\in\players$}.
If the equilibrium variation of the game is $\tvar(\nRuns) = \bigoh(\nRuns^{\vexp})$ for some $\vexp>0$,
\cref{alg:prox} enjoys the bound
\begin{equation}
\label{eq:err-bound-powers}
\exof{\err(\nRuns)}
	= \bigoh\parens[\big]{\nRuns^{1-\pexp_{\min}} + \nRuns^{2\pexp_{\max}+\vexp}}.
\end{equation}
where $\pexp_{\min} = \min_{\play} \pexp_{\play}$ and $\pexp_{\max} = \max_{\play} \pexp_{\play}$. 
In particular, if each player runs \cref{alg:prox} with $\step_{\play,\run} \propto 1/\run^{(1-\vexp)/3}$, then
\begin{equation}
\label{eq:err-bound-tuned}
\exof{\err(\nRuns)}
	= \bigoh\parens[\big]{\nRuns^{\frac{2+\vexp}{3}}}.
\end{equation}
\end{corollary}

\cref{thm:track} is our basic equilibrium tracking result, so we proceed with some remarks:

%----------------------------------------------------------------------
\para{Step-size requirements and tuning}
If the players' gradient oracle is unbiased and bounded in mean square ($\bexp_{\play} = \infty$ and $\sexp_{\play}=0$ for all $\play\in\players$), \cref{cor:track} shows that equilibrium tracking is possible as long as
\begin{equation}
\label{eq:step-allowed}
\pexp_{\play}
	< \frac{1-\vexp}{2}
	\quad
	\text{for all $\play\in\players$}.
\end{equation}
Comparing this condition with the step-size requirements for equilibrium convergence (\cf \cref{thm:conv,cor:convergence}), we may infer that equilibrium tracking is more lightweight in terms of prerequisites:
specifically, since \cref{thm:track} does not require the step-size compatibility condition \eqref{eq:step2}, each player can pick $\pexp_{\play}$ independently of one another.
The reason for this difference has to do with the fact that equilibrium tracking focuses on the players' \emph{average} behavior over the horizon of play;
by contrast, the convergence of the sequence of play depends on the entire tail of $\step_{\play,\run}$, so the asymptotic behavior of the players' step-size policies cannot be too different.
%\envend
%\end{remark}

%----------------------------------------------------------------------
\para{Equilibrium tracking and dynamic regret minimization: similarities}

In our setup, the dynamic regret incurred by the $\play$-th player up to time $\nRuns$ under the sequence of play $\state_{\run} \in \points$, $\run=\running$, can be defined as
\begin{equation}
\label{eq:reg-dyn-BR}
\dynreg_{\play}(\nRuns)
	= \sum_{\run=\start}^{\nRuns}
		\bracks{\pay_{\play,\run}(\test_{\play,\run};\state_{-\play,\run}) - \pay_{\play,\run}(\state_{\run})}
	= \sum_{\run=\start}^{\nRuns}
		\bracks{\tilde\pay_{\play,\run}(\test_{\play,\run}) - \tilde\pay_{\play,\run}(\state_{\play,\run})}
\end{equation}
where $\tilde\pay_{\play,\run} \defeq \pay_{\play}(\cdot;\state_{-\play,\run})$ denotes the \emph{effective} payoff function encountered by player $\play\in\players$ at stage $\run$ given the chosen action profile $\state_{-\play,\run}$ of all other players, and
\begin{equation}
\label{eq:BR}
\test_{\play,\run}
	\in \argmax_{\point_{\play}\in\points_{\play}}
		\pay_{\play,\run}(\point_{\play};\state_{-\play,\run})
	= \argmax_{\point_{\play}\in\points_{\play}}
		\tilde\pay_{\play,\run}(\point_{\play})
\end{equation}
denotes the $\play$-th player's \emph{best response} to $\state_{-\play,\run}$ in the game $\game_{\run}$ (the latter assumed fixed as a sequence but otherwise arbitrary and unknown to the players).
\endrev
%----------------------------------------------------------------------
Obviously, if there are no other players in the game, $\test_{\run}$ coincides with the \acl{NE} of the $\run$-th stage game against nature, so a natural question that arises is whether the equilibrium tracking guarantees of \cref{thm:track} can be related to a dynamic regret bound.

%Expanding on this idea, an agent's ``shifting regret'' compares an agent's cumulative payoff to that of an \emph{arbitrary} sequence of actions and focuses either on the number of ``hard shifts'' in the comparator or the total variation thereof \citep{HW98}.
%Another closely related notion is the notion of ``adaptive regret'' of \citet{HazSes09} which compounds the player's regret over contiguous time intervals.
%Albeit stronger as a performance target, the algorithms developed for dynamic/shifting regret minimization also perform well relative to adaptive regret measures, so these notions are often treated concurrently.
%%We will briefly revisit all this in the sequel;
%For an introduction to this literature, we refer the reader to the excellent book by \citet{CBL06}, as well as the original works of \citet{CBGLS12}, \citet{GLL12}, \citet{HW13,HW15}, \citet{BGZ15}, \citet{JRSS15}, and \citet{GS16}.

In this regard, a slight modification of the proof of \cref{thm:track} yields the following:
if an agent with a convex compact action set $\points$ runs \cref{alg:prox} with step-size $\step_{\run}\propto 1/\run^{\pexp}$ against a stream of concave \textendash\ though not necessarily \emph{strongly} concave \textendash\ payoff functions $\pay_{\run}\from\points\to\R$ with drift $\tvar(\nRuns)$, then
\begin{equation}
\label{eq:reg-mean-dyn}
\txs
\exof{\dynreg(\nRuns)}
%	= \bigoh(\nRuns^{1+\pexp-\qexp} + \nRuns^{1-\bexp} + \nRuns^{1+2\sexp-\pexp} + \nRuns^{\qexp} \tvar(\nRuns)).
	= \bigoh\parens*{\nRuns^{1+2\sexp-\pexp} + \nRuns^{1-\bexp} + \nRuns^{2\pexp-2\sexp}\tvar(\nRuns)}.
\end{equation}
In particular, if $\tvar(\nRuns) = \bigoh(\nRuns^{\vexp})$ and the player's oracle feedback is unbiased and bounded in mean square ($\bexp=\infty$, $\sexp=0$), the choice $\pexp = (1-\vexp)/3$ guarantees
\begin{equation}
\label{eq:reg-mean-dyn-opt}
\exof{\dynreg(\nRuns)}
	= \bigoh\parens[\big]{\nRuns^{\frac{2+\vexp}{3}}}.
\end{equation}
For a precise statement and proof, we refer the reader to \cref{sec:regret}.%
\footnote{The guarantee \eqref{eq:reg-mean-dyn} is not a consequence of \cref{thm:track} because it concerns function values and it makes no strong concavity assumptions for the payoff functions faced by the agent;
the proof, however, is similar.}

%----------------------------------------------------------------------
\para{Equilibrium tracking and dynamic regret minimization: differences}

\revise{%
Going back to the multi-agent case, the sequence $\test_{\run} = (\test_{\play,\run})_{\play\in\players}$ with $\test_{\play,\run}$ given by \eqref{eq:BR} may be very different from the \acl{NE} sequence $\eq_{\run}$:
the former best responds to the actual sequence of play $\state_{\run}$, while the latter best responds to itself (so it depends only on $\game_{\run}$ and is otherwise \emph{independent} of $\state_{\run}$).}
As we saw above, this distinction is redundant in the single-player case, but it is crucial in the multi-agent one:
the sequence $\test_{\run}$ may vary rapidly even if $\eq_{\run}$ is constant.
For example, even if the sequence of base payoff functions $\pay_{\play,\run}$ does not depend on $\run$ \emph{explicitly} (\ie $\pay_{\play,\run} \equiv \pay_{\play}$ for all $\run$), the \emph{effective} payoff functions $\tilde\pay_{\play,\run} \defeq \pay_{\play}(\cdot;\state_{-\play,\run})$ encountered individually by each agent depend on $\run$ \emph{implicitly} via $\state_{-\play,\run}$.
%As a result, the bound \eqref{eq:reg-mean-dyn} does not a priori apply:
%%$\dynreg_{\play}(\nRuns)$ is no longer bounded by \eqref{eq:reg-mean-dyn}:
%since $\norm{\state_{-\play,\run+1} - \state_{-\play,\run}} = \bigoh(\step_{\run}) = \bigoh(\run^{-\pexp})$, the variation of $\tilde\pay_{\play,\run}$ over a window of length $\nRuns$ under \cref{alg:prox} could be as high as $\Theta(\nRuns^{1-\pexp})$, in which case the bound \eqref{eq:reg-mean-dyn} is no longer applicable.

This subtlety is also reflected on the strong monotonicity assumption in \cref{thm:track} which invites the question whether the bound \eqref{eq:err-bound} is tight.
To wit, when faced with a sequence of \emph{strongly} concave payoff functions, \citet{BGZ15} showed that an adversary can always impose $\dynreg(\nRuns) = \Omega(\tvar(\nRuns)^{1/2}\nRuns^{1/2})$.
This bound is strictly better than the $\bigoh(\nRuns^{1-\pexp} + \tvar(\nRuns)\nRuns^{2\pexp})$ guarantee of \cref{cor:track}, suggesting that there may be room for improvement.
Nevertheless, there are two important roadblocks to achieve this:
\begin{enumerate}
\addtolength{\itemsep}{\smallskipamount}
\item
First, in the single-agent case, the key to attaining faster regret minimization is the basic inequality
\begin{equation}
\label{eq:pay-strong}
\pay_{\run}(\eq_{\run}) - \pay_{\run}(\point)
	\leq \braket{\vecfield_{\run}(\point)}{\eq_{\run} - \point}
	- \frac{\strong}{2} \norm{\point - \eq_{\run}}^{2}
\end{equation}
where $\eq_{\run}$ denotes the (necessarily unique) maximizer of $\pay_{\run}$.
As a result, the growth of $\gap(\nRuns)$ \textendash\ which is driven by gradient terms of the form $\braket{\vecfield_{\run}(\state_{\run})}{\eq_{\run} - \state_{\run}}$ \textendash\ is mitigated by the quadratic correction terms:
by balancing these two terms, it is possible to obtain sharper bounds for $\dynreg(\nRuns)$ when each $\pay_{\run}$ is strongly concave.

On the other hand, in a multi-agent, game-theoretic setting, \eqref{eq:pay-strong} becomes
\begin{equation}
\label{eq:pay-strong-multi}
\pay_{\play,\run}(\eq_{\play,\run};\point_{-\play}) - \pay_{\play,\run}(\point)
	\leq \braket{\vecfield_{\play,\run}(\point)}{\eq_{\play,\run} - \point_{\play}}
	- \frac{\strong}{2} \norm{\point_{\play} - \eq_{\play,\run}}^{2}
\end{equation}
where $\eq_{\run}$ now denotes the (necessarily unique) \acl{NE} of the strongly monotone stage game $\game_{\run} \equiv \game_{\run}(\players,\points,\pay_{\run})$.
Arguing as in the single-agent setting would indeed yield a sharper bound on the quantity
\begin{equation}
\sum_{\run\in\runs} \sum_{\play\in\players}
	\bracks{\pay_{\play,\run}(\eq_{\play,\run};\state_{-\play,\run}) - \pay_{\play,\run}(\state_{\run})}
\end{equation}
but, in general, the minimization of this quantity does not provide a certificate that $\state_{\run}$ is in any way close to equilibrium.
In particular, in contrast to the single-agent case, \eqref{eq:pay-strong-multi} could be either positive or negative, so it cannot act as a merit function for tracking an evolving equilibrium.

\item
Second, the optimal \emph{static} regret minimization rate in strongly convex problems is attained when $\step_{\run} \propto 1/\run$.
However, \citet{BGZ15} provide a counterexample where this step-size policy produces \emph{linear} dynamic regret.
In view of this,
\revise{achieving an $\bigoh(\tvar(\nRuns)^{1/2}\nRuns^{1/2})$ dynamic regret minimization rate would seem to require a different approach and/or assumptions \textendash\ for example, an adaptive policy in the spirit of \citet{JRSS15} in the case of perfect gradient feedback.}
\end{enumerate}

We mention the above to emphasize that
%even though the obtained guarantees look similar,
bounding the equilibrium tracking error $\err(\nRuns)$ is significantly different than bounding the dynamic regret of an individual agent in the unilateral setting (even though the obtained guarantees look similar).
\revise{It is reasonable to conjecture that the bound \eqref{eq:err-bound} may be improved in games that admit a strongly concave potential function,
%(perhaps through the use of a finely tuned restart mechanism),
but the general case seems considerably more difficult.}

%----------------------------------------------------------------------
\beginrev
\para{Legendre \acp{DGF}}

We should also note that the reciprocity condition \eqref{eq:RC} has been replaced in the statement of \cref{thm:track} by the stronger requirement $\sup_{\point_{\play}}\dnorm{\nabla\hreg_{\play}(\point_{\play})} < \infty$ which rules out Legendre-like \acp{DGF} (such as the entropic setup of \cref{ex:prox-entropic}).
This condition is needed in \cref{prop:gap},
%an intermediate step in the proof of \cref{thm:track},
which requires a finite Bregman diameter $\bdiam_{\play} \defeq \sup_{\point_{\play},\alt\point_{\play}} \breg_{\play}(\point_{\play},\alt\point_{\play})$ to bound the ``regret-like'' quantity $\sum_{\run=\start}^{\nRuns} \braket{\vecfield_{\play,\run}(\state_{\run})}{\point_{\play} - \state_{\play,\run}}$.
\citet{OP18} recently showed that \eqref{eq:MD} may incur linear regret when run with a variable step-size in problems with infinite Bregman diameter, so this requirement is not an artifact of the analysis.

That being said, there are several ways to overcome this hurdle:
First, the players could run \eqref{eq:MD} with a constant step-size over windows of a specified length and use a restart mechanism to achieve a sublinear equilibrium tracking error;
this approach was proposed by \citet{BGZ15} for the minimization of dynamic regret and we discuss it in more detail in \cref{sec:regret}.
Another way is to add an ``anchoring term'' in the definition of the prox-mapping $\proxmap_{\play}$ and play the so-called \emph{dual-stabilized} \acl{MD} policy
\begin{equation}
\label{eq:DS-MD}
\tag{DS-MD}
\state_{\play,\run+1}
	= \argmin_{\point_{\play}\in\points_{\play}}
		\braces*{
			\step_{\play,\run}\braket{\signal_{\play,\run}}{\state_{\play,\run} - \point_{\play}}
			+ \breg_{\play}(\point_{\play},\state_{\play,\run})
			+ \parens{\step_{\play,\run+1}^{-1} - \step_{\play,\run}^{-1}} \breg_{\play}(\point_{\play},\state_{\play,\start})
	}
\end{equation}
This policy was introduced by \citet{FHPF20} who showed that \eqref{eq:DS-MD} achieves sublinear regret even in domains with an infinite Bregman diameter.
Finally, another \textendash\ and arguably simpler \textendash\ approach is to switch to the \acl{DA} policy of \citet{Nes09} which instead prescribes
\begin{equation}
\label{eq:DA}
\tag{DA}
\state_{\play,\run+1}
	= \argmax_{\point_{\play}\in\points_{\play}}
		\braces*{
			\sum_{\runalt=\start}^{\run} \braket{\signal_{\play,\runalt}}{\point_{\play}}
			- \step_{\play,\run}\hreg_{\play}(\point_{\play})
		}.
\end{equation}
This algorithm has the advantage of attaining order-optimal regret guarantees with the Bregman diameter $\bdiam_{\play}$ replaced by the range $\mathcal{R}_{\play} \defeq \max\hreg_{\play} - \min\hreg_{\play}$ of $\hreg_{\play}$ (which is always finite since $\points_{\play}$ is compact and the domain of $\hreg_{\play}$ contains $\points_{\play}$).
Either of these algorithmic tweaks would ultimately yield a sublinear tracking error in domains with an infinite Bregman diameter, but the details lie beyond the scope of our work so we do not discuss them here.

\endrev
%----------------------------------------------------------------------

%----------------------------------------------------------------------
%%% Nash convergence
%----------------------------------------------------------------------
\subsection{Proof of \cref{thm:conv}\afterhead}
\label{sec:convergence}

%----------------------------------------------------------------------
\beginrev

The rest of this section is devoted to proving the results stated above, starting with the proof of \cref{thm:conv}.
The first key step in this direction is the definition of a suitable ``energy-like'' function that is \textendash\ on average and up to small, second-order errors \textendash\ decreasing along the trajectory of play $\state_{\run}$.
In the analysis of \acl{MD} algorithms, this role is usually played by the Bregman divergence relative to the target point under study (in our case, the \acl{NE} of $\game$).
However, because each player $\play\in\players$ now learns at a different pace (as determined by their individual step-size policy $\step_{\play,\run}$), the definition of a suitable energy function for \cref{alg:prox} is not as straightforward.

To that end (and with a fair amount of hindsight), we begin by introducing the player-specific weights
\begin{equation}
\label{eq:weight}
\weight_{\play}
	= \parens*{\prod\nolimits_{\playalt\in\players} \weight_{\play\playalt}}^{1/\nPlayers}
	\quad
	\text{for all $\play\in\players$},
\end{equation}
with $\weight_{\play\playalt} > 0$, $\play,\playalt\in\players$, given by the mutual compatibility condition \eqref{eq:step2}.
As we show below, these weights enjoy a decomposition property that is key for the sequel:

\begin{lemma}
\label{lem:steps}
Suppose that $\step_{\play,\run}$ satisfies \eqref{eq:step1} and \eqref{eq:step2}.
Then $\weight_{\play\playalt} = \weight_{\play} / \weight_{\playalt}$ for all $\play,\playalt\in\players$.
\end{lemma}

\begin{Proof}
Our proof relies on the two intermediate claims below:
\begin{enumerate}
[left=0pt,label={\bfseries Claim \arabic*:}]
\item
\emph{The weights $\weight_{\play\playalt}$ are uniquely defined.}
Indeed, suppose that \eqref{eq:step2} holds also with $\alt\weight_{\play\playalt} \neq \weight_{\play\playalt}$ for some $\play,\playalt\in\players$.
Then, for all $\run=\start$, we have:
\begin{align}
\abs{\weight_{\play\playalt} - \alt\weight_{\play\playalt}} \step_{\playalt,\run}
	= \abs{\weight_{\play\playalt} \step_{\playalt,\run} - \alt\weight_{\play\playalt} \step_{\playalt,\run}}
%	\notag\\
	\leq \abs{\step_{\play,\run} - \weight_{\play\playalt}\step_{\playalt,\run}}
		+ \abs{\step_{\play,\run} - \alt\weight_{\play\playalt}\step_{\playalt,\run}}
\end{align}
so $\abs{\weight_{\play\playalt} - \alt\weight_{\play\playalt}} \step_{\playalt,\run}$ is summable given that both $\abs{\step_{\play,\run} - \weight_{\play\playalt}\step_{\playalt,\run}}$ and $\abs{\step_{\play,\run} - \alt\weight_{\play\playalt}\step_{\playalt,\run}}$ are summable (by assumption).
This contradicts \eqref{eq:step1} so our claim follows.

\item
\emph{The weights $\weight_{\play\playalt}$ satisfy the chain rule $\weight_{\play\playaltalt} = \weight_{\play\playalt} \weight_{\playalt\playaltalt}$ for all $\play,\playalt,\playaltalt\in\players$.}
Indeed:
\begin{align}
\sum_{\run=\start}^{\infty}
	\abs{\step_{\play,\run}
			- \weight_{\play\playalt}\weight_{\playalt\playaltalt} \step_{\playaltalt,\run}}
	&= \sum_{\run=\start}^{\infty}
		\abs{\step_{\play,\run}
			- \weight_{\play\playalt} \step_{\playalt,\run}
			+ \weight_{\play\playalt} \step_{\playalt,\run}
			- \weight_{\play\playalt}\weight_{\playalt\playaltalt} \step_{\playaltalt,\run}}
	\notag\\
	&\leq \sum_{\run=\start}^{\infty}
		\abs{\step_{\play,\run}
			- \weight_{\play\playalt} \step_{\playalt,\run}}
	+ \weight_{\play\playalt} \sum_{\run=\start}^{\infty}
		\abs{\step_{\playalt,\run}
			- \weight_{\playalt\playaltalt} \step_{\playaltalt,\run}}
	\notag\\
	&< \infty
\end{align}
with the last inequality following from \eqref{eq:step2}.
Our claim then follows from the definition of $\weight_{\play\playaltalt}$ and our uniqueness claim above.
\end{enumerate}
Thus, with these two claims in hand, we readily obtain
\begin{equation}
\frac{\weight_{\play}}{\weight_{\playalt}}
	= \frac
		{\parens*{\prod_{\playaltalt\in\players} \weight_{\play\playaltalt}}^{1/\nPlayers}}
		{\parens*{\prod_{\playaltalt\in\players} \weight_{\playalt\playaltalt}}^{1/\nPlayers}}
%	= \prod_{\playaltalt=1}^{\nPlayers} \parens{\weight_{\play\playaltalt} \weight_{\playaltalt\playalt}}^{1/\nPlayers}
	= \prod_{\playaltalt\in\players} \parens{\weight_{\play\playaltalt} / \weight_{\playalt\playaltalt}}^{1/\nPlayers}
	= \prod_{\playaltalt\in\players} \weight_{\play\playalt}^{1/\nPlayers}
	= \weight_{\play\playalt}
\end{equation}
where, in the third step, we used the chain rule above to write $\weight_{\play\playalt} = \weight_{\play\playaltalt} / \weight_{\playalt\playaltalt}$.
This establishes our assertion and completes our proof.
\end{Proof}

This lemma shows that \eqref{eq:step2} can be rewritten as
%$\sum_{\run=\start}^{\infty} \abs{\frac{\step_{\play,\run}}{\weight_{\play}} - \frac{\step_{\playalt,\run}}{\weight_{\playalt}}} < \infty$
$\sum_{\run=\start}^{\infty} \abs{\step_{\play,\run}/\weight_{\play} - \step_{\playalt,\run}/\weight_{\playalt}} < \infty$,
which in turn implies that $\weight_{\play}$ can be interpreted as the relative ``learning speed'' of player $\play\in\players$.
In view of this, we will consider the effective step-size
\begin{equation}
\label{eq:step}
\step_{\run}
	= \frac{1}{\nPlayers} \sum_{\play\in\players} \frac{\step_{\play,\run}}{\weight_{\play}}
\end{equation}
and the energy function
\begin{equation}
\label{eq:energy}
\energy(\point)
	= \sum_{\play\in\players} \frac{\breg_{\play}(\eq_{\play},\point_{\play})}{\weight_{\play}}
\end{equation}
where
$\eq_{\play}\in\points_{\play}$ denotes the $\play$-th component of the \acl{NE} $\eq$ of $\game$.
We then have the following quasi-descent inequality for $\energy$ under \eqref{eq:MD}:

\begin{lemma}
\label{lem:energy}
Suppose that each player $\play\in\players$ runs \cref{alg:prox} with a step-size policy $\step_{\play,\run}$ satisfying \eqref{eq:step1} and \eqref{eq:step2}.
Then the iterates $\energy_{\run} \defeq \energy(\state_{\run})$ of $\energy$ under $\state_{\run}$ enjoy the bound
\begin{align}
\label{eq:energy-bound}
\energy_{\run+1}
	\leq \energy_{\run}
	+ \step_{\run} \braket{\vecfield(\state_{\run})}{\state_{\run} - \eq}
	&+ \sum_{\play\in\players} \frac{\step_{\play,\run}}{\weight_{\play}}
		\braket{\diff_{\play,\run} + \error_{\play,\run}}{\state_{\play,\run} - \eq_{\play}}
	+ \sum_{\play\in\players} \frac{\step_{\play,\run}^{2}}{2\weight_{\play}\hstr_{\play}} \dnorm{\signal_{\play,\run}}^{2}
	\notag\\
	&+ \frac{\max_{\play} \vbound_{\play} \diam(\points_{\play})}{\nPlayers} \sum_{\play,\playalt\in\players}
		\abs*{\frac{\step_{\play,\run}}{\weight_{\play}} - \frac{\step_{\playalt,\run}}{\weight_{\playalt}}}
\end{align}
with $\diff_{\play,\run} = \vecfield_{\play,\run}(\state_{\run}) - \vecfield_{\play}(\state_{\run})$.
\end{lemma}

\begin{Proof}
By \cref{lem:template} in \cref{app:Bregman}, the Bregman divergence $\breg_{\play,\run} \defeq \breg_{\play}(\eq_{\play},\state_{\play,\run})$ satisfies the inequality
\begin{equation}
\label{eq:template-play}
\breg_{\play,\run+1}
	\leq \breg_{\play,\run}
		+ \step_{\play,\run} \braket{\signal_{\play,\run}}{\state_{\play,\run} - \eq_{\play}}
		+ \frac{\step_{\play,\run}^{2}}{2\hstr_{\play}} \dnorm{\signal_{\play,\run}}^{2}.
\end{equation}
Therefore, with $\signal_{\play,\run} = \vecfield_{\play,\run}(\state_{\run}) + \error_{\play,\run} = \vecfield_{\play}(\state_{\run}) + \diff_{\play,\run} + \error_{\play,\run}$ and $\energy_{\run} = \sum_{\play\in\players} \weight_{\play}^{-1} \breg_{\play,\run}$, we get
\begin{subequations}
\label{eq:energy-temp}
\begin{align}
\energy_{\run+1}
	\leq \energy_{\run}
%	= \sum_{\play\in\players} \frac{\breg_{\play,\run+1}}{\weight_{\play}}
%	\leq \sum_{\play\in\players} \frac{\breg_{\play,\run}}{\weight_{\play}}
		+ \sum_{\play\in\players} \frac{\step_{\play,\run}}{\weight_{\play}}
			\braket{\error_{\play,\run} + \diff_{\play,\run}}{\state_{\play,\run} - \eq_{\play}}
		&+ \sum_{\play\in\players} \frac{\step_{\play,\run}^{2}}{2\weight_{\play}\hstr_{\play}} \dnorm{\signal_{\play,\run}}^{2}.
	\\
\label{eq:energy-step}
		&+ \sum_{\play\in\players} \frac{\step_{\play,\run}}{\weight_{\play}} \braket{\vecfield_{\play}(\state_{\run})}{\state_{\play,\run} - \eq_{\play}}
\end{align}
\end{subequations}
so it suffices to upper bound the term \eqref{eq:energy-step} of the above inequality.
To that end, we have
\begin{align}
\eqref{eq:energy-step}
	&= \step_{\run} \braket{\vecfield(\state_{\run})}{\state_{\run} - \eq}
		+ \sum_{\play\in\players}
			\parens*{\frac{\step_{\play,\run}}{\weight_{\play}} - \step_{\run}}
			\braket{\vecfield_{\play}(\state_{\run})}{\state_{\play,\run} - \eq_{\play}}
	\notag\\
	&\leq \step_{\run} \braket{\vecfield(\state_{\run})}{\state_{\run} - \eq}
		+ \sum_{\play\in\players}
			\abs*{\frac{\step_{\play,\run}}{\weight_{\play}} - \step_{\run}}
				\cdot \vbound_{\play} \diam(\points_{\play})
	\notag\\
	&= \step_{\run} \braket{\vecfield(\state_{\run})}{\state_{\run} - \eq}
		+ \sum_{\play\in\players}
			\frac{\vbound_{\play} \diam(\points_{\play})}{\nPlayers}
			\abs*{\sum_{\playalt\in\players}
				\parens*{\frac{\step_{\play,\run}}{\weight_{\play}} - \frac{\step_{\playalt,\run}}{\weight_{\playalt}}}}
	\notag\\
	&\leq \step_{\run} \braket{\vecfield(\state_{\run})}{\state_{\run} - \eq}
		+ \frac{\max_{\play} \vbound_{\play} \diam(\points_{\play})}{\nPlayers}
			\sum_{\play,\playalt\in\players}
				\abs*{\frac{\step_{\play,\run}}{\weight_{\play}} - \frac{\step_{\playalt,\run}}{\weight_{\playalt}}}.
\end{align}
Our claim then follows by substituting this bound back in \eqref{eq:energy-temp}.
\end{Proof}

The importance of the energy-like bound \eqref{eq:energy-bound} lies in that the ``drift term'' $\step_{\run} \braket{\vecfield(\state_{\run})}{\state_{\run} - \eq}$ provides a leading negative contribution to $\energy_{\run}$ (since $\eq$ is a \acl{NE} of $\game$), while all other terms become vanishingly small over time.
The proposition below formalizes this idea and shows that $\energy_{\run}$ converges to some (random) finite value:

%Subsequently, we show that $\state_{\run}$ cannot remain at uniformly positive distance away from $\eq$ for all sufficiently large $\run$.
%Combining these results will show that $\state_{\run}$ can only converge to the zero-level set of the Bregman divergence, \ie $\lim_{\run\to\infty} \state_{\run} = \eq$.

\begin{proposition}
\label{prop:quasiFejer}
Suppose that each player $\play\in\players$ runs \cref{alg:prox} with a step-size $\step_{\play,\run}$ satisfying \eqref{eq:step1}, \eqref{eq:step2} and \eqref{eq:step3}.
Then $\energy_{\run}$ converges \as to a random variable $\energy_{\infty}$ with $\exof{\energy_{\infty}} < \infty$.
\end{proposition}

\begin{Proof}
We begin by decomposing each player's oracle signal as
\begin{equation}
\signal_{\play,\run}
	= \vecfield_{\play,\run}(\state_{\run})
		+ \bias_{\play,\run}
		+ \noise_{\play,\run}
	= \vecfield_{\play}(\state_{\run})
		+ \diff_{\play,\run}
		+ \bias_{\play,\run}
		+ \noise_{\play,\run}
\end{equation}
and we set respectively
\begin{subequations}
\label{eq:errors}
\begin{alignat}{2}
\sdiff_{\play,\run}
	&= \braket{\diff_{\play,\run}}{\state_{\play,\run} - \eq_{\play}}
	&\qquad
\sdiff_{\run}
	&= \sum_{\play\in\players} \frac{\step_{\play,\run}}{\weight_{\play} \step_{\run}} \sdiff_{\play,\run}
	\\
\sbias_{\play,\run}
	&= \braket{\bias_{\play,\run}}{\state_{\play,\run} - \eq_{\play}}
	&\qquad
\sbias_{\run}
	&= \sum_{\play\in\players} \frac{\step_{\play,\run}}{\weight_{\play} \step_{\run}} \sbias_{\play,\run}
\shortintertext{and}
\snoise_{\play,\run}
	&= \braket{\noise_{\play,\run}}{\state_{\play,\run} - \eq_{\play}}
	&\qquad
\snoise_{\run}
	&= \sum_{\play\in\players} \frac{\step_{\play,\run}}{\weight_{\play} \step_{\run}} \snoise_{\play,\run}
\end{alignat}
\end{subequations}
with
$\step_{\run}$ given by \eqref{eq:step}
and
$\diff_{\play,\run} = \vecfield_{\play,\run}(\state_{\play,\run}) - \vecfield_{\play}(\state_{\run})$ defined as in \cref{lem:energy}.
The energy inequality \eqref{eq:energy-bound} then gives
\begin{align}
\label{eq:energy-temp2}
\energy_{\run+1}
	\leq \energy_{\run}
	+ \step_{\run} \braket{\vecfield(\state_{\run})}{\state_{\run} - \eq}
		+ \step_{\run} \parens{\sdiff_{\run} + \sbias_{\run} + \snoise_{\run}}
		+ \stepgap_{\run}
		+ \sum_{\play\in\players} \frac{\step_{\play,\run}^{2}}{2\weight_{\play}\hstr_{\play}} \dnorm{\signal_{\play,\run}}^{2}
\end{align}
where we set
\begin{equation}
\label{eq:step-gap}
\stepgap_{\run}
	= \frac{\max_{\play} \vbound_{\play} \diam(\points_{\play})}{\nPlayers}
		\sum_{\play,\playalt\in\players}
			\abs*{\frac{\step_{\play,\run}}{\weight_{\play}} - \frac{\step_{\playalt,\run}}{\weight_{\playalt}}}.
\end{equation}
Therefore,
%given that $\braket{\vecfield(\state_{\run})}{\state_{\run} - \eq} \leq 0$ (since $\eq$ is a \acl{NE} of $\game$),
conditioning on the history $\filter_{\run}$ of $\state_{\run}$ up to stage $\run$ (inclusive) and taking expectations, we get
\begin{align}
\label{eq:energy-temp3}
\exof{\energy_{\run+1} \given \filter_{\run}}
	&\leq \exof*{
		\energy_{\run}
			+ \step_{\run} \braket{\vecfield(\state_{\run})}{\state_{\run} - \eq}
			+ \step_{\run} \parens{\sdiff_{\run} + \sbias_{\run} + \snoise_{\run}}
			+ \stepgap_{\run}
			+ \sum_{\play\in\players} \frac{\step_{\play,\run}^{2} \dnorm{\signal_{\play,\run}}^{2}}{2\weight_{\play}\hstr_{\play}}
		\given \filter_{\run}}
	\notag\\
	&\leq \energy_{\run}
		+ \step_{\run} \parens{\sdiff_{\run} + \sbias_{\run}}
		+ \stepgap_{\run}
		+ \sum_{\play\in\players} \frac{\step_{\play,\run}^{2}}{2\weight_{\play}\hstr_{\play}} \sbound_{\play,\run}^{2}
\end{align}
where we used the definition \eqref{eq:sbound} of $\sbound_{\play,\run}$ and the fact that
\begin{enumerate*}
[\itshape a\upshape)]
\item
$\eq$ is a \acl{NE} of $\game$ (so $\braket{\vecfield(\state_{\run})}{\state_{\run} - \sol} \leq 0$);
\item
$\sdiff_{\run}$ and $\sbias_{\run}$ are both $\filter_{\run}$-measurable (by definition);
and
\item
$\exof{\snoise_{\run} \given \filter_{\run}} = \braket{\exof{\noise_{\run} \given \filter_{\run}}}{\state_{\run} - \sol} = 0$.
\end{enumerate*}

To proceed, note that
\begin{equation}
\sdiff_{\play,\run}
	= \braket{\diff_{\play,\run}}{\state_{\play,\run} - \eq_{\play}}
	\leq \dnorm{\diff_{\play,\run}} \norm{\state_{\play,\run} - \eq_{\play}}
	\leq \diam(\points_{\play}) \difbound_{\play,\run}
\end{equation}
and, similarly, $\sbias_{\play,\run} \leq \diam(\points_{\play}) \bbound_{\play,\run}$.
The bound \eqref{eq:energy-temp3} may then be written as
%\begin{equation}
\(
\exof{\energy_{\run+1} \given \filter_{\run}}
	\leq \energy_{\run}
	+ \eps_{\run}
\)
%\end{equation}
where
\begin{equation}
\eps_{\run}
	= \sum_{\play\in\players} \bracks*{
		\frac{\step_{\play,\run}}{\weight_{\play}}
			\diam(\points_{\play})
			\cdot (\difbound_{\play,\run} + \bbound_{\play,\run})
		+\frac{\step_{\play,\run}^{2}}{2\weight_{\play}\hstr_{\play}} \sbound_{\play,\run}^{2}}.
\end{equation}
\endrev
%----------------------------------------------------------------------
Consider now the auxiliary process $\zeta_{\run} = \energy_{\run+1} + \sum_{\runalt=\run+1}^{\infty} \eps_{\runalt}$.
Taking expectations yields
\begin{equation}
\exof{\zeta_{\run} \given \filter_{\run}}
	\leq \energy_{\run} + \eps_{\run} + \sum_{\runalt=\run+1}^{\infty} \eps_{\runalt}
	= \energy_{\run} + \sum_{\runalt=\run}^{\infty} \eps_{\runalt}
	= \zeta_{\run-1},
\end{equation}
\ie $\zeta_{\run}$ is a supermartingale relative to $\filter_{\run}$.
Moreover, since $\sum_{\run=\start}^{\infty} \eps_{\run} < \infty$ by \eqref{eq:step3} and \cref{lem:steps}, we also get $\exof{\zeta_{\run}} \leq \exof{\zeta_{\start}} < \infty$, \ie $\zeta_{\run}$ is bounded in $L^{1}$.
Therefore, by Doob's (sub)martingale convergence theorem \citep[Theorem~2.5]{HH80}, it follows that $\zeta_{\run}$ converges almost surely to some random variable $\zeta$ that is itself finite (almost surely and in $L^{1}$).
Since $\energy_{\run} = \zeta_{\run-1} - \sum_{\runalt=\run}^{\infty} \eps_{\runalt}$ and $\lim_{\run\to\infty} \sum_{\runalt=\run}^{\infty} \eps_{\runalt} = 0$,
we conclude that $\energy_{\run}$ converges \as to $\zeta$ and our proof is complete.
\end{Proof}

\endrev
%----------------------------------------------------------------------

Moving forward, our next result shows that we can extract a subsequence of $\state_{\run}$ that converges to a \acl{NE} of the limit game $\game$:

%----------------------------------------------------------------------
\beginrev

\begin{proposition}
\label{prop:subseq}
%\Acl{wp1}, there exists a \textpar{random} subsequence $\state_{\run_{k}}$ of $\state_{\run}$ which converges to $\eq$.
With assumptions as in \cref{prop:quasiFejer}, we have $\liminf_{\run} \norm{\state_{\run} - \sol} = 0$ \as.
\end{proposition}

\begin{Proof}
We begin by showing that, for all $\eps>0$, the hitting time
\begin{equation}
\label{eq:hit}
\tau_{\eps}
	= \inf\setdef{\run\in\N}{\norm{\state_{\run} - \eq} \geq \eps}
\end{equation}
is finite \acl{wp1};
formally, we will show that the event $\event_{\eps} = \{\tau_{\eps} = \infty\}$ has $\probof{\event_{\eps}} = 0$ for all $\eps>0$.

To do so, fix some $\eps>0$ and let $\const_{\eps} = -\inf\setdef{\braket{\vecfield(\point)}{\point - \eq}}{\norm{\point-\eq} \geq \eps}$, so $\const_{\eps} > 0$ by the strict monotonicity of $\game$ and the fact that $\vecfield$ is continuous and $\points$ is compact.
Then, with notation as in the proof of \cref{prop:quasiFejer}, telescoping the bound \eqref{eq:energy-temp2} yields
\begin{align}
\label{eq:energy-run}
\energy_{\run+1}
	&\leq \energy_{\start}
		- \const_{\eps} \sum_{\runalt=\start}^{\run} \step_{\runalt}
		+ \underbrace{
			\sum_{\runalt=\start}^{\run} \step_{\runalt} (\sdiff_{\runalt} + \sbias_{\runalt})
			+ \sum_{\runalt=\start}^{\run} \stepgap_{\runalt}
			}_{\termOne_{\run}}
		+ \underbrace{
			\sum_{\runalt=\start}^{\run} \step_{\runalt} \snoise_{\runalt}
			}_{\termTwo_{\run}}
		+ \underbrace{
			\sum_{\runalt=\start}^{\run}
			\sum_{\play\in\players}
				\frac{\step_{\play,\runalt}^{2}}{2\weight_{\play}\hstr_{\play}}
				\dnorm{\signal_{\play,\runalt}}^{2}
				}_{\termThree_{\run}}
\end{align}
for all $\run \leq \tau_{\eps}$.
%Hence, setting $\runtime_{\run} = \sum_{\runalt=\start}^{\run} \step_{\runalt}$ and using \eqref{eq:farbound}, we get the bound
%\begin{equation}
%\label{eq:energy-run}
%\breg_{\run+1}
%	\leq \breg_{\start}
%	- \runtime_{\run} \bracks*{
%		\const
%		- \frac{\sum_{\runalt=\start}^{\run} \step_{\runalt} (\sdiff_{\runalt} + \sbias_{\runalt})}{\runtime_{\run}}
%%		- \frac{\sum_{\runalt=\start}^{\run} \step_{\runalt} \sbias_{\runalt}}{\runtime_{\run}}
%		- \frac{\sum_{\runalt=\start}^{\run} \step_{\runalt} \snoise_{\runalt}}{\runtime_{\run}}
%		- \frac{1}{2\hstr} \frac{\sum_{\runalt=\start}^{\run} \step_{\runalt}^{2} \dnorm{\signal_{\runalt}}^{2}}{\runtime_{\run}}
%	}.
%\end{equation}
We now proceed to bound each of the underscored terms above:

\begin{enumerate}[left=\parindent,label={\arabic*.}]
\addtolength{\itemsep}{\smallskipamount}

%\item
%The first sum in \eqref{eq:energy-run} can be expressed for convenience as $-\const_{\eps}\runtime_{\run}$ where $\runtime_{\run} = \sum_{\runalt=\start}^{\run} \step_{\run}$ is the algorithm's ``running length'' up to time $\run$.

\item
First, for the term $\termOne_{\run}$, we have shown in the proof of \cref{prop:quasiFejer} that
\begin{equation}
\step_{\run} (\sdiff_{\run} + \sbias_{\run})
	\leq \sum_{\play\in\players}
		\frac{\step_{\play,\run}}{\weight_{\play}}
			\diam(\points_{\play})
			\cdot (\difbound_{\play,\run} + \bbound_{\play,\run})
\end{equation}
so $\sum_{\run=\start}^{\infty} \step_{\run} (\sdiff_{\run} + \sbias_{\run}) < \infty$ by \eqref{eq:step3}.
Condition \eqref{eq:step2} further gives $\sum_{\run=\start}^{\infty} \stepgap_{\run} < \infty$, so $\termOne_{\run}$ is uniformly bounded from above by $\termOne_{\infty} \defeq \sum_{\run=\start}^{\infty} \bracks{\step_{\run}(\sdiff_{\run} + \sbias_{\run}) + \stepgap_{\run}} < \infty$.

\item
For the noise term $\termTwo_{\run} = \sum_{\runalt=\start}^{\run} \step_{\runalt} \snoise_{\runalt}$, we have $\exof{\snoise_{\run} \given \filter_{\run}} = 0$, so $\termTwo_{\run}$ is a martingale.
Furthermore, by \eqref{eq:variance} and the step-size assumption \eqref{eq:step3} of \cref{thm:conv}, we have
\begin{align}
\sum_{\run=\start}^{\infty} \step_{\play,\run}^{2} \exof{\snoise_{\play,\run}^{2} \given \filter_{\run}}
	&\leq \sum_{\run=\start}^{\infty} \step_{\play,\run}^{2} \norm{\state_{\play,\run} - \eq_{\play}}^{2} \exof{\dnorm{\noise_{\play,\run}}^{2} \given \filter_{\run}}
	\notag\\
%	&\leq 2\diam(\points_{\play})^{2} \sum_{\run=\start}^{\infty} \step_{\play,\run}^{2} \exof{ \dnorm{\signal_{\play,\run}}^{2} + \dnorm{\vecfield_{\play}(\state_{\run})}^{2} \given \filter_{\run}}
%	\notag\\
	&\leq \diam(\points_{\play})^{2} \sum_{\run=\start}^{\infty} \step_{\play,\run}^{2} \sdev_{\play,\run}^{2}
	< \infty.
\end{align}
In turn, this implies that $\sum_{\run=\start}^{\infty} \step_{\run}^{2} \exof{\snoise_{\run}^{2} \given \filter_{\run}} < \infty$ so, by the law of large numbers for martingale difference sequences \citep[Theorem~2.18]{HH80}, we conclude that $\sum_{\runalt=\start}^{\run} \step_{\runalt} \snoise_{\runalt} \big/ \sum_{\runalt=\start}^{\run} \step_{\runalt} \to 0$ \as.

\item
Finally, for the last term, let $\Psi_{\play,\run} = \sum_{\runalt=\start}^{\run} \step_{\play,\runalt}^{2} \dnorm{\signal_{\play,\runalt}}^{2}$ so $\termThree_{\run} = \sum_{\play\in\players} (2\weight_{\play}\hstr_{\play})^{-1} \Psi_{\play,\run}$.
We then have
\begin{align}
\exof{\Psi_{\play,\run} \given \filter_{\run}}
	&= \exof*{%
		\sum_{\runalt=\start}^{\run-1} \step_{\play,\runalt}^{2} \dnorm{\signal_{\play,\runalt}}^{2}
		+ \step_{\play,\run}^{2} \dnorm{\signal_{\play,\run}}^{2} \given \filter_{\run}
		}
	\notag\\
	&= \Psi_{\play,\run-1} + \step_{\play,\run}^{2} \exof{\dnorm{\signal_{\play,\run}}^{2} \given \filter_{\run}}
	\geq \Psi_{\play,\run-1},
\end{align}
\ie $\Psi_{\play,\run}$ is a submartingale relative to $\filter_{\run}$ (recall that $\signal_{\run}$ is generated after $\state_{\run}$ so it is not $\filter_{\run}$-measurable).
Furthermore, by the law of total expectation, we also have
\begin{equation}
\exof{\Psi_{\play,\run}}
	= \exof{\exof{\Psi_{\play,\run} \given \filter_{\run}}}
%	\leq \sbound^{2} \sum_{\runalt=\start}^{\run} \step_{\runalt}^{2}
	\leq \sum_{\run=\start}^{\infty} \step_{\play,\run}^{2} \sbound_{\play,\run}^{2}
	< \infty.
\end{equation}
This shows that $\Psi_{\play,\run}$ is uniformly bounded in $L^{1}$ so, by Doob's (sub)martingale convergence theorem \citep[Theorem~2.5]{HH80}, it follows that $\Psi_{\play,\run}$ converges to some (almost surely finite) random variable $\Psi_{\play,\infty}$ with $\exof{\Psi_{\play,\infty}} < \infty$.
We thus conclude that $\termThree_{\run}$ is likewise bounded from above by $\termThree_{\infty} = \sum_{\play\in\players} (2\weight_{\play}\hstr_{\play})^{-1} \Psi_{\play,\infty} < \infty$ \as.
\end{enumerate}

Suppose now that $\probof{\event_{\eps}} = \probof{\tau_{\eps} = \infty} > 0$.
Then there exists a realization of $\state_{\run}$ such that
%If this were the case, some realization of $\state_{\run}$ would satisfy
\begin{equation}
\energy_{\run+1}
	\leq \energy_{\run}
		- \bracks*{
			\const_{\eps} 
			- \frac{\termOne_{\run} + \termTwo_{\run} + \termThree_{\run}}{\sum_{\runalt=\start}^{\run} \step_{\runalt}}}
		\cdot \sum_{\runalt=\start}^{\run} \step_{\runalt}
	\quad
	\text{for all $\run=\running$}
\end{equation}
and, in addition, $(\termOne_{\run} + \termTwo_{\run} + \termThree_{\run}) \big/ \sum_{\runalt=\start}^{\run} \step_{\runalt} \to 0$ (since we have shown that this last event occurs \ac{wp1}).
However, by \eqref{eq:step1}, this gives $\lim_{\run\to\infty} \energy_{\run} = -\infty$, a contradiction which shows that $\tau_{\eps} < \infty$ \ac{wp1} for all $\eps>0$.
Hence, given that each $\event_{1/k}$ is a zero-probability event and there is a countable number thereof, we conclude that
\begin{align}
\probof{\liminf\nolimits_{\run} \norm{\state_{\run} - \eq} = 0}
	&= \probof{\tau_{1/k} < \infty\;\textrm{for all $k=1,\dotsc,\infty$}}
	\notag\\
	&= \probof*{\intersect\nolimits_{k=1}^{\infty}\{\tau_{1/k} < \infty\}}
	= 1 - \probof*{\union\nolimits_{k=1}^{\infty} \event_{1/k}}
	= 1
\end{align}
and our proof is complete.
\end{Proof}

With these two intermediate results at hand, we are finally in a position to prove \cref{thm:conv}:

\begin{Proof}[Proof of \cref{thm:conv}]
By \cref{prop:subseq}, $\state_{\run}$ admits a (possibly random) subsequence $\state_{\run_{k}}$ such that $\state_{\run_{k}}\to\eq$ \as.
By the reciprocity condition \eqref{eq:RC}, this further implies that $\liminf_{\run\to\infty} \energy_{\run} = 0$ \as.
However, since $\lim_{\run\to\infty} \energy_{\run}$ exists (by \cref{prop:quasiFejer}), we conclude that
\begin{equation}
\probof*{\lim_{\run\to\infty} \state_{\run} = \eq}
	= \probof*{\lim_{\run\to\infty} \energy_{\run} = 0}
	= \probof*{\liminf_{\run\to\infty} \energy_{\run} = 0}
	= 1
%\lim_{\run\to\infty} \energy_{\run}
%	= \liminf_{\run\to\infty} \energy_{\run}
%	= 0
\end{equation}
and our proof is complete.
%\ie $\state_{\run}$ converges to $\eq$.
\end{Proof}

%----------------------------------------------------------------------
%%% Equilibrium tracking
%----------------------------------------------------------------------
\subsection{Proof of \cref{thm:track}\afterhead}
\label{sec:track}
We now proceed to prove the equilibrium tracking guarantees of \cref{alg:prox}.
To that end, given
a sequence of action profiles $\state_{\run} \in \points$, $\run=\running$,
and
a window of interest $\runs = \window{\runstart}{\runend}$, it will be useful to consider the gap functions
\begin{subequations}
\label{eq:gap}
\begin{alignat}{2}
\label{eq:gap-point}
\gap_{\point_{\play}}(\runs)
	&= \sum_{\run\in\runs} \braket{\vecfield_{\play,\run}(\state_{\run})}{\point_{\play} - \state_{\play,\run}}
	&\qquad
\gap_{\point}(\runs)
	&= \sum_{\play\in\players} \gap_{\point_{\play}}(\runs)
\shortintertext{and}
\label{eq:gap-max}
\gap_{\play}(\runs)
	&= \max_{\point_{\play}\in\points_{\play}} \gap_{\point_{\play}}(\runs)
	&\qquad
\gap(\runs)
	&= \sum_{\play\in\players} \gap_{\play}(\runs)
\end{alignat}
\end{subequations}
and we will likewise write $\gap_{\point_{\play}}(\nRuns)$, $\gap_{\point}(\nRuns)$, etc. when the window of interest is of the form $\runs = \window{\start}{\nRuns}$.
By the strong monotonicity of $\game_{\run}$, we have $\strong\norm{\state_{\run}-\eq_{\run}}^{2} \leq \braket{\vecfield_{\run}(\state_{\run})}{\eq_{\run} - \state_{\run}}$, so $\gap(\nRuns)$ will act as a surrogate for bounding the equilibrium tracking error $\err(\nRuns)$ of \cref{alg:prox}.
In view of this, we begin with a technical bound for the gap under \eqref{eq:MD}:

\begin{proposition}
\label{prop:gap}
Suppose that player $\play\in\players$ runs \cref{alg:prox} with
step-size $\step_{\play,\run}$
%a Lipschitz \ac{DGF},
and
oracle feedback of the form \eqref{eq:signal}.
Then, for any window of the form $\runs = \window{\runstart}{\runend}$, we have:
\begin{equation}
\label{eq:gap-bound}
\gap_{\point_{\play}}(\runs)
	\leq \sum_{\run\in\runs} \parens*{\frac{1}{\step_{\play,\run}} - \frac{1}{\step_{\play,\run-1}}} \breg_{\play}(\point_{\play},\state_{\play,\run})
	+ \sum_{\run\in\runs} \braket{\error_{\play,\run}}{\state_{\play,\run} - \point_{\play}}
	+ \frac{1}{2\hstr_{\play}} \sum_{\run\in\runs} \step_{\play,\run} \dnorm{\signal_{\play,\run}}^{2}
\end{equation}
with the convention $\step_{\play,\runstart-1} = \infty$ in the sum above.
In addition, if $\step_{\play,\run}$ is non-increasing, then
\begin{equation}
\label{eq:gap-mean-bound}
\exof{\gap_{\play}(\runs)}
	\leq \frac{2\bregbound_{\play}(\point_{\play})}{\step_{\play,\runend}}
	+ 2\diam(\points_{\play}) \sum_{\run\in\runs} \bbound_{\play,\run}
	+ \frac{1}{2\hstr_{\play}} \sum_{\run\in\runs} \step_{\play,\run}
		\totbound_{\play,\run}^{2}
%		(\sbound_{\play,\run}^{2} + \sdev_{\play,\run}^{2})
\end{equation}
where $\bregbound_{\play}(\point_{\play}) = \sup_{\alt\point_{\play}\in\points_{\hreg_{\play}}} \breg(\point_{\play},\alt\point_{\play})$.
%$\bregbound_{\play,\run} = \exof{\max_{\point_{\play}\in\points_{\play}} \breg_{\play}(\point_{\play},\state_{\play,\run})}$.
\end{proposition}
\endrev
%----------------------------------------------------------------------

\begin{Proof}
We first focus on the pointwise bound \eqref{eq:gap-bound}.
To that end, since $\state_{\play,\run+1} = \prox{\state_{\play,\run}}{\step_{\play,\run}\signal_{\play,\run}}$ for all $\run=\running$, invoking \cref{lem:template} with $\dstate_{\run} \gets \step_{\play,\run}\signal_{\play,\run}$ and $\temp_{\run} \gets 1/\step_{\play,\run}$
%(and the convention $\temp_{\runstart-1}=0$) readily
yields
\begin{equation}
\label{eq:template-signal}
\sum_{\run\in\runs} \braket{\signal_{\play,\run}}{\point_{\play} - \state_{\play,\run}}
	\leq \sum_{\run\in\runs} \parens*{\frac{1}{\step_{\play,\run}} - \frac{1}{\step_{\play,\run-1}}} \breg_{\play}(\point_{\play},\state_{\play,\run})
	+ \frac{1}{2\hstr_{\play}} \sum_{\run\in\runs} \step_{\play,\run} \dnorm{\signal_{\play,\run}}^{2}.
\end{equation}
By the feedback model \eqref{eq:signal}, we have $\signal_{\play,\run} = \vecfield_{\play,\run}(\state_{\run}) + \error_{\play,\run}$ so
\begin{equation}
\label{eq:gap-signal}
\gap_{\point_{\play}}(\runs)
	= \sum_{\run\in\runs} \braket{\vecfield_{\play,\run}(\state_{\run})}{\point_{\play} - \state_{\play,\run}}
%	= \sum_{\run\in\runs} \braket{\signal_{\run} - \error_{\run}}{\point - \state_{\run}}.
	= \sum_{\run\in\runs} \braket{\signal_{\play,\run}}{\point_{\play} - \state_{\play,\run}}
	+ \sum_{\run\in\runs} \braket{\error_{\play,\run}}{\state_{\play,\run} - \point_{\play}}.
\end{equation}
Our claim then follows by adding \eqref{eq:template-signal} and \eqref{eq:gap-signal}.

For the bound \eqref{eq:gap-mean-bound}, maximizing over $\point_{\play}\in\points_{\play}$ in \eqref{eq:gap-bound} and taking expectations, we get
\begin{subequations}
\begin{align}
\exof{\gap_{\play}(\runs)}
	= \exof*{\max_{\point_{\play}\in\points_{\play}} \gap_{\point_{\play}}(\runs)}
	&\label{eq:gap-mean-breg}
	\leq \exof*{\sum_{\run\in\runs} \parens*{\frac{1}{\step_{\play,\run}} - \frac{1}{\step_{\play,\run-1}}} \breg_{\play}(\point_{\play},\state_{\play,\run})}
	\\
%	\leq \sum_{\run\in\runs} \frac{\breg(\point,\state_{\run+1}) - \breg(\point,\state_{\run})}{\step_{\run}}
	&\label{eq:gap-mean-second}
	+ \frac{1}{2\hstr_{\play}} \sum_{\run\in\runs} \step_{\play,\run} \exof{\dnorm{\signal_{\play,\run}}^{2}}
	\\
	&\label{eq:gap-mean-noise}
	+ \exof*{\max_{\point_{\play}\in\points_{\play}} \sum_{\run\in\runs} \braket{\error_{\play,\run}}{\state_{\play,\run} - \point_{\play}}}
\end{align}
\end{subequations}
%----------------------------------------------------------------------
\beginrev
With $\step_{\play,\run}$ non-increasing, the first two terms above are readily bounded as
\begin{subequations}
\begin{align}
\eqref{eq:gap-mean-breg}
	&\leq \sum_{\run\in\runs}
		\parens*{\frac{1}{\step_{\play,\run}} - \frac{1}{\step_{\play,\run-1}}}
		\bregbound_{\play}(\point_{\play})
	\leq \frac{\bregbound_{\play}(\point_{\play})}{\step_{\play,\runend}}
\shortintertext{and}
\eqref{eq:gap-mean-second}
	&\leq \frac{\hstr_{\play}}{2} \sum_{\run\in\runs} \step_{\play,\run} \sbound_{\play,\run}^{2}
\end{align}
\end{subequations}
so we are left to bound \eqref{eq:gap-mean-noise}.
\endrev
%----------------------------------------------------------------------
To that end, introduce the auxiliary process
\begin{equation}
\aux_{\play,\run+1}
	= \prox{\aux_{\play,\run}}{-\step_{\play,\run}\noise_{\play,\run}}
\end{equation}
with $\aux_{\start} = \state_{\start}$.
We then have
\begin{flalign}
\label{eq:error-bound}
\hspace{-1ex}
\sum_{\run\in\runs} \braket{\error_{\play,\run}}{\state_{\play,\run} - \point_{\play}}
	&= \sum_{\run\in\runs} \braket{\error_{\play,\run}}{(\state_{\play,\run} - \aux_{\play,\run}) + (\aux_{\play,\run} - \point_{\play})}
	\notag\\
	&= \sum_{\run\in\runs} \braket{\error_{\play,\run}}{\state_{\play,\run} - \aux_{\play,\run}}
	+ \sum_{\run\in\runs} \braket{\bias_{\play,\run}}{\aux_{\play,\run} - \point_{\play}}
	+ \sum_{\run\in\runs} \braket{\noise_{\play,\run}}{\aux_{\play,\run} - \point_{\play}}
%	\hspace{-1ex}
\end{flalign}
so it suffices to derive a bound for each of these terms.
This can be done as follows:

\begin{enumerate}
[left=1em]
\item
The first term of \eqref{eq:error-bound} does not depend on $\point_{\play}$, so we have
\begin{align}
\exof*{\max_{\point_{\play}\in\points_{\play}} \sum_{\run\in\runs} \braket{\error_{\play,\run}}{\state_{\play,\run} - \aux_{\play,\run}}}
	&= \sum_{\run\in\runs}
		\exof*{\exof{\braket{\error_{\play,\run}}{\state_{\play,\run} - \aux_{\play,\run}} \given \filter_{\run}}}
	\notag\\
	&= \sum_{\run\in\runs} \exof{\braket{\bias_{\play,\run}}{\state_{\play,\run} - \aux_{\play,\run}}}
%	\notag\\
	\leq \diam(\points_{\play}) \bbound_{\play,\run}
\end{align}
where, in the last step, we used the definition \eqref{eq:bbound} of $\bbound_{\play,\run}$ and the bound
\begin{equation}
\braket{\bias_{\play,\run}}{\state_{\play,\run} - \aux_{\play,\run}}
	\leq \norm{\state_{\play,\run} - \aux_{\play,\run}} \dnorm{\bias_{\play,\run}}
	\leq \diam(\points_{\play}) \dnorm{\bias_{\play,\run}}.
\end{equation}

\item
The second term of \eqref{eq:error-bound} can be bounded in a similar way as
\begin{align}
\exof*{\max_{\point_{\play}\in\points_{\play}} \sum_{\run\in\runs} \braket{\bias_{\play,\run}}{\aux_{\play,\run} - \point_{\play}}}
	\leq \exof*{\diam(\points_{\play}) \dnorm{\bias_{\play,\run}}}
	\leq \diam(\points_{\play}) \bbound_{\play,\run}.
\end{align}

\item
%----------------------------------------------------------------------
\beginrev
Finally, for the last term, \cref{lem:template} with $\dstate_{\run} \gets -\step_{\play,\run}\noise_{\play,\run}$ and $\temp_{\run} = 1/\step_{\play,\run}$ gives
%(always with the convention $\temp_{\runstart-1}=0$),
\begin{align}
\sum_{\run\in\runs} \braket{\noise_{\play,\run}}{\aux_{\play,\run} - \point_{\play}}
	&= \sum_{\run\in\runs} \temp_{\play,\run} \braket{-\step_{\play,\run}\noise_{\play,\run}}{\point_{\play} - \aux_{\play,\run}}
	\notag\\
	&\leq \sum_{\run\in\runs} \parens*{\frac{1}{\step_{\play,\run}} - \frac{1}{\step_{\play,\run-1}}} \breg(\point_{\play},\aux_{\play,\run})
	+ \frac{1}{2\hstr_{\play}} \sum_{\run\in\runs} \step_{\play,\run} \dnorm{\noise_{\play,\run}}^{2}.
\end{align}
Thus, after taking expectations and telescoping, we obtain
\begin{equation}
\exof*{\max_{\point_{\play}\in\points_{\play}} \braket{\noise_{\play,\run}}{\aux_{\play,\run} - \point_{\play}}}
	\leq \frac{\bregbound_{\play}(\point_{\play})}{\step_{\play,\runend}}
%	\leq \sum_{\run\in\runs}
%		\parens*{\frac{1}{\step_{\play,\run}} - \frac{1}{\step_{\play,\run-1}}}
%		\bregbound_{\play}(\point_{\play})
	+ \frac{1}{2\hstr_{\play}} \sum_{\run\in\runs} \step_{\play,\run} \sdev_{\play,\run}^{2}.
\end{equation}
\end{enumerate}
\endrev
%----------------------------------------------------------------------
The bound \eqref{eq:gap-mean-bound} then follows by plugging back all of the above in \eqref{eq:gap-mean-noise}.
\end{Proof}

We are now in a position to prove our equilibrium tracking result.
Our proof strategy will be to leverage the gap minimization guarantees of \cref{alg:prox} (as encoded in \cref{prop:gap}) together with a batch comparison idea due to \citet{BGZ15}.

\begin{Proof}[Proof of \cref{thm:track}]
For the sake of the analysis (and only the analysis), partition the horizon of play $\runs = \window{1}{\nRuns}$ in $\nBatches$ contiguous batches $\runs_{\iBatch}$, $\iBatch=1,\dotsc,\nBatches$, each of length $\batch$ (except possibly the $\nBatches$-th one, which might be smaller).
We will prove the error bound \eqref{eq:err-bound} by linking $\err(\runs_{\iBatch})$ to $\gap(\runs_{\iBatch}) = \sum_{\play\in\players} \gap_{\play}(\runs_{\iBatch})$ for all $\iBatch=1,\dotsc,\nBatches = \ceil{\nRuns/\batch}$.

More explicitly, take the batch length to be of the form $\batch = \ceil{\nRuns^{\qexp}}$ for some constant $\qexp\in[0,1]$ to be determined later.
In this way, the number of batches is $\nBatches = \ceil{\nRuns/\batch} = \Theta(\nRuns^{1-\qexp})$ and the $\iBatch$-th batch will be of the form $\runs_{\iBatch} = \window{(\iBatch-1)\batch+1}{\iBatch\batch}$ for all $\iBatch = 1,\dotsc,\nBatches-1$ (the value $\iBatch = \nBatches$ is excluded as the $\nBatches$-th batch might be smaller).
%To resolve any ambiguities arising from the passage to the multi-agent setting, $\gap_{\point}(\runs)$ and $\gap(\runs)$ are respectively defined via \cref{eq:gap-point,eq:gap-max}, with $\vecfield_{\run}(\point) = (\vecfield_{\play,\run}(\point))_{\play\in\players}$.
Then, to bound the players' equilibrium tracking error within $\runs_{\iBatch}$, the strong monotonicity property \eqref{eq:DC-strong} for $\game_{\run}$ gives
\begin{equation}
\strong\norm{\state_{\run}-\eq_{\run}}^{2}
	\leq \braket{\vecfield_{\run}(\state_{\run})}{\eq_{\run} - \state_{\run}}
	= \braket{\vecfield_{\run}(\state_{\run})}{\test - \state_{\run}}
	+ \braket{\vecfield_{\run}(\state_{\run})}{\eq_{\run} - \test}
\end{equation}
for every reference action profile $\test\in\points$ and all $\run\in\runs$.
We thus obtain the batch bound
\begin{align}
\label{eq:err=gap+var}
\strong\err(\runs_{\iBatch})
	= \strong \sum_{\run\in\runs_{\iBatch}} \norm{\state_{\run}-\eq_{\run}}^{2}
	&\leq \sum_{\run\in\runs_{\iBatch}} \braket{\vecfield_{\run}(\state_{\run})}{\eq_{\run} - \state_{\run}}
	\notag\\
	&= \sum_{\run\in\runs_{\iBatch}} \braket{\vecfield_{\run}(\state_{\run})}{\test - \state_{\run}}
	+ \sum_{\run\in\runs_{\iBatch}}\braket{\vecfield_{\run}(\state_{\run})}{\eq_{\run}-\test}
	\notag\\
	&\leq \gap(\runs_{\iBatch})
	+ \sum_{\run\in\runs_{\iBatch}} \braket{\vecfield_{\run}(\state_{\run})}{\eq_{\run} - \test}.
\end{align}
%where, as a reminder, we set $\gap(\runs_{\iBatch}) \defeq \sum_{\play\in\players} \gap_{\play}(\runs_{\iBatch})$.

To proceed, pick a batch-specific reference action $\test_{\iBatch}\in\points$ for each $\iBatch=1,\dotsc,\nBatches$, and write
\begin{equation}
C_{\iBatch}
	= \sum_{\run\in\runs_{\iBatch}}\braket{\vecfield_{\run}(\state_{\run})}{\eq_{\run} - \test_{\iBatch}},
\end{equation}
for the last term of \eqref{eq:err=gap+var}.
A meaningful bound for $C_{\iBatch}$ can then be obtained by taking $\test_{\iBatch}$ to be the (unique) \acl{NE} of the first game encountered in the batch $\runs_{\iBatch}$, \ie setting $\test_{\iBatch} = \eq_{\min\runs_{\iBatch}}$.
Doing this, we obtain the series of estimates:
\begin{flalign}
%\sum_{\run\in\runs_{\iBatch}} \braket{\vecfield_{\run}(\state_{\run})}{\eq_{\run} - \test_{\iBatch}}
C_{\iBatch}
	&\leq \sum_{\run\in\runs_{\iBatch}}\norm{\vecfield_{\run}(\state_{\run})}_{\ast}\cdot\norm{\eq_{\run} - \test_{\iBatch}}
	\tag*{\small\{by Cauchy-Schwarz\}}
	\\[\smallskipamount]
	&\leq \sum_{\run\in\runs_{\iBatch}} \vbound \norm{\eq_{\run} - \test_{\iBatch}}
	\tag*{\small\{by \cref{asm:payv}\}}
	\\[\smallskipamount]
	&\leq \vbound\batch\max_{\run\in\runs_{\iBatch}}\norm{\eq_{\run} - \test_{\iBatch}}
	\tag*{\small\{term-by-term bound\}}
	\\[\smallskipamount]
	&\leq \vbound\batch\sum_{\run\in\runs_{\iBatch}}\norm{\eq_{\run+1} - \eq_{\run}}
	\tag*{\small\{by definition of $\test_{\iBatch}$\}}
	\\[\smallskipamount]
	&= \vbound\batch\tvar(\runs_{\iBatch}).
\end{flalign}
Thus, plugging everything back in \eqref{eq:err=gap+var} and summing over all batches $\iBatch = \start,\dotsc,\nBatches$, we get the total bound
\begin{equation}
\label{eq:err=gap+var-tot}
\exof{\err(\nRuns)}
	\leq \frac{1}{\strong} \exof{\gap(\nRuns)}
		+ \frac{\vbound\batch}{\strong} \tvar(\nRuns).
\end{equation}

%----------------------------------------------------------------------
\beginrev

With this estimate in hand, let $\bdiam_{\play} \defeq \sup_{\point_{\play},\alt\point_{\play}} \breg_{\play}(\point_{\play},\alt\point_{\play}) = \max_{\point_{\play}\in\points_{\play}} \bregbound_{\play}(\point_{\play})$, so $\bdiam_{\play}<\infty$ by \cref{lem:nonsteep}.
Then, with $\step_{\play,\run}$ decreasing, summing the second part of \cref{prop:gap} over all $\play\in\players$ yields
\begin{align}
\label{eq:gap-mean3}
\sum_{\iBatch=1}^{\nBatches} \exof{\gap(\runs_{\iBatch})}
	&\leq \sum_{\play\in\players}
		\bracks*{
			\sum_{\iBatch=1}^{\nBatches} \frac{2\bdiam_{\play}}{\step_{\play,\iBatch\batch}}
			+ 2\diam(\points_{\play}) \sum_{\run=\start}^{\nRuns} \bbound_{\play,\run}
			+ \frac{1}{2\hstr_{\play}} \sum_{\run=\start}^{\nRuns} \step_{\play,\run} \totbound_{\play,\run}^{2}
		}
	\notag\\
	&= \bigoh\parens*{
		\batch^{\max_{\play}\pexp_{\play}} \sum_{\iBatch=1}^{\nBatches} \iBatch^{\max_{\play}\pexp_{\play}}
		+ \sum_{\run=\start}^{\nRuns} \run^{-\min_{\play}\bexp_{\play}}
		+ \sum_{\run=\start}^{\nRuns} \run^{-\min_{\play}(\pexp_{\play} - 2\sexp_{\play})}
		} % end parens
	\notag\\
%	&= \bigoh(\nRuns^{1-\pexp+2\sexp} + \nRuns^{1-\bexp} + \nRuns^{1+2\sexp-\pexp}),
	&= \bigoh\parens*{
		\batch^{\max_{\play}\pexp_{\play}} \nBatches^{1+\max_{\play}\pexp_{\play}}
		+ \nRuns^{1-\min_{\play}\bexp_{\play}}
		+ \nRuns^{1-\min_{\play}(\pexp_{\play} - 2\sexp_{\play})}
		} % end parens
\end{align}
where, in the second line, we used the fact that $\step_{\play,\run} = \Theta(1/\run^{\pexp_{\play}})$.
%Since $\batch^{\pexp}\nBatches^{1+\pexp} = \bigoh(\nRuns^{\qexp\pexp} \nRuns^{(1-\qexp)(1+\pexp)}) = \bigoh(\nRuns^{1+\pexp-\qexp})$, the bound \eqref{eq:reg-mean-dyn} follows by setting the batch length exponent equal to $\qexp = 2\pexp - 2\sexp$ in \eqref{eq:reg-gap-dyn2}.
Since $\batch = \bigoh(\nRuns^{\qexp})$ and $\nBatches = \bigoh(\nRuns/\batch) = \bigoh(\nRuns^{1-\qexp})$, we get
\begin{align}
\label{eq:err-batch}
\batch^{\max_{\play}\pexp_{\play}} \nBatches^{1 + \max_{\play}\pexp_{\play}}
%	&= \bigoh((\nBatches\batch)^{\max_{\play}\pexp_{\play}} \, \nBatches)
	= \bigoh(\nRuns^{\qexp\max_{\play}\pexp_{\play}} \nRuns^{(1 - \qexp)(1 + \max_{\play}\pexp_{\play})}
%	\notag\\
	&= \bigoh(\nRuns^{1+\max_{\play}\pexp_{\play} - \qexp})
\end{align}
%where we set $\pexp_{\max} = \max_{\play}\pexp_{\play}$.
In turn, this yields the error bound
\begin{equation}
\exof{\err(\nRuns)}
	= \bigoh\parens*{
		\nRuns^{1 + \max_{\play}\pexp_{\play}-\qexp}
		+ \nRuns^{1 - \min_{\play}\bexp_{\play}}
		+ \nRuns^{1 - \min_{\play}(\pexp_{\play} - 2\sexp_{\play})}
		+ \nRuns^{\qexp}\tvar(\nRuns)} % end parens
\end{equation}
so the guarantee \eqref{eq:err-bound} follows by setting $\qexp = \max_{\play}\pexp_{\play} + \min_{\play}(\pexp_{\play} - 2\sexp_{\play})$.
\end{Proof}

\endrev
%----------------------------------------------------------------------

%% file: Bandit.tex
%----------------------------------------------------------------------
%%% BANDIT
%----------------------------------------------------------------------
% !TEX root = ./Main.tex

In this section, we proceed to examine a ``payoff-based'' learning scheme, \ie a method that relies only on observations of the players' realized, in-game payoffs (the so-called ``bandit setting'').
The first step will be to introduce a payoff-based \acl{SFO}
in the spirit of \citet{Spa92,Spa97};
%that can be cast in the general framework of \cref{sec:signal};
subsequently, by mapping this oracle to the general feedback model of \cref{sec:setup}, we will leverage the analysis of \cref{sec:results} to derive the algorithm's properties in time-varying games.

%----------------------------------------------------------------------
%%% SPSA
%----------------------------------------------------------------------
\subsection{Payoff-based feedback and estimation of payoff gradients\afterhead}
\label{sec:SPSA}

Heuristically, the main idea of the player's gradient estimation process is easiest to describe in one-dimensional environments.
In particular, suppose that an agent wishes to estimate the derivative of an unknown function $\obj\from\R\to\R$ at some point $\pivot\in\R$.
Then, by definition, given an accuracy target $\mix$, the derivative of $\obj$ at $\pivot$ can be approximated by two queries of $\obj$ as
\begin{equation}
\obj'(\pivot)
	\approx \frac{\obj(\pivot+\mix) - \obj(\pivot-\mix)}{2\mix}.
\end{equation}
Building on this idea, $\obj'(\pivot)$ can be estimated from a \emph{single} function evaluation
as follows:
%at either of the test points $\pivot + \mix$ or $\pivot - \mix$, chosen at random.
%Specifically,
let $\unitvec$ be a random variable taking the value $+1$ or $-1$ with probability $1/2$, and consider the estimator
\begin{equation}
\label{eq:SPSA-1dim}
\signal
	= \frac{\obj(\pivot + \mix\unitvec)}{\mix} \unitvec.
\end{equation}
In expectation, this gives:
\begin{equation}
\exof{\signal}
	= \frac{1}{2\mix} \obj(\pivot+\mix) - \frac{1}{2\mix} \obj(\pivot-\mix).
\end{equation}
Thus, if $\obj'$ is Lipschitz continuous, we readily get $\exof{\signal - \obj'(\pivot)} = \bigoh(\mix)$, \ie the estimator \eqref{eq:SPSA-1dim} is accurate up to $\bigoh(\delta)$.

This idea is the starting point of the so-called \acdef{SPSA} method that was pioneered by \citet{Spa92,Spa97}.
Its extension to a multi-dimensional setting is straightforward:
If an agent seeks to estimate the gradient of a function $\obj\from\R^{\vdim}\to\R$, it suffices to sample a perturbation direction $\unitvec$ uniformly at random from $\bvecs \equiv \{\pm\bvec_{1},\dotsc,\pm\bvec_{\vdim}\}$ and consider the estimator
\begin{equation}
\label{eq:SPSA-ndim}
\signal
	= \frac{\vdim}{\mix} \obj(\pivot + \mix\unitvec) \, \unitvec.
\end{equation}
The only difference between \eqref{eq:SPSA-1dim} and \eqref{eq:SPSA-ndim} is the dimensional scaling factor $\vdim$ which compensates for the fact that each principal direction of $\R^{\vdim}$ is sampled with probability $1/\vdim$. 
Then the same reasoning as above shows that $\exof{\norm{\signal - \nabla\obj(\pivot)}} = \bigoh(\mix)$.
% \ie $\vecfield^{\mix}$ is an $\bigoh(\mix)$-accurate estimate of $\nabla\obj$.

In the presence of constraints, a caveat that arises is that the query point $\query = \pivot + \mix\unitvec$ must remain feasible.
To guarantee this, let $\cvx$ be a convex body in $\R^{\vdim}$, and let $\obj\from\cvx\to\R$ be a function whose gradient we want to estimate at some point $\pivot\in\cvx$.
To avoid the occurrence $\pivot+\mix\unitvec \notin \cvx$, we first transfer $\pivot$ towards the interior of $\cvx$ by a homothetic transformation of the form
\begin{equation}
\label{eq:safety}
\pivot
	\mapsto \pivot^{\mix}
	\equiv \pivot - \frac{\mix}{\radius} (\pivot - \base)
\end{equation}
where $\base \in \intr(\cvx)$ is an interior point of $\cvx$ and $\radius>0$ is such that
\begin{enumerate*}
[\itshape a\upshape)]
\item
the ball $\ball_{\radius}(\base)$ is entirely contained in $\cvx$;
and
\item
$\mix/\radius < 1$.
\end{enumerate*}
Taken together, these conditions ensure that the query point
\begin{equation}
\label{eq:query}
\query
	= \pivot^{\mix} + \mix\unitvec
	= (1 - \mix/\radius) \pivot
		+ (\mix/\radius) (\base + \radius \, \unitvec)
\end{equation}
belongs itself to $\cvx$
(simply note that $\base + \radius \unitvec \in \ball_{\radius}(\base) \subseteq \cvx$).
%\footnote{Simply note that $\base + \radius\mix \, \unitvec \in \cvx$ (by the condition $\ball_{\radius}(\base) \subseteq \cvx)$.}
%Then, querying $\obj$ at $\query$, we obtain the gradient estimator
%\begin{equation}
%\signal
%%\vecfield^{\mix}(\pivot)
%	= \frac{\vdim}{\mix} \obj(\pivot^{\mix} + \mix\unitvec) \, \unitvec.
%\end{equation}

%----------------------------------------------------------------------
\beginrev

With all this in mind, we obtain the following process for estimating individual payoff gradients in the context of a continuous game $\game \equiv \game(\players,\points,\pay)$:
\begin{enumerate}
[left=2em,label={\arabic*)}]
\item
Every player $\play\in\players$ selects a \emph{pivot point} $\point_{\play} \in \points_{\play}$ and draws a \emph{perturbation vector} $\unitvec_{\play}$ uniformly at random from $\bvecs_{\play} \defeq \{\pm\bvec_{1},\dotsc,\pm\bvec_{\vdim_{\play}}\}$.
Subsequently, each player plays
\begin{equation}
\label{eq:query-multi}
\query_{\play}
	= \point_{\play} + \mix_{\play} \unitvec_{\play} + (\mix_{\play}/\radius_{\play})(\base_{\play} - \point_{\play})
\end{equation}
and they receive the associated payoffs $\est\pay_{\play} \defeq \pay_{\play}(\query_{1},\dotsc,\query_{\nPlayers})$, $\play\in\players$.

\item
Each player constructs the \acl{SPSA} estimate
\begin{equation}
\label{eq:SPSA}
\signal_{\play}
	= \frac{\vdim_{\play}}{\mix_{\play}} \est\pay_{\play} \cdot \unitvec_{\play}
%\signal_{\play,\run}
%	= \frac{\vdim_{\play}}{\mix_{\play,\run}} \est\pay_{\play,\run} \, \unitvar_{\play,\run},
\end{equation}
and the process repeats.
\end{enumerate}
In the above,
the sampling radius $\mix_{\play}$
and
the homothety parameters $\base_{\play}\in\points_{\play}$, $\radius_{\play} > 0$,
are chosen arbitrarily by each player $\play\in\players$, only subject to the requirements $\ball_{\radius_{\play}}(\base_{\play}) \subseteq \points_{\play}$ and $\mix_{\play} / \radius_{\play} < 1$ (to guarantee that $\query_{\play}$ is a feasible action).
Also, when unfolding over the course of a learning process, we will assume that players employ a variable sampling radius $\mix_{\play,\run}$ (similar to the players' individual step-size policy $\step_{\play,\run}$).
In this way, the estimator \eqref{eq:SPSA} can be seen as a payoff-based oracle which can be coupled with \cref{alg:prox} to generate a new candidate action and continue playing.
For a pseudocode implementation of the resulting policy, see \cref{alg:SPSA}.

\endrev
%----------------------------------------------------------------------

\begin{remark*}
Throughout this section, we tacitly assume that the players' action spaces are convex bodies, \ie they have nonempty topological interior.
This assumption is only made for convenience:
if this is not the case, it suffices to replace the basis vectors $\{\pm\bvec_{\coord}\}$ with a basis of the affine hull of each player's action space and proceed in the same way.
\hfill
\envend
\end{remark*}

%----------------------------------------------------------------------
%% SPSA description begins here

\begin{algorithm}[t]
\beginrev
\small
\ttfamily
\caption{Payoff-based learning via \acl{MD}}
\input{Algorithms/SPSA}
\endrev
\label{alg:SPSA}
\end{algorithm}

%% SPSA description ends here
%----------------------------------------------------------------------

%----------------------------------------------------------------------
%%% Analysis
%----------------------------------------------------------------------
\subsection{Analysis and results\afterhead}
\label{sec:SPSA-results}

The first step in the analysis of \cref{alg:SPSA} consists of quantifying the statistics of the players' gradient estimation process:

%----------------------------------------------------------------------
\beginrev

\begin{lemma}
\label{lem:SPSA}
The \ac{SPSA} estimator \eqref{eq:SPSA} satisfies:
\begin{equation}
\dnorm{\exof{\signal_{\play} - \vecfield_{\play}(\point)}}
	= \bigoh(\mix_{\max}^{2}/\mix_{\play})
	\quad
	\text{and}
	\quad
\exof{\dnorm{\signal_{\play}}^{2}}
	= \bigoh(1/\mix_{\play}^{2}).
\end{equation}
where $\mix_{\max} = \max_{\play} \mix_{\play}$.
\end{lemma}

\begin{Proof}
The second moment bound $\exof{\dnorm{\signal_{\play}}^{2}} = \bigoh(1/\mix_{\play}^{2})$ follows trivially from the definition \eqref{eq:SPSA} of $\signal$ and the boundedness of $\pay_{\play}$.
As for our first claim, let
\begin{equation}
\label{eq:perturb}
\perturb_{\play}
	= \query_{\play} - \point_{\play}
	= \mix_{\play} \unitvec_{\play} + (\mix_{\play}/\radius_{\play}) (\base_{\play} - \point_{\play}).
\end{equation}
and set $\perturb = (\perturb_{\play})_{\play\in\players}$.
Then, by the smoothness of $\pay_{\play}$, a first-order Taylor expansion with integral remainder gives
\begin{subequations}
\label{eq:Taylor}
\begin{align}
\label{eq:Taylor1}
\signal_{\play}
	= \frac{\vdim_{\play}}{\mix_{\play}} \pay_{\play}(\query) \cdot \unitvec_{\play}
	= \frac{\vdim_{\play}}{\mix_{\play}} \pay_{\play}(\point) \cdot \unitvec_{\play}
		&+ \frac{\vdim_{\play}}{\mix_{\play}} \sum_{\playalt\in\players} \braket{\nabla_{\point_{\playalt}}\pay_{\play}(\point)}{\perturb_{\playalt}}
		\, \unitvec_{\play}
	\\
\label{eq:Taylor2}
		&+ \sum_{\playalt,\playaltalt\in\players}
			\int_{0}^{1} 
				(1-t) \,
				\perturb_{\playalt}^{\top}
				\nabla_{\point_{\playalt}\point_{\playaltalt}}^{2} \pay_{\play}(\point + t\perturb) \,
				\perturb_{\playaltalt}
				\dd t \cdot
		\unitvec_{\play}
\end{align}
\end{subequations}
Hence, taking expectations, the first term above becomes
\begin{align}
\label{eq:Taylor1-temp}
\exof{\eqref{eq:Taylor1}}
	&= \frac{\vdim_{\play}}{\mix_{\play}}
		\exof{\braket{\vecfield_{\play}(\point)}{\perturb_{\play}} \,
			\unitvec_{\play}}
	+ \frac{\vdim_{\play}}{\mix_{\play}}
		\sum_{\playalt\neq\play}
		\braket{\nabla_{\point_{\playalt}} \pay_{\play}(\point)}{\exof{\perturb_{\playalt}}} \,
		\exof{\unitvec_{\play}}
	\notag\\
	&= \vdim_{\play} \exof{\braket{\vecfield_{\play}(\point)}{\unitvec_{\play}} \, \unitvec_{\play}}
	= \vdim_{\play} \cdot
		\frac{1}{2\vdim_{\play}}
		\sum_{\ell=1}^{\vdim_{\play}}
			\bracks{\vecfield_{\play\ell}(\point) \bvec_{\ell} - \vecfield_{\play\ell}(\point) (-\bvec_{\ell})}
	\notag\\
	&= \vecfield_{\play}(\point)
\end{align}
where we used the fact that $\exof{\unitvec_{\play}} = 0$ for all $\play\in\players$ and that $\unitvec_{\play}$ and $\unitvec_{\playalt}$ are independent for all $\play,\playalt\in\players$, $\play\neq\playalt$.
As for the second term, we have
\begin{align}
\label{eq:Taylor2-temp}
\exof{\eqref{eq:Taylor2}}
	&= \frac{\vdim_{\play}}{\mix_{\play}}
		\sum_{\playalt,\playaltalt\in\players}
			\mix_{\playalt} \mix_{\playaltalt}
			\exof*{
			\int_{0}^{1} 
				(1-t) \,
				\perturb_{\playalt}^{\top}
				\nabla_{\point_{\playalt}\point_{\playaltalt}}^{2} \pay_{\play}(\point + t\perturb) \,
				\perturb_{\playaltalt}
				\dd t \cdot
				\unitvec_{\play}}
%	= \bigoh\parens*{\frac{(\sum_{\playalt\in\players}\mix_{\playalt})^{2}}{\mix_{\play}}}
%	\notag\\
	= \bigoh\parens*{\mix_{\max}^{2} / \mix_{\play}},
\end{align}
where we used the fact that $\points$ is compact and $\pay_{\play}$ is $C^{2}$-smooth over $\points$.
Our claim then follows by combining the bounds \eqref{eq:Taylor1-temp} and \eqref{eq:Taylor2-temp}.
\end{Proof}

We are now in a position to state and prove our main result for the payoff-based learning policy outlined in \cref{alg:SPSA}:

\begin{theorem}
\label{thm:SPSA}
Let $\game_{\run}$ be a time-varying game satisfying \cref{asm:payv}.
Suppose further that each player $\play\in\players$ runs \cref{alg:prox} with
%a \ac{DGF} satisfying \eqref{eq:RC},
step-size $\step_{\play,\run} \propto \run^{-\pexp_{\play}}$
and
sampling radius $\mix_{\play,\run} \propto \run^{-\qexp_{\play}}$
for some $\pexp_{\play},\qexp_{\play}\in(0,1]$.
Then:
\begin{enumerate}
[left=0pt,label={\upshape\textpar{\arabic*\hspace*{.5pt}}}]
\addtolength{\itemsep}{\smallskipamount}

\item
\label{itm:SPSA-stable}
If $\game_{\run}$ stabilizes to a strictly monotone game $\game$ at a rate $\difbound_{\play,\run} = \bigoh(1/\run^{\vexp_{\play}})$, $\vexp_{\play}>0$, and $\pexp_{\play} = \pexp > \max\{1 - \vexp_{\play},1 + \qexp_{\play} - 2\qexp_{\min},1/2 + \qexp_{\play}\}$ for all $\play\in\players$, the sequence of chosen actions $\est\state_{\run}$, $\run=\running$, converges to the \acl{NE} of $\game$ with probability $1$.
In particular, convergence to \acl{NE} is guaranteed under the choice $\pexp_{\play} = 1$, $\qexp_{\play} = 1/3$.

\item
\label{itm:SPSA-track}
If $\game_{\run}$ is strongly monotone and its drift is bounded as $\tvar(\nRuns) = \bigoh(\nRuns^{\vexp})$ for some $\vexp<1$, the sequence of chosen actions $\est\state_{\run}$, $\run=\running$, enjoys the equilibrium tracking guarantee:
\begin{equation}
\label{eq:err-SPSA}
\exof*{\sum_{\run=\start}^{\nRuns} \norm{\est\state_{\run} - \eq_{\run}}^{2}}
	= \bigoh\parens*{
		\nRuns^{1 - \min_{\play}(\pexp_{\play} - 2\qexp_{\play})}
		+ \nRuns^{1 + \qexp_{\max} - 2\qexp_{\min}}
		+ \nRuns^{\vexp + \pexp_{\max} + \min_{\play}(\pexp_{\play} - 2\qexp_{\play})}
%		\nRuns^{1 - \min_{\play}(\pexp_{\play} - 2\qexp_{\play})}
%		+ \nRuns^{1 +\max_{\play}\qexp_{\play} - 2\min_{\play}\qexp_{\play}}
%		+ \nRuns^{\vexp + \max_{\play}\pexp_{\play} + \min_{\play}(\pexp_{\play} - 2\qexp_{\play})}
		}
\end{equation}
where $\eq_{\run}$ denotes the \textpar{necessarily unique} \acl{NE} of $\game_{\run}$, and we set $\pexp_{\min/\max} = \min/\max_{\play} \pexp_{\play}$ and $\qexp_{\min/\max} = \min/\max_{\play} \qexp_{\play}$.
In particular, for $\pexp_{\play} = 3(1-\vexp)/5$ and $\qexp_{\play} = (1-\vexp)/5$, we get the optimized tracking guarantee:
\begin{equation}
\label{eq:err-opt-SPSA}
\exof*{\sum_{\run=\start}^{\nRuns} \norm{\est\state_{\run} - \eq_{\run}}^{2}}
	= \bigoh\parens*{\nRuns^{\frac{4+\vexp}{5}}}.
\end{equation}
\end{enumerate}
\end{theorem}

\endrev
%----------------------------------------------------------------------

\cref{thm:SPSA} combines two regimes:
Part \ref{itm:SPSA-stable} treats time-varying games that stabilize to a well-defined limit,
while
Part \ref{itm:SPSA-track} concerns the case where the game evolves without converging.
This is in direct analogy to \cref{thm:track,thm:conv} for the case of generic \ac{SFO} feedback and, indeed, \cref{thm:SPSA} draws heavily on these results.
However, there is now a discrepancy between the actions $\est\state_{\run}$ chosen by the players and the candidate actions $\state_{\run}$ on which the \ac{SPSA} estimator \eqref{eq:SPSA} returns feedback.
We explain this difference in the proof of \cref{thm:SPSA} below:

\begin{Proof}[Proof of \cref{thm:SPSA}]
Let $\state_{\run}$, $\run=\running$, be the sequence of pivot points generated by \cref{alg:SPSA}:
specifically, $\state_{\run}$ is given by \eqref{eq:MD}, but the players' realized action profile $\est\state_{\run}$ is given by \eqref{eq:query}.
Then, by \cref{lem:SPSA}, it follows that the \ac{SPSA} estimator $\signal_{\run}$ of \eqref{eq:SPSA} returns feedback of the form \eqref{eq:signal} on $\state_{\run}$ with bias and variance bounded as $\bbound_{\run} = \bigoh(\mix_{\play,\run}) = \bigoh(1/\run^{\qexp_{\play}})$ and $\sbound_{\run}^{2} = \bigoh(1/\mix_{\play,\run}^{2}) = \bigoh(\run^{2\qexp_{\play}})$ respectively.
Since the sequence $\state_{\run}$ is generated via the prox-rule $\state_{\run+1} = \prox{\state_{\run}}{\step_{\run}\signal_{\run}}$ of \cref{alg:prox}, we have:

%----------------------------------------------------------------------
\beginrev

\begin{enumerate}
[left=0pt,label={(\arabic*)}]
\addtolength{\itemsep}{\smallskipamount}

\item
If $\game_{\run}$ stabilizes to a strictly monotone game $\game$, invoking  \cref{cor:convergence} with $\bexp_{\play} = \sexp_{\play} = \qexp_{\play}$ shows that the sequence of pivot points $\state_{\run}$ converges \as to the (necessarily unique) equilibrium of $\game$ as long as $\pexp_{\play} = \pexp > \max\{1-\vexp_{\play},1-\qexp_{\play},1/2+\qexp_{\play}\}$ for all $\play\in\players$.
Since $\norm{\est\state_{\run} - \state_{\run}} = \bigoh(\mix_{\play,\run})$ and $\mix_{\play,\run}\to0$, our claim follows.

\item
If $\game_{\run}$ is strongly monotone with drift $\tvar(\nRuns) = \bigoh(\nRuns^{\vexp})$, \cref{thm:track} gives
\begin{equation}
\exof{\err(\nRuns)}
	= \bigoh\parens*{
		\nRuns^{1-\min_{\play}(\pexp_{\play}-2\qexp_{\play})}
		+ \nRuns^{1 + \qexp_{\max} - 2\qexp_{\min}}
		+ \nRuns^{\pexp_{\max} + \min_{\play}(\pexp_{\play} - 2\qexp_{\play})}
			\tvar(\nRuns)}
%	= \bigoh\parens*{\nRuns^{1+2\qexp-\pexp} + \nRuns^{1-\qexp} + \nRuns^{2\pexp-2\qexp+\vexp}},
\end{equation}
where, by virtue of \cref{lem:SPSA}, we set $\sexp_{\play} = \qexp_{\play}$ and $\bexp_{\play} = 2\qexp_{\min} - \qexp_{\play}$ in \eqref{eq:err-bound}.
However, by \eqref{eq:query} and the compactness of $\points$, we also have $\norm{\est\state_{\run} - \state_{\run}} = \bigoh(\mix_{\play,\run}) = \bigoh(1/\run^{\qexp_{\min}})$, implying in turn that
\begin{align}
\frac{1}{2}\sum_{\run=\start}^{\nRuns} \norm{\est\state_{\run} - \eq_{\run}}^{2}
	&\leq \sum_{\run=\start}^{\nRuns} \norm{\est\state_{\run} - \state_{\run}}^{2}
		+ \sum_{\run=\start}^{\nRuns} \norm{\state_{\run} - \eq_{\run}}^{2}
	\notag\\
	&= \bigoh(\nRuns^{1-2\qexp_{\min}})
		+ \sum_{\run=\start}^{\nRuns} \norm{\state_{\run} - \eq_{\run}}^{2}.
\end{align}
\end{enumerate}
Putting all this together, we conclude that $\exof*{\sum_{\run=\start}^{\nRuns} \norm{\est\state_{\run} - \eq_{\run}}^{2}}$ is bounded as per \eqref{eq:err-SPSA},
%\begin{equation}
%\exof*{\sum_{\run\in\runs} \norm{\est\state_{\run} - \eq_{\run}}^{2}}
%	= \bigoh(\nRuns^{1-2\qexp})
%		+ \bigoh\parens*{\nRuns^{1+2\qexp-\pexp} + \nRuns^{1-\qexp} + \nRuns^{2\pexp-2\qexp+\vexp}},
%\end{equation}
and our proof is complete.
\end{Proof}

\endrev
%----------------------------------------------------------------------

As a special case, Part \ref{itm:SPSA-stable} of \cref{thm:SPSA} implies that the sequence of play induced by \cref{alg:SPSA} in a \emph{fixed} strictly monotone game $\game_{\run} \equiv \game$ converges to \acl{NE} with probability $1$ as long as $\pexp > \max\{1-\qexp,1/2+\qexp\}$.
In this way, we recover a recent result by \citet{BLM18} who used a different form of the \ac{SPSA} estimator \eqref{eq:SPSA} to establish the convergence of payoff-based no-regret learning in \emph{constant}, monotone games.
It is also possible to undertake a finer analysis for the method's rate of convergence in the case where the limit game $\game$ is strongly monotone, but this lies beyond the scope of this work.

%% file: Algorithms/SPSA.tex
%----------------------------------------------------------------------
%%% SPSA ALGORITHM
%----------------------------------------------------------------------
% !TEX root = ../Main.tex

\begin{algorithmic}[1]
\Require
	step-size $\step_{\play,\run} > 0$;
	sampling radius $\mix_{\play,\run} > 0$;
	homothety parameters $\base_{\play} \in \points_{\play}$, $\radius_{\play}  >0$
\smallskip

\State
	initialize $\state_{\play,\start} \in \points_{\hreg_{\play}}$
	\Comment{initialize pivot}%
\For{$\run=\running$}{ simultaneously for all $\play=1,\dotsc,\nPlayers$}%
		\vspace{2pt}
	\State
		draw $\unitvar_{\play,\run}$ uniformly from $\{\pm\bvec_{1},\dotsc,\pm\bvec_{\vdim_{\play}}\}$
		\Comment{random perturbation}%
		\vspace{2pt}
	\State
		play $\est\state_{\play,\run} = \state_{\play,\run} + \mix_{\play,\run} \unitvar_{\play,\run} + (\mix_{\play,\run} / \radius_{\play,\run}) (\base_{\play} - \state_{\play,\run})$
		\Comment{select action}%
		\vspace{2pt}
	\State
		receive $\est\pay_{\play,\run} \equiv \pay_{\play,\run}(\est\state_{\play,\run};\est\state_{-\play,\run})$
		\Comment{get payoff}%
		\vspace{2pt}
	\State
		set $\signal_{\play,\run} = (\vdim_{\play}/\mix_{\play,\run}) \, \est\pay_{\play,\run} \unitvar_{\play,\run}$
		\Comment{estimate gradient}%
		\vspace{2pt}
	\State
		set $\state_{\play,\run+1} \leftarrow \proxplay{\state_{\play,\run}}{\step_{\play,\run}\signal_{\play,\run}}$
		\Comment{update pivot}
\EndFor
\end{algorithmic}

%% file: Discussion.tex
%----------------------------------------------------------------------
%%% DISCUSSION
%----------------------------------------------------------------------
% !TEX root = ./Main.tex

In this section, we proceed to discuss some extensions and applications of our results that would have otherwise disrupted the flow of our paper.

%----------------------------------------------------------------------
%%% Random games
%----------------------------------------------------------------------
\beginrev
\subsection{Games with randomly evolving payoffs\afterhead}
\label{sec:random}

We begin by discussing some applications of our results to games that evolve randomly over time \textendash\  \ie when $\game_{\run}$ is determined by some randomly drawn parameter $\sample_{\run}$ describing the ``state of the world''.
%Time-varying games of this type arise quite often in practice:
Randomly evolving games of this type are commonly referred to as \emph{stochastic Nash games} in the mathematical optimization, control and engineering literatures \cite{RavSha11,Cui:2021ud}, where they are sometimes analyzed within a more general framework featuring joint coupling constraints.
For example, in the wireless communications problem we described earlier (\cref{ex:power} in \cref{sec:setup}), this would correspond to the case where the users' channel gains $g_{\play,\run}$ fluctuate randomly between transmission frames \textendash\ the so-called ``fast-fading'' channel model \cite{TV05}.

To define this game-theoretic setting in detail, suppose that the players' utilities are determined by an ensemble of random functions of the form $\tilde\pay_{\play}\from\points\times\samples\to\R$ where
$\samples$ has the structure of a complete probability space
and
each $\tilde\pay_{\play}(\point;\sample)$ is assumed to be
\begin{enumerate*}
[\itshape a\upshape)]
\item
measurable in $\sample$;
\item
$C^{2}$-smooth in $\point$ with uniformly bounded derivatives;
and
\item
individually concave in the $\play$-th component of $\point$.
\end{enumerate*}
Then, at each stage $\run=\running$, an \acs{iid} state variable $\sample_{\run}$ is drawn from $\samples$ according to $\prob$, and the players face the game $\game_{\run}$ with payoff functions
\begin{equation}
\pay_{\play,\run}(\point)
	= \tilde\pay_{\play}(\point;\sample_{\run})
	\quad
	\text{for all $\play\in\players$}.
\end{equation}

Given the randomness involved, it is meaningful to consider the associated mean game $\game \equiv \game(\players,\points,\pay)$ with payoff functions 
\begin{equation}
\label{eq:pay-mean}
\pay_{\play}(\point)
	= \exof{\tilde\pay_{\play}(\point;\sample)}
	\quad
	\text{for all $\play\in\players$}
\end{equation}
where the expectation $\exof{\cdot}$ is taken relative to the (common) law of the state variables $\sample_{\run}$.
It is then natural to ask whether the players' behavior under \cref{alg:prox} approaches a \acl{NE} of the mean $\game$ as the game unfolds.
Our next result provides a positive result in this direction under the assumption that the players' individual payoff gradients have finite variance, \ie
\begin{equation}
\exof{
	\dnorm{
		\nabla_{\point_{\play}}\tilde\pay_{\play}(\point;\sample)
		- \nabla_{\point_{\play}} \pay_{\play}(\point)}^{2}}
	\leq \gamevar
	\quad
	\text{for all $\point\in\points$}.
\end{equation}
Under this assumption, we have the following equilibrium convergence guarantee:

\begin{theorem}
\label{thm:conv-random}
Let $\game_{\run}$, $\run=\running$, be a sequence of random games as above, and assume that the mean game $\game$ is strictly monotone.
Suppose further that each player $\play\in\players$ runs \cref{alg:prox} with
a \ac{DGF} satisfying \eqref{eq:RC} and a step-size policy satisfying \eqref{eq:step1}, \eqref{eq:step2}, and
\begin{equation}
\label{eq:step3-alt}
\tag{\ref*{eq:step3}$'$}
\sum_{\run=\start}^{\infty} \step_{\play,\run} \bbound_{\play,\run}
	< \infty
	\quad
	\text{and}
	\quad
\sum_{\run=\start}^{\infty} \step_{\play,\run}^{2} \totbound_{\play,\run}^{2}
	< \infty.
\end{equation}
Then, \acl{wp1}, the sequence of realized actions $\state_{\run}$ converges to the \textpar{necessarily unique} \acl{NE} $\eq$ of $\game$.
\end{theorem}

\begin{remark*}
There are two distinct and conditionally independent sources of stochasticity in \cref{thm:conv-random}:
\begin{enumerate*}
[\itshape a\upshape)]
\item
the randomness coming from $\sample_{\run}$ (which determines the $\run$-th stage game $\game_{\run}$);
and
\item
the randomness in the players' oracle feedback.
\end{enumerate*}
In particular, we tacitly assume here that the filtration $\filter_{\run}$ underlying the definition \eqref{eq:SFO-stats} of the players' feedback process refers to the joint history of $\state_{\run}$ and $\sample_{\run}$, and the statement ``\acl{wp1}'' likewise refers to both sources of randomness taken together.
\end{remark*}

\begin{Proof}
%Let $\tilde\filter_{\run}$ denote the natural filtration of the state variable $\sample_{\run}$, so $\tilde\filter_{\run} \subseteq \filter_{\run}$ since actions are selected in $\game_{\run}$ only after $\sample_{\run}$ has been drawn.
%Let $\vecfield_{\play,\run}(\point) = \nabla_{\point_{\play}}\tilde\pay_{\play}(\point;\sample_{\run})$ denote the individual gradient field of player $\play\in\players$ at the $\run$-th stage of the process.
Let $\tilde\vecfield_{\play}(\point;\sample) = \nabla_{\point_{\play}}\tilde\pay_{\play}(\point;\sample)$ and $\vecfield_{\play}(\point) = \nabla_{\point_{\play}}\pay_{\play}(\point)$.
Then, by differentiating under the integral sign, we have $\exof{\tilde\vecfield_{\play}(\point;\sample)} = \exof{\nabla_{\point_{\play}}\tilde\pay_{\play}(\point;\sample)} = \nabla_{\point_{\play}} \exof{\tilde\vecfield_{\play}(\point;\sample)} = \vecfield_{\play}(\point)$, so the players' oracle signal may be decomposed as
\begin{equation}
\signal_{\play,\run}
	= \vecfield_{\play}(\state_{\run};\sample_{\run})
		+ \noise_{\play,\run}
		+ \bias_{\play,\run}
	= \vecfield_{\play}(\state_{\run})
		+ \bar\noise_{\play,\run}
		+ \bias_{\play,\run}
\end{equation}
where $\bar\noise_{\play,\run} = \noise_{\play,\run} + \vecfield_{\play}(\state_{\run};\sample_{\run}) - \vecfield_{\play}(\state_{\run})$.
Then, in a slight abuse of notation, we obtain
\begin{equation}
\exof{\bar\noise_{\play,\run} \given \state_{\run},\dotsc,\state_{\start}}
	= \exof{\exof{\bar\noise_{\play,\run} \given \filter_{\run}}}
	= 0 + \exof{\vecfield_{\play}(\state_{\run};\sample_{\run}) - \vecfield_{\play}(\state_{\run})}
	= 0
\end{equation}
and, furthermore
\begin{align}
\exof{\dnorm{\bar\noise_{\play,\run}}^{2} \given \state_{\run},\dotsc,\state_{\start}}
	&\leq 2\exof{
		\dnorm{\noise_{\play,\run}}^{2}
		+ \dnorm{\vecfield_{\play}(\state_{\run};\sample_{\run}) - \vecfield_{\play}(\state_{\run})}^{2}
		\given \state_{\run},\dotsc,\state_{\start}}
	\notag\\
	&\leq 2\sdev_{\play,\run}^{2}
		+ 2 \gamevar
	= \bigoh(\totbound_{\play,\run}^{2}).
\end{align}
Finally, letting $\bar\bias_{\play,\run} = \exof{\bias_{\play,\run}}$, we also get $\dnorm{\bar\bias_{\play,\run}} \leq \bbound_{\play,\run}$ by definition.
Accordingly, given that $\exof{\signal_{\play,\run} \given \state_{\run},\dotsc,\state_{\start}} = \vecfield_{\play}(\state_{\run}) + \bar\bias_{\play,\run}$, our claim follows by applying \cref{thm:conv} to the sequence of (strictly monotone) games $\bar\game_{\run} \equiv \game$ for all $\run\geq\start$.
\end{Proof}

Even though \cref{thm:conv} plays a major role in the proof of \cref{thm:conv-random}, the latter is conceptually distinct from the former because it provides an equilibrium convergence result in a setting where the sequence of stage games does not stabilize over time.
Analogous results for equilibrium tracking or payoff-based learning
(in the direction of \cref{thm:track} or \cref{thm:SPSA} respectvely) can also be derived, but this would take us too far afield, so we do not carry out the detailed analysis here.

\endrev
\subsection{Regret bounds\afterhead}
\label{sec:regret}

We close this section with a precise statement and derivation of the dynamic regret bound \eqref{eq:reg-mean-dyn} that was alluded to in \cref{sec:track}.

\begin{proposition}
\label{prop:reg-dyn}
Suppose that a single player runs \cref{alg:prox}
against a sequence of concave payoff functions $\pay_{\run}\from\points\to\R$
with
a Lipschitz \ac{DGF}
and
step-size and oracle feedback parameters as in \cref{thm:track}.
Then the player's dynamic regret is bounded as
\begin{equation}
\tag{\ref*{eq:reg-mean-dyn}, redux}
\txs
\exof{\dynreg(\nRuns)}
%	= \bigoh(\nRuns^{1+\pexp-\qexp} + \nRuns^{1-\bexp} + \nRuns^{1+2\sexp-\pexp} + \nRuns^{\qexp} \tvar(\nRuns)).
	= \bigoh\parens*{\nRuns^{1+2\sexp-\pexp} + \nRuns^{1-\bexp} + \nRuns^{2\pexp-2\sexp}\tvar(\nRuns)}.
\end{equation}
In particular, if
$\tvar(\nRuns) = \bigoh(\nRuns^{\vexp})$
and
the algorithm's feedback is unbiased and bounded in mean square \textpar{$\bexp = \infty$, $\sexp = 0$}, the player enjoys the bound
%\begin{equation}
%\label{eq:reg-mean-dyn-tuned}
\(
\exof{\dynreg(\nRuns)}
	= \bigoh\parens[\big]{\nRuns^{1-\pexp} + \nRuns^{2\pexp+\vexp}}.
\)
%\end{equation}
Hence, for $\pexp = (1-\vexp)/3$, the player achieves
\begin{equation}
\tag{\ref*{eq:reg-mean-dyn-opt}, redux}
\exof{\dynreg(\nRuns)}
	= \bigoh\parens[\big]{\nRuns^{\frac{2+\vexp}{3}}}.
\end{equation}
\end{proposition}

\begin{Proof}[Proof of \cref{prop:reg-dyn}]
As in the proof of \cref{thm:track}, partition the horizon of play $\runs = \window{1}{\nRuns}$ in $\nBatches$ contiguous batches $\runs_{\iBatch}$, $\iBatch=1,\dotsc,\nBatches$, each of length $\batch$ (except possibly the $\nBatches$-th one, which might be smaller).
We then have
\begin{align}
\label{eq:dynreg=gap+var}
\dynreg(\runs_{\iBatch})
	= \sum_{\run\in\runs_{\iBatch}} \bracks{\pay_{\run}(\eq_{\run}) - \pay_{\run}(\state_{\run})}
	&\leq \sum_{\run\in\runs_{\iBatch}} \braket{\vecfield_{\run}(\state_{\run})}{\eq_{\run} - \state_{\run}}
	\notag\\
	&= \sum_{\run\in\runs_{\iBatch}} \braket{\vecfield_{\run}(\state_{\run})}{\test_{\iBatch} - \state_{\run}}
		+ \sum_{\run\in\runs_{\iBatch}}\braket{\vecfield_{\run}(\state_{\run})}{\eq_{\run}-\test_{\iBatch}}
	\notag\\
	&\leq \gap(\runs_{\iBatch})
		+ \sum_{\run\in\runs_{\iBatch}} \braket{\vecfield_{\run}(\state_{\run})}{\eq_{\run} - \test_{\iBatch}},
\end{align}
where $\test_{\iBatch}\in\points$ is a test action specific to each batch $\iBatch=1,\dotsc,\nBatches$.
Then, repeating the series of arguments leading up to \eqref{eq:err=gap+var-tot}, we get the dynamic regret bound
\begin{equation}
\label{eq:dynreg=gap+var-tot}
\exof{\dynreg(\nRuns)}
	\leq \exof{\gap(\nRuns)}
		+ \vbound\batch \tvar(\nRuns)
\end{equation}
and our claim by invoking the bounds \eqref{eq:gap-mean3} and \eqref{eq:err-batch}.
\end{Proof}

Dynamic regret guarantees of the form \eqref{eq:reg-mean-dyn-opt} already exist in the literature.
Specifically, \citet{BGZ15} obtained a similar bound by exploiting the following meta-principle:\
\begin{enumerate*}
%[\itshape a\upshape)]
[(\itshape i\hspace*{.5pt}\upshape)]
\item
first, break the horizon of play into batches of size $\batch$;
\item
over each batch, run an algorithm that guarantees low static regret relative to $\batch$;
then
\item
finetune these steps in terms of the horizon $\nRuns$ and the variation $\tvar(\nRuns)$ of the agent's payoff functions in order to get low dynamic regret.
\end{enumerate*}
In our setting, if \cref{alg:prox} is rebooted every $\batch \sim \bracks{\nRuns / \tvar(\nRuns)}^{2/3}$ iterations and is run with \emph{constant} step-size $\step \sim 1/\sqrt{\batch}$ between reboots, the meta-principle of \citet{BGZ15} guarantees the dynamic regret bound
\begin{equation}
\exof{\dynreg(\nRuns)}
	= \bigoh(\nRuns^{2/3}\tvar(\nRuns)^{1/3}).
\end{equation}
\citet{BGZ15} further showed that this bound is unimprovable under the blanket feedback model \eqref{eq:signal},
%\footnote{In particular, an informed adversary can always impose $\dynreg(\nRuns) = \Omega(\tvar(\nRuns)^{1/3} \nRuns^{2/3})$ in this case.}
so \eqref{eq:reg-mean-dyn-opt} is tight in this regard.%
\footnote{Strictly speaking, \citet{BGZ15} define $\tvar(\nRuns)$ as $\tvar(\nRuns) = \sum_{\run=\start}^{\nRuns} \norm{\pay_{\run+1} - \pay_{\run}}_{\infty}$, but this distinction is not important for our purposes.}
%by contrast,
%\citet{CBGLS12} and \citet{JRSS15} replace $\eq_{\run}$ by an arbitrary comparator sequence,
%while
%\citet{CYLM+12} consider the total variation of $\nabla\pay_{\run}$.
%These measures are closely related
%%for our purposes, \eqref{eq:eqvar} is the most adapted definition,
%so we will not dwell on this distinction.}
%The main difference between the strategy of \citet{BGZ15} and \cref{alg:prox} is that the former requires a succession of restarts whereas the latter does not.

%----------------------------------------------------------------------
\beginrev

A disadvantage of this restart approach is that\
\begin{enumerate*}
[(\itshape i\hspace*{1pt}\upshape)]
\item
the batch length $\batch$ must be chosen carefully relative to the total variation of the sequence of payoff functions encountered;
and
\item
at every reboot, the algorithm begins \emph{tabula rasa}, essentially forgetting all knowledge it had accumulated up to the point in question.
%\footnote{Of course, this disadvantage also applies to any policy that relies on a doubling trick to achieve an anytime regret bound.}
\end{enumerate*}
\citet{BGZ15} already discuss some possible ways to avoid restarts, and we are aware of at least two related approaches in the literature:
\endrev
%----------------------------------------------------------------------
\citet{JOWW17} proposed a meta-aggregator based on coin betting,
while
\citet{JRSS15} and \citet{SJ18} take an approach based on optimistic \acl{MD}.
Importantly, both policies achieve $\dynreg(\nRuns) = \bigoh(\tvar(\nRuns)^{1/2}\nRuns^{1/2})$ \emph{without} prior knowledge of $\tvar(\nRuns)$:
since $\tvar(\nRuns)^{1/2} \nRuns^{1/2} = \tvar(\nRuns)^{1/3} \tvar(\nRuns)^{1/6} \nRuns^{1/2} = o(\tvar(\nRuns)^{1/3}\nRuns^{2/3})$ whenever $\tvar(\nRuns) = o(\nRuns)$, these guarantees would seem to contradict the optimality of the bound $\bigoh(\nRuns^{2/3}\tvar(\nRuns)^{1/3})$.
%\footnote{\revise{We thank one of the anonymous reviewers for bringing this point to our attention.}}
The resolution of this apparent incongruity is that \citet{JOWW17} and \citet{JRSS15} assume access to a \emph{perfect} gradient oracle, while the discussion above only assumes access to a \emph{stochastic} one.

%In view of all this, the bound \eqref{eq:reg-mean-dyn-opt} should be seen as a ``middle ground'' between the analysis of \citet{BGZ15} and that of \citet{JRSS15} and \citet{JOWW17}:
%albeit not adaptive to $\tvar(\nRuns)$, the bound \eqref{eq:reg-mean-dyn-opt} does not require access to a perfect gradient oracle, nor does it involve a periodic restart schedule.
To the best of our knowledge, the perfect oracle requirement cannot be relaxed:
if the players' gradient feedback is noisy, successive oracle calls cannot provide reliable information about the variation of the agent's payoff functions from one stage to the next, so the learning process cannot adapt to $\tvar(\nRuns)$.
Designing a policy that provably interpolates between the stochastic and deterministic regimes is a very fruitful question for further research, but one which lies beyond the scope of this paper.

%% file: Conclusion.tex
%----------------------------------------------------------------------
%%% CONCLUSION
%----------------------------------------------------------------------
% !TEX root = ./Main.tex

There are many interesting points for future research.
A particularly promising one is to bridge the gap between the step-size policies that guarantee an optimal equilibrium tracking error and the policies that guarantee convergence to a \acl{NE} in the case where $\game_{\run}$ stabilizes to a well-defined limit.
As we saw, these considerations are not always in tune:
when the rules of the game fluctuate constantly, players can use very different step-sizes, and still track the game's equilibrium on average;
by contrast, when the game stabilizes, convergence to \acl{NE} requires a certaing compatibility between the players' step-size policies (and requires finer tuning).
Balancing these two objectives in an adaptive, context-agnostic manner is a rich and promising direction for future research.

While on the topic of adaptivity, it should be recalled that players with access to perfect gradient information can achieve better rates of dynamic regret minimization, without any prior knowledge of the game's drift over time \citep{JRSS15,SJ18}.
Whether this is still possible in the stochastic (or, worse, bandit) case is another fruitful open question for further research.

%% file: App-Bregman.tex
%----------------------------------------------------------------------
%%%  APP: BREGMAN
%----------------------------------------------------------------------
% !TEX root = ./Main.tex

In this appendix we collect some basic technical facts on \aclp{DGF} and prox-mappings.
These results are not new, but given the range of conventions and definitions in the literature, we find it useful to provide here precise statements and proofs.
For a detailed discussion, we refer the reader to \citet{NJLS09,JNT11}, and references therein.
\smallskip

In what follows, $\hreg$ will denote a \acl{DGF} on a compact convex subset $\cvx$ of an $\vdim$-dimensional normed space $\vecspace\cong\R^{\vdim}$ with dual $\dspace = \dual\vecspace$, as per \cref{def:Bregman}.
We begin with a basic subgradient comparison lemma:

\begin{lemma}
\label{lem:subgrads}
For all $\base\in\cvx$ and all $\dpoint\in\subd\hreg(\point)$, $\point\in\subcvx$, we have:
\begin{equation}
\label{eq:subgrads}
\braket{\nabla \hreg(\point)}{\point-\base}
	\leq \braket{\dpoint}{\point-\base}.
\end{equation}
\end{lemma}

\begin{Proof}
By continuity, it suffices to show that \eqref{eq:subgrads} holds for all $\base\in\relint\cvx$.
To show this, fix $\base\in\relint\cvx$, and let
\begin{equation}
\phi(t)
	= \hreg(\point+t(\base-\point))
	- [\hreg(\point)+\braket{\dpoint}{\point+t(\base-\point)}]
	\quad
	\text{for all $t\in[0,1]$}.
\end{equation}
Given that $\hreg$ is strongly convex and $\dpoint\in \partial \hreg(\point)$, it follows that $\phi(t)\geq 0$ with equality if and only if $t=0$.
Since $\psi(t)=\braket{\nabla \hreg(\point+t(\base-x))-\dpoint}{\base-\point}$ is a continuous selection of subgradients of $\phi$ and both $\phi$ and $\psi$ are continuous over $[0,1]$, it follows that $\phi$ is continuously differentiable with $\phi'=\psi$ on $[0,1]$.
Hence, with $\phi$ convex and $\phi(t)\geq 0=\phi(0)$ for all $t \in [0,1]$, we conclude that $\phi'(0)=\braket{\nabla \hreg(x)-\dpoint}{\base-\point}\geq 0$, and our proof is complete.
\end{Proof}

We continue with a basic property of Bregman divergences known as the ``three-point identity'' \cite{CT93}:

\begin{lemma}[$3$-point identity]
\label{lem:3point}
For all $\base \in \cvx$ and all $\point,\pointalt \in \subcvx$, we have:
\begin{equation}
\label{eq:3point}
\breg(\base,\point)
	= \breg(\base,\pointalt)
	+ \breg(\pointalt,\point)
	+ \braket{\nabla \hreg(\point)
	- \nabla \hreg(\pointalt)}{\pointalt - \base}.
\end{equation}
\end{lemma}

%\begin{Proof}
%By definition, we get:
%\begin{equation}
%\begin{aligned}
%\breg(\base,\pointalt)
%	&= \hreg(\base) - \hreg(\pointalt) - \braket{\nabla\hreg(\pointalt)}{\base - \pointalt}
%	\\
%\breg(\base,\point)\hphantom{'}
%	&= \hreg(\base) - \hreg(\point) - \braket{\nabla\hreg(\point)}{\base - \point}
%	\\
%\breg(\point,\pointalt)
%	&= \hreg(\point) - \hreg(\pointalt) - \braket{\nabla\hreg(\pointalt)}{\point - \pointalt}.
%\end{aligned}
%\end{equation}
%The lemma then follows by adding the two last lines and subtracting the first.
%\end{Proof}

The proof of this lemma is a straightforward expansion, so we omit it.
Below, we employ this identity to estimate the Bregman divergence relative to a base point $\base\in\cvx$ before and after a prox-step.

\begin{lemma}
\label{lem:update}
Fix some $\base\in\cvx$ and consider the recursive update rule
\begin{equation}
\label{eq:update}
\new\point
	= \prox{\point}{\dpoint}
%	\quad
%	\text{for $\point\in\subcvx$, $\dpoint\in\dpoints$}.
\end{equation}
for $\point\in\subcvx$, $\dpoint\in\dpoints$.
Then:
\begin{subequations}
\label{eq:update}
\begin{align}
\breg(\base,\new\point)
\label{eq:update-sharp}
	&\leq \breg(\base,\point)
	- \breg(\new\point,\point)
	+ \braket{\dpoint}{\new\point - \base}
	\\
\label{eq:update-strong}
	&\leq \breg(\base,\point)
	+ \braket{\dpoint}{\point - \base}
	+ \frac{1}{2\hstr} \dnorm{\dpoint}^{2}.
\end{align}
\end{subequations}
\end{lemma}

\begin{Proof}
By the definition \eqref{eq:proxmap} of $\proxmap$, we have $\dpoint + \nabla\hreg(\point) \in \subd\hreg(\new\point)$.
This means that $\new\point \in \dom\subd\hreg \equiv \subcvx$, so the three-point identity (\cref{lem:3point}) applies.
We thus get
\begin{equation}
\breg(\base,\point)
	= \breg(\base,\new\point)
	+ \breg(\new\point,\point)
	+ \braket{\nabla\hreg(\point) - \nabla \hreg(\new\point)}{\new\point - \base}
\end{equation}
or, after rearranging:
\begin{equation}
\label{eq:update-exact}
\breg(\base,\new\point)
	= \breg(\base,\point)
	- \breg(\new\point,\point)
	+ \braket{\nabla \hreg(\new\point) - \nabla \hreg(\point)}{\new\point - \base}.
\end{equation}
Since $\nabla \hreg(\point)+\dpoint \in \partial \hreg (\new\point)$, \cref{lem:subgrads} yields $\braket{\nabla\hreg(\new\point)}{\new\point - \base} \leq \braket{\dpoint + \nabla\hreg(\point)}{\new\point - \base}$, so \eqref{eq:update-sharp} follows by plugging this bound back to \eqref{eq:update-exact}.

For the second part of the lemma, first rewrite \eqref{eq:update-sharp} as
\begin{equation}
\label{eq:update-temp}
\breg(\base,\new\point)
	\leq \breg(\base,\point)
	+ \braket{\dpoint}{\point - \base}
	+ \braket{\dpoint}{\new\point-\point}
	- \breg(\new\point,\point).
\end{equation}
By Young's inequality, we also have
\begin{equation}
\label{eq:Young}
\braket{\dpoint}{\new\point-\point}
	\leq \frac{1}{2\hstr} \dnorm{\dpoint}^{2}
	+ \frac{\hstr}{2} \norm{\new\point - \point}^{2}
\end{equation}
so \eqref{eq:update-temp} becomes
\begin{align}
\breg(\base,\new\point)
	&\leq \breg(\base,\point)
	+ \braket{\dpoint}{\point - \base}
	+ \frac{1}{2\hstr} \dnorm{\dpoint}^{2}
	+ \frac{\hstr}{2} \norm{\new\point - \point}^{2}
	- \breg(\new\point,\point).
%	\notag\\
%	&\leq \breg(\base,\point)
%	+ \braket{\dpoint}{\point - \base}
%	+ \frac{1}{2\hstr} \dnorm{\dpoint}^{2},
\end{align}
Then, by the strong convexity of $\hreg$, we obtain $\breg(\new\point,\point) = \hreg(\new\point) - \hreg(\point) - \braket{\nabla\hreg(\point)}{\new\point - \point} \geq (\hstr/2) \norm{\new\point - \point}^{2}$, and our claim follows.
\end{Proof}

This basic lemma allows us to derive the following ``template inequality'' for processes of the general form \eqref{eq:update}:

\begin{lemma}
\label{lem:template}
Consider a sequence of dual vectors $\dstate_{\run}\in\dpoints$, $\run=\running$,
%defined over the interval $\run \in \runs = \window{\runstart}{\nRuns}$,
and let
\begin{equation}
\label{eq:state}
\state_{\run+1}
	= \prox{\state_{\run}}{\dstate_{\run}}
\end{equation}
with $\state_{\start} \in \subcvx$ initialized arbitrarily.
Then, for all $\point\in\cvx$ and every nonnegative sequence $\temp_{\run}\geq0$ defined over the window $\runs = \window{\runstart}{\runend}$, we have
\begin{equation}
\label{eq:template}
\sum_{\run\in\runs} \temp_{\run} \braket{\dstate_{\run}}{\point - \state_{\run}}
%	\leq \temp_{\start} \breg(\point,\state_{\start})
	\leq \sum_{\run\in\runs} (\temp_{\run} - \temp_{\run-1}) \breg(\point,\state_{\run})
	+ \frac{1}{2\hstr} \sum_{\run\in\runs} \temp_{\run} \dnorm{\dstate_{\run}}^{2},
\end{equation}
with the convention that $\temp_{\runstart-1} = 0$ in the above sum.
\end{lemma}

\begin{Proof}
Let $\breg_{\run} = \breg(\point,\state_{\run})$.
Then \eqref{eq:update-strong} readily yields
\begin{equation}
\breg_{\run+1}
	\leq \breg_{\run}
	+ \braket{\dstate_{\run}}{\state_{\run} - \point}
	+ \frac{1}{2\hstr} \dnorm{\dstate_{\run}}^{2}
\end{equation}
so, after multiplying by $\temp_{\run} \geq 0$ and rearranging, we get
\begin{equation}
\temp_{\run} \braket{\dstate_{\run}}{\point - \state_{\run}}
	\leq \temp_{\run} (\breg_{\run} - \breg_{\run+1})
	+ \frac{\temp_{\run}}{2\hstr} \dnorm{\dstate}^{2}.
\end{equation}
Therefore, by bringing $\braket{\dstate_{\run}}{\state_{\run} - \point}$ to the \acl{LHS} and summing over $\run\in\runs$, we get
\begin{align}
\sum_{\run\in\runs} \temp_{\run} \braket{\dstate_{\run}}{\point - \state_{\run}}
	&\leq \sum_{\run\in\runs} \temp_{\run} (\breg_{\run} - \breg_{\run+1})
	+ \frac{1}{2\hstr} \sum_{\run\in\runs} \temp_{\run} \dnorm{\dstate}^{2}
	\notag\\
	&= \sum_{\run\in\runs} (\temp_{\run}-\temp_{\run-1}) \breg_{\run}
	- \temp_{\runend} \breg_{\runend+1}
	+ \frac{1}{2\hstr} \sum_{\run\in\runs} \temp_{\run} \dnorm{\dstate}^{2}.
\end{align}
Since $\breg_{\runend+1}\geq0$, our claim follows.
\end{Proof}

Finally, we will make frequent use of the following straightforward result:

\begin{lemma}
\label{lem:nonsteep}
Suppose that $\hreg$ is Lipschitz.
Then $\sup_{\point\in\cvx,\pointalt\in\subcvx} \breg(\point,\pointalt) < \infty$.
\end{lemma}

\begin{Proof}
For all $\point\in\cvx$ and all $\pointalt\in\subcvx$, we have:
\begin{equation}
\breg(\point,\pointalt)
	= \hreg(\point) - \hreg(\pointalt) - \braket{\nabla\hreg(\pointalt)}{\point - \pointalt}
	\leq \hreg(\point) - \hreg(\pointalt) + \dnorm{\nabla\hreg(\pointalt)} \norm{\point - \pointalt}.
\end{equation}
By assumption, $L \equiv \sup_{\pointalt} \dnorm{\nabla\hreg(\pointalt)} < \infty$.
Hence, with $\cvx$ compact, we readily get
\begin{equation}
\breg(\point,\pointalt)
	\leq \hreg(\point) - \hreg(\pointalt) + L \diam(\cvx).
%	\leq \depth + L \diam(\cvx)
\end{equation}
Since $\hreg(\point) - \hreg(\pointalt) \leq \max\hreg - \min\hreg < \infty$, our assertion follows.
\end{Proof}

%% file: Thanks.tex
%----------------------------------------------------------------------
%%% THANKS
%----------------------------------------------------------------------
% !TEX root = ./Main.tex
%
%
This research was partially supported by the COST Action CA16228 ``European Network for Game Theory'' (GAMENET).
P.~Mertikopoulos is grateful for financial support by
the French National Research Agency (ANR) in the framework of
the ``Investissements d'avenir'' program (ANR-15-IDEX-02),
the LabEx PERSYVAL (ANR-11-LABX-0025-01),
MIAI@Grenoble Alpes (ANR-19-P3IA-0003),
and the grants ORACLESS (ANR-16-CE33-0004) and ALIAS (ANR-19-CE48-0018-01).

%% file: Main.bbl
\begin{thebibliography}{67}
\expandafter\ifx\csname natexlab\endcsname\relax\def\natexlab#1{#1}\fi
\expandafter\ifx\csname url\endcsname\relax
  \def\url#1{{\tt #1}}\fi
\expandafter\ifx\csname urlprefix\endcsname\relax\def\urlprefix{URL }\fi
\expandafter\ifx\csname urlstyle\endcsname\relax
  \expandafter\ifx\csname doi\endcsname\relax
  \def\doi#1{doi:\discretionary{}{}{}#1}\fi \else
  \expandafter\ifx\csname doi\endcsname\relax
  \def\doi{doi:\discretionary{}{}{}\begingroup \urlstyle{rm}\Url}\fi \fi

\bibitem[{Abernethy et~al.(2008)Abernethy, Bartlett, Rakhlin, and
  Tewari}]{ABRT08}
Abernethy, Jacob, Peter~L. Bartlett, Alexander Rakhlin, Ambuj Tewari. 2008.
\newblock Optimal strategies and minimax lower bounds for online convex games.
\newblock {\it COLT '08: Proceedings of the 21st Annual Conference on Learning
  Theory\/}.

\bibitem[{Arora et~al.(2012)Arora, Hazan, and Kale}]{AHK12}
Arora, Sanjeev, Elad Hazan, Satyen Kale. 2012.
\newblock The multiplicative weights update method: A meta-algorithm and
  applications.
\newblock {\it Theory of Computing\/} {\bf 8}(1) 121--164.

\bibitem[{Auer et~al.(1995)Auer, Cesa-Bianchi, Freund, and Schapire}]{ACBFS95}
Auer, Peter, Nicol{\`o} Cesa-Bianchi, Yoav Freund, Robert~E. Schapire. 1995.
\newblock Gambling in a rigged casino: The adversarial multi-armed bandit
  problem.
\newblock {\it Proceedings of the 36th Annual Symposium on Foundations of
  Computer Science\/}.

\bibitem[{Beck and Teboulle(2003)}]{BecTeb03}
Beck, Amir, Marc Teboulle. 2003.
\newblock Mirror descent and nonlinear projected subgradient methods for convex
  optimization.
\newblock {\it Operations Research Letters\/} {\bf 31}(3) 167--175.

\bibitem[{Beggs(2005)}]{Beg05}
Beggs, Alan~W. 2005.
\newblock On the convergence of reinforcement learning.
\newblock {\it Journal of Economic Theory\/} {\bf 122} 1--36.

\bibitem[{Bena{\"\i}m(1999)}]{Ben99}
Bena{\"\i}m, Michel. 1999.
\newblock Dynamics of stochastic approximation algorithms.
\newblock Jacques Az{\'e}ma, Michel {\'E}mery, Michel Ledoux, Marc Yor, eds.,
  {\it S{\'e}minaire de Probabilit{\'e}s XXXIII\/}, {\it Lecture Notes in
  Mathematics\/}, vol. 1709. Springer Berlin Heidelberg, 1--68.

\bibitem[{Bena{\"\i}m et~al.(2005)Bena{\"\i}m, Hofbauer, and Sorin}]{BHS05}
Bena{\"\i}m, Michel, Josef Hofbauer, Sylvain Sorin. 2005.
\newblock Stochastic approximations and differential inclusions.
\newblock {\it SIAM Journal on Control and Optimization\/} {\bf 44}(1)
  328--348.

\bibitem[{Benveniste et~al.(1990)Benveniste, M{\'e}tivier, and
  Priouret}]{BMP90}
Benveniste, Albert, Michel M{\'e}tivier, Pierre Priouret. 1990.
\newblock {\it Adaptive Algorithms and Stochastic Approximations\/}.
\newblock Springer.

\bibitem[{Bervoets et~al.(2020)Bervoets, Bravo, and Faure}]{BBF20}
Bervoets, Sebastian, Mario Bravo, Mathieu Faure. 2020.
\newblock Learning with minimal information in continuous games.
\newblock {\it Theoretical Economics\/} {\bf 15} 1471--1508.

\bibitem[{Besbes et~al.(2015)Besbes, Gur, and Zeevi}]{BGZ15}
Besbes, Omar, Yonatan Gur, Assaf Zeevi. 2015.
\newblock Non-stationary stochastic optimization.
\newblock {\it Operations Research\/} {\bf 63}(5) 1227--1244.

\bibitem[{Borkar(1997)}]{Bor97}
Borkar, Vivek~S. 1997.
\newblock Stochastic approximation with two time scales.
\newblock {\it Systems \& Conrol Letters\/} {\bf 29} 291--294.

\bibitem[{Bravo et~al.(2018)Bravo, Leslie, and Mertikopoulos}]{BLM18}
Bravo, Mario, David~S. Leslie, Panayotis Mertikopoulos. 2018.
\newblock Bandit learning in concave ${N}$-person games.
\newblock {\it NeurIPS '18: Proceedings of the 32nd International Conference of
  Neural Information Processing Systems\/}.

\bibitem[{Bubeck and Cesa-Bianchi(2012)}]{BCB12}
Bubeck, S{\'e}bastien, Nicol{\`o} Cesa-Bianchi. 2012.
\newblock Regret analysis of stochastic and nonstochastic multi-armed bandit
  problems.
\newblock {\it Foundations and Trends in Machine Learning\/} {\bf 5}(1) 1--122.

\bibitem[{Chen and Teboulle(1993)}]{CT93}
Chen, Gong, Marc Teboulle. 1993.
\newblock Convergence analysis of a proximal-like minimization algorithm using
  {Bregman} functions.
\newblock {\it SIAM Journal on Optimization\/} {\bf 3}(3) 538--543.

\bibitem[{Cominetti et~al.(2010)Cominetti, Melo, and Sorin}]{CMS10}
Cominetti, Roberto, Emerson Melo, Sylvain Sorin. 2010.
\newblock A payoff-based learning procedure and its application to traffic
  games.
\newblock {\it Games and Economic Behavior\/} {\bf 70}(1) 71--83.

\bibitem[{Coucheney et~al.(2015)Coucheney, Gaujal, and Mertikopoulos}]{CGM15}
Coucheney, Pierre, Bruno Gaujal, Panayotis Mertikopoulos. 2015.
\newblock Penalty-regulated dynamics and robust learning procedures in games.
\newblock {\it Mathematics of Operations Research\/} {\bf 40}(3) 611--633.

\bibitem[{Cui et~al.(2021)Cui, Franci, Grammatico, Shanbhag, and
  Staudigl}]{Cui:2021ud}
Cui, Shisheng, Barbara Franci, Sergio Grammatico, Uday~V Shanbhag, Mathias
  Staudigl. 2021.
\newblock A relaxed-inertial forward-backward-forward algorithm for stochastic
  generalized nash equilibrium seeking.
\newblock {\it arXiv preprint arXiv:2103.13115\/} .

\bibitem[{D'Oro et~al.(2015)D'Oro, Mertikopoulos, Moustakas, and
  Palazzo}]{DMMP15}
D'Oro, Salvatore, Panayotis Mertikopoulos, Aris~L. Moustakas, Sergio Palazzo.
  2015.
\newblock Interference-based pricing for opportunistic multi-carrier cognitive
  radio systems.
\newblock {\it {IEEE} Trans. Wireless Commun.\/} {\bf 14}(12) 6536--6549.

\bibitem[{Erev and Roth(1998)}]{ER98}
Erev, Ido, Alvin~E. Roth. 1998.
\newblock Predicting how people play games: Reinforcement learning in
  experimental games with unique, mixed strategy equilibria.
\newblock {\it American Economic Review\/} {\bf 88} 848--881.

\bibitem[{Facchinei and Kanzow(2007)}]{FK07}
Facchinei, Francisco, Christian Kanzow. 2007.
\newblock Generalized {Nash} equilibrium problems.
\newblock {\it 4OR\/} {\bf 5}(3) 173--210.

\bibitem[{Facchinei and Pang(2003)}]{FP03}
Facchinei, Francisco, Jong-Shi Pang. 2003.
\newblock {\it Finite-Dimensional Variational Inequalities and Complementarity
  Problems\/}.
\newblock Springer Series in Operations Research, Springer.

\bibitem[{Fang et~al.(2020)Fang, Harvey, Portella, and Friedlander}]{FHPF20}
Fang, Huang, Nick Harvey, Victor Portella, Michael Friedlander. 2020.
\newblock Online mirror descent and dual averaging: {Keeping} pace in the
  dynamic case.
\newblock {\it ICML '20: Proceedings of the 37th International Conference on
  Machine Learning\/}.

\bibitem[{Flaxman et~al.(2005)Flaxman, Kalai, and McMahan}]{FKM05}
Flaxman, Abraham~D., Adam~Tauman Kalai, H.~Brendan McMahan. 2005.
\newblock Online convex optimization in the bandit setting: gradient descent
  without a gradient.
\newblock {\it SODA '05: Proceedings of the 16th annual ACM-SIAM Symposium on
  Discrete Algorithms\/}. 385--394.

\bibitem[{Freund and Schapire(1999)}]{FS99}
Freund, Yoav, Robert~E. Schapire. 1999.
\newblock Adaptive game playing using multiplicative weights.
\newblock {\it Games and Economic Behavior\/} {\bf 29} 79--103.

\bibitem[{Hall and Heyde(1980)}]{HH80}
Hall, P., C.~C. Heyde. 1980.
\newblock {\it Martingale Limit Theory and Its Application\/}.
\newblock Probability and Mathematical Statistics, Academic Press, New York.

\bibitem[{Hart and Mas-Colell(2003)}]{HMC03}
Hart, Sergiu, Andreu Mas-Colell. 2003.
\newblock Uncoupled dynamics do not lead to {Nash} equilibrium.
\newblock {\it American Economic Review\/} {\bf 93}(5) 1830--1836.

\bibitem[{Hofbauer and Sandholm(2009)}]{HS09}
Hofbauer, Josef, William~H. Sandholm. 2009.
\newblock Stable games and their dynamics.
\newblock {\it Journal of Economic Theory\/} {\bf 144}(4) 1665--1693.

\bibitem[{Jadbabaie et~al.(2015)Jadbabaie, Rakhlin, Shahrampour, and
  Sridharan}]{JRSS15}
Jadbabaie, Ali, Alexander Rakhlin, Shahin Shahrampour, Karthik Sridharan. 2015.
\newblock Online optimization: Competing with dynamic comparators.
\newblock {\it AISTATS '15: Proceedings of the 18th International Conference on
  Artificial Intelligence and Statistics\/}.

\bibitem[{Juditsky et~al.(2011)Juditsky, Nemirovski, and Tauvel}]{JNT11}
Juditsky, Anatoli, Arkadi~Semen Nemirovski, Claire Tauvel. 2011.
\newblock Solving variational inequalities with stochastic mirror-prox
  algorithm.
\newblock {\it Stochastic Systems\/} {\bf 1}(1) 17--58.

\bibitem[{Jun et~al.(2017)Jun, Orabona, Wright, and Willett}]{JOWW17}
Jun, Kwang-Sung, Francesco Orabona, Stephen~J. Wright, Rebecca~M. Willett.
  2017.
\newblock Improved strongly adaptive online learning using coin betting.
\newblock {\it AISTATS '17: Proceedings of the 20th International Conference on
  Artificial Intelligence and Statistics\/}.

\bibitem[{Kelly et~al.(1998)Kelly, Maulloo, and Tan}]{KMT98}
Kelly, Frank~P., Aman~K. Maulloo, David K.~H. Tan. 1998.
\newblock Rate control for communication networks: shadow prices, proportional
  fairness and stability.
\newblock {\it Journal of the Operational Research Society\/} {\bf 49}(3)
  237--252.

\bibitem[{Kivinen and Warmuth(1997)}]{KW97}
Kivinen, Jyrki, Manfred~K. Warmuth. 1997.
\newblock Exponentiated gradient versus gradient descent for linear predictors.
\newblock {\it Information and Computation\/} {\bf 132}(1) 1--63.

\bibitem[{Krichene et~al.(2015)Krichene, Drigh{\`e}s, and Bayen}]{KDB15}
Krichene, Walid, Benjamin Drigh{\`e}s, Alexandre~M. Bayen. 2015.
\newblock Online learning of {Nash} equilibria in congestion games.
\newblock {\it SIAM Journal on Control and Optimization\/} {\bf 53}(2)
  1056--1081.

\bibitem[{Laraki et~al.(2019)Laraki, Renault, and Sorin}]{LRS19}
Laraki, Rida, J{\'e}r{\^o}me Renault, Sylvain Sorin. 2019.
\newblock {\it Mathematical Foundations of Game Theory\/}.
\newblock Universitext, Springer.

\bibitem[{Leslie and Collins(2003)}]{LC03}
Leslie, David~S., E.~J. Collins. 2003.
\newblock Convergent multiple-timescales reinforcement learning algorithms in
  normal form games.
\newblock {\it The Annals of Applied Probability\/} {\bf 13}(4) 1231--1251.

\bibitem[{Leslie and Collins(2005)}]{LC05}
Leslie, David~S., E.~J. Collins. 2005.
\newblock Individual {$Q$}-learning in normal form games.
\newblock {\it SIAM Journal on Control and Optimization\/} {\bf 44}(2)
  495--514.

\bibitem[{Littlestone and Warmuth(1994)}]{LW94}
Littlestone, Nick, Manfred~K. Warmuth. 1994.
\newblock The weighted majority algorithm.
\newblock {\it Information and Computation\/} {\bf 108}(2) 212--261.

\bibitem[{Mertikopoulos and Moustakas(2016)}]{MM16}
Mertikopoulos, Panayotis, Aris~L. Moustakas. 2016.
\newblock Learning in an uncertain world: {MIMO} covariance matrix optimization
  with imperfect feedback.
\newblock {\it {IEEE} Trans. Signal Process.\/} {\bf 64}(1) 5--18.

\bibitem[{Mertikopoulos and Zhou(2019)}]{MZ19}
Mertikopoulos, Panayotis, Zhengyuan Zhou. 2019.
\newblock Learning in games with continuous action sets and unknown payoff
  functions.
\newblock {\it Mathematical Programming\/} {\bf 173}(1-2) 465--507.

\bibitem[{Monderer and Shapley(1996)}]{MS96}
Monderer, Dov, Lloyd~S. Shapley. 1996.
\newblock Potential games.
\newblock {\it Games and Economic Behavior\/} {\bf 14}(1) 124 -- 143.

\bibitem[{Nemirovski et~al.(2009)Nemirovski, Juditsky, Lan, and
  Shapiro}]{NJLS09}
Nemirovski, Arkadi~Semen, Anatoli Juditsky, Guanghui Lan, Alexander Shapiro.
  2009.
\newblock Robust stochastic approximation approach to stochastic programming.
\newblock {\it SIAM Journal on Optimization\/} {\bf 19}(4) 1574--1609.

\bibitem[{Nemirovski et~al.(2010)Nemirovski, Onn, and Rothblum}]{NOR10}
Nemirovski, Arkadi~Semen, Shmuel Onn, Uriel~G. Rothblum. 2010.
\newblock Accuracy certificates for computational problems with convex
  structure.
\newblock {\it Mathematics of Operations Research\/} {\bf 35}(1) 52--78.

\bibitem[{Nemirovski and Yudin(1983)}]{NY83}
Nemirovski, Arkadi~Semen, David~Berkovich Yudin. 1983.
\newblock {\it Problem Complexity and Method Efficiency in Optimization\/}.
\newblock Wiley, New York, NY.

\bibitem[{Nesterov(2009)}]{Nes09}
Nesterov, Yurii. 2009.
\newblock Primal-dual subgradient methods for convex problems.
\newblock {\it Mathematical Programming\/} {\bf 120}(1) 221--259.

\bibitem[{Orabona and P{\'a}l(2018)}]{OP18}
Orabona, Francesco, D{\'a}vid P{\'a}l. 2018.
\newblock Scale-free online learning.
\newblock {\it Theoretical Computer Science\/} {\bf 716} 50--69.

\bibitem[{Orda et~al.(1993)Orda, Rom, and Shimkin}]{ORShi93}
Orda, Ariel, Raphael Rom, Nahum Shimkin. 1993.
\newblock Competitive routing in multi-user communication networks.
\newblock {\it {IEEE/ACM} Trans. Netw.\/} {\bf 1}(5) 614--627.

\bibitem[{Perkins and Leslie(2014)}]{PL14}
Perkins, Steven, David~S. Leslie. 2014.
\newblock Stochastic fictitious play with continuous action sets.
\newblock {\it Journal of Economic Theory\/} {\bf 152} 179--213.

\bibitem[{Rakhlin and Sridharan(2013)}]{RS13-NIPS}
Rakhlin, Alexander, Karthik Sridharan. 2013.
\newblock Optimization, learning, and games with predictable sequences.
\newblock {\it NIPS '13: Proceedings of the 27th International Conference on
  Neural Information Processing Systems\/}.

\bibitem[{Ravat and Shanbhag(2011)}]{RavSha11}
Ravat, U., U.~Shanbhag. 2011.
\newblock On the characterization of solution sets of smooth and nonsmooth
  convex stochastic nash games.
\newblock {\it SIAM Journal on Optimization\/} {\bf 21}(3) 1168--1199.
\newblock \doi{10.1137/100792644}.
\newblock \urlprefix\url{https://doi.org/10.1137/100792644}.

\bibitem[{Rosen(1965)}]{Ros65}
Rosen, J.~B. 1965.
\newblock Existence and uniqueness of equilibrium points for concave
  ${N}$-person games.
\newblock {\it Econometrica\/} {\bf 33}(3) 520--534.

\bibitem[{Sandholm(2015)}]{San15}
Sandholm, William~H. 2015.
\newblock Population games and deterministic evolutionary dynamics.
\newblock H.~Peyton Young, Shmuel Zamir, eds., {\it Handbook of Game Theory
  {IV}\/}. Elsevier, 703--778.

\bibitem[{Scutari et~al.(2010)Scutari, Facchinei, Palomar, and Pang}]{SFPP10}
Scutari, Gesualdo, Francisco Facchinei, Daniel~P{\'e}rez Palomar, Jong-Shi
  Pang. 2010.
\newblock Convex optimization, game theory, and variational inequality theory
  in multiuser communication systems.
\newblock {\it {IEEE} Signal Process. Mag.\/} {\bf 27}(3) 35--49.

\bibitem[{Shahrampour and Jadbabaie(2018)}]{SJ18}
Shahrampour, Shahin, Ali Jadbabaie. 2018.
\newblock Distributed online optimization in dynamic environments using mirror
  descent.
\newblock {\it {IEEE} Trans. Autom. Control\/} {\bf 63}(3) 714--725.

\bibitem[{Shalev-Shwartz(2011)}]{SS11}
Shalev-Shwartz, Shai. 2011.
\newblock Online learning and online convex optimization.
\newblock {\it Foundations and Trends in Machine Learning\/} {\bf 4}(2)
  107--194.

\bibitem[{Shalev-Shwartz and Singer(2006)}]{SSS06}
Shalev-Shwartz, Shai, Yoram Singer. 2006.
\newblock Convex repeated games and {Fenchel} duality.
\newblock {\it NIPS' 06: Proceedings of the 19th Annual Conference on Neural
  Information Processing Systems\/}. MIT Press, 1265--1272.

\bibitem[{Shannon and Weaver(1949)}]{SW49}
Shannon, Claude~E., Warren Weaver. 1949.
\newblock {\it The Mathematical Theory of Communication\/}.
\newblock University of Illinois Press.

\bibitem[{Sorin and Wan(2016)}]{SW16}
Sorin, Sylvain, Cheng Wan. 2016.
\newblock Finite composite games: Equilibria and dynamics.
\newblock {\it Journal of Dynamics and Games\/} {\bf 3}(1) 101--120.

\bibitem[{Spall(1992)}]{Spa92}
Spall, James~C. 1992.
\newblock Multivariate stochastic approximation using a simultaneous
  perturbation gradient approximation.
\newblock {\it {IEEE} Trans. Autom. Control\/} {\bf 37}(3) 332--341.

\bibitem[{Spall(1997)}]{Spa97}
Spall, James~C. 1997.
\newblock A one-measurement form of simultaneous perturbation stochastic
  approximation.
\newblock {\it Automatica\/} {\bf 33}(1) 109--112.

\bibitem[{Syrgkanis et~al.(2015)Syrgkanis, Agarwal, Luo, and Schapire}]{SALS15}
Syrgkanis, Vasilis, Alekh Agarwal, Haipeng Luo, Robert~E. Schapire. 2015.
\newblock Fast convergence of regularized learning in games.
\newblock {\it NIPS '15: Proceedings of the 29th International Conference on
  Neural Information Processing Systems\/}. 2989--2997.

\bibitem[{Tatarenko and Kamgarpour(2019{\natexlab{a}})}]{TK19}
Tatarenko, Tatiana, Maryam Kamgarpour. 2019{\natexlab{a}}.
\newblock Learning generalized {Nash} equilibria in a class of convex games.
\newblock {\it {IEEE} Trans. Autom. Control\/} {\bf 64}(4) 1426--1439.

\bibitem[{Tatarenko and Kamgarpour(2019{\natexlab{b}})}]{TK19-CDC}
Tatarenko, Tatiana, Maryam Kamgarpour. 2019{\natexlab{b}}.
\newblock Learning {Nash} equilibria in monotone games.
\newblock {\it CDC '19: Proceedings of the 58th IEEE Annual Conference on
  Decision and Control\/}.
\newblock \doi{10.1109/CDC40024.2019.9029659}.

\bibitem[{Teboulle(2018)}]{Teb18}
Teboulle, Marc. 2018.
\newblock A simplified view of first order methods for optimization.
\newblock {\it Mathematical Programming\/} {\bf 170} 67--96.

\bibitem[{Tse and Viswanath(2005)}]{TV05}
Tse, David, Pramod Viswanath. 2005.
\newblock {\it Fundamentals of Wireless Communication\/}.
\newblock Cambridge University Press, Cambridge, UK.

\bibitem[{Tullock(1980)}]{Tul80}
Tullock, Gordon. 1980.
\newblock Efficient rent seeking.
\newblock J.~M. Buchanan R.~D. Tollison, Gordon Tullock, eds., {\it Toward a
  theory of the rent-seeking society\/}. Texas A\&M University Press.

\bibitem[{Viossat and Zapechelnyuk(2013)}]{VZ13}
Viossat, Yannick, Andriy Zapechelnyuk. 2013.
\newblock No-regret dynamics and fictitious play.
\newblock {\it Journal of Economic Theory\/} {\bf 148}(2) 825--842.

\bibitem[{Vovk(1990)}]{Vov90}
Vovk, Vladimir~G. 1990.
\newblock Aggregating strategies.
\newblock {\it COLT '90: Proceedings of the 3rd Workshop on Computational
  Learning Theory\/}. 371--383.

\end{thebibliography}
